\documentclass[12pt,a4paper,notitlepage]{article}

\usepackage[T1]{fontenc}
\usepackage[latin1]{inputenc}
\usepackage{lmodern}
\usepackage{indentfirst}
\usepackage{nonumonpart}
\usepackage{amsmath}
\usepackage{amssymb}
\usepackage{mathrsfs}
\usepackage{amsthm}
\usepackage{bm}
\usepackage[bf]{caption}
\usepackage{dcolumn}
\usepackage{booktabs}
\usepackage{pgfplots}
\usepackage{enumerate}
\usepackage{subfig}
\usepackage{tikz}
\usepackage{syntonly}
\usepackage{listings}
\usepackage{bookmark}
\usepackage{tensor}
\usepackage{tabularx}
\usepackage{frontespizio}
\usepackage{graphicx}
\usepackage{emptypage}
\usepackage{float}
\usetikzlibrary{angles,arrows.meta,intersections,patterns,plotmarks,quotes,shapes}
\pgfplotsset{/pgf/number format/use comma,compat=newest,legend style={font=\scriptsize}}
\usepackage{hyperref}
\usepackage[a4paper,top=3cm,bottom=3cm,left=2.2cm,right=2.2cm]{geometry}
\usepackage{cite}

\theoremstyle{definition}

\newtheorem{lemma}{Lemma}

\theoremstyle{plain}
\newtheorem{theorem}{Theorem}

\begin{document}
\title{The First Law of Black Hole Mechanics}
\author{Lorenzo Rossi\footnote{Please, send comments and corrections to lr437@cantab.net \,.}\\
\small Based on the essay written during Part III of the Mathematical Tripos\\
\small Department of Applied Mathematics and Theoretical Physics (DAMTP)\\
\small University of Cambridge\\
\small Wilberforce Road, Cambridge CB3 0WA, United Kingdom}
\date{May 5 2017}

\maketitle
\thispagestyle{empty}
    \begin{abstract}
The first law of black hole mechanics has been the main motivation for investigating thermodynamic properties of black holes. The first version of this law was proved in \cite{Bardeen:1973gs} by considering perturbations of 
an asymptotically flat, stationary black hole spacetime to other stationary black hole spacetimes. This result was then extended to fully general perturbations, first in the context of Einstein-Maxwell theory in \cite{Sudarsky:1992ty},\cite{Wald:1993ki}, and then in the context of a general diffeomorphism invariant theory of gravity with an arbitrary number of matter fields in \cite{Wald:1993nt},\cite{Iyer:1994ys}. Here a review of these two generalizations of the first law is presented, with particular attention to outlining the necessary formalisms and calculations in an explicit and thorough way, understandable at a graduate level. The open problem of defining the entropy for a dynamical black hole that satisfies a form of the second law of black hole mechanics is briefly discussed.
\end{abstract}

	\cleardoublepage

\setcounter{page}{1}

\tableofcontents

\section{Introduction}
\label{sec:intro}

In 1973 Bardeen, Carter and Hawking \cite{Bardeen:1973gs} proposed three laws of black hole mechanics which had an evident resemblance with the laws of thermodynamics. This suggested that black holes are thermodynamic objects which emit radiation, although they are classically (i.e. non-quantum mechanically) defined as regions of spacetimes from which nothing can escape. A semi-classical study (i.e. a calculation that does not consider the interaction of the gravitational field with itself) of quantum field theory in a curved spacetimes with a black hole region \cite{Hawking:1974sw} (see \cite{Reall:2017bhnotes} for a review) showed that these thermodynamic properties of black holes arise when we take into account also their quantum behaviour. This discovery is the main reason to believe that understanding the quantum physics of black holes may be the key step to obtain a complete theory of quantum gravity.

A discussion about the zeroth, first and second law of black hole mechanics can be found, e.g., in \cite{Reall:2017bhnotes}. We collect here the main ideas. The zeroth law ensures that, under certain hypotheses, the surface gravity of a stationary black hole is constant over the future event horizon.
The first law of black hole mechanics is a formula relating the variations of the energy $\mathcal{E}$, angular momentum $\mathcal{J}$ and entropy $S[B]$ of a stationary black hole spacetime, with bifurcate Killing horizon and bifurcation surface $B$\footnote{The Killing horizon is composed of 4 different parts, defined in appendix \ref{app:binsur}. Except in this appendix, we will not specify which part we are referring to, assuming that this is clear from the context. In most cases, we will be mostly interested in physical quantities and physical phenomena on the portion $\mathcal{H}^+_R$, i.e. the portion of the future event horizon in the causal future of $B$.}, induced by a perturbation of the metric and matter fields. This is usually written in the form
\begin{equation}
\label{eq:intro}
\frac{\kappa}{2\pi}\delta S[B]=\delta\mathcal{E}-\Omega\delta\mathcal{J},
\end{equation}
where $\Omega$ is interpreted as the angular velocity of the black hole. There exist different versions of this law. The original one was given in \cite{Bardeen:1973gs}, and it is proved by considering $\delta S[B],\delta\mathcal{E}$ and $\delta\mathcal{J}$ as variations induced by a variation of the metric (and of the matter fields, if there are any) of the stationary, axisymmetric black hole to the metric (and matter fields) of another stationary, axisymmetric black hole. (To be precise, there exists an earlier version, proved by Hartle and Hawking in \cite{Hawking:1972hy}, whose proof takes into consideration the perturbation of the metric and matter fields (if there are any) of a stationary black hole induced by the flow inside the black hole of a small amount of matter with energy $\delta\mathcal{E}$ and angular momentum $\delta\mathcal{J}$. Eventually, the black hole settles down to a new stationary state. For this proof, \eqref{eq:intro} holds but with $\delta\mathcal{E}$ and $\delta\mathcal{J}$ given by the energy and angular momentum of the incoming matter, not the variation of the energy and angular momentum of the spacetime.) 

A second, more general, version is proved by Sudarski and Wald in \cite{Sudarsky:1992ty},\cite{Wald:1993ki} for perturbations of an Einstein-Maxwell stationary black hole to an arbitrary (not necessarily stationary) spacetime (in \cite{Sudarsky:1992ty} the more general case of Einstein-Yang-Mills theory is considered) and it will be presented in detail in the first part of this essay \ref{sec:first1}. To this end, in section \ref{sec:Hamfor}, the initial value problem of Einstein-Maxwell theory is briefly summarised and the Hamiltonian formulation of Einstein-Maxwell theory is developed. This leads, in particular, to the definition of the ``true'' Hamiltonian of the theory. In section \ref{sec:enchan}, the true Hamiltonian is used to define the canonical energy and canonical angular momentum of a spacetime. The variations of these quantities are related to the variation of the black hole entropy by the first law of black hole mechanics for Einstein-Maxwell stationary black holes with bifurcate Killing horizon. To conclude the first part, in section \ref{sec:appfirst} we review an application of the first law that shows that Einstein-Maxwell stationary black holes are either static or axisymmetric~\cite{Wald:1993ki}. This result closed a gap in the black holes uniqueness theorems.

The strong resemblance of equation \eqref{eq:intro} with the first law of thermodynamics, $dE=T dS+\mu dJ$ (where $E$ is the internal energy of the thermodynamical system, $T$ is its temperature, $S$ its entropy, $J$ its angular momentum, and $\mu$ is the chemical potential) suggests, as anticipated above, that the black hole has a temperature proportional to $\kappa$ (which, in fact, has been confirmed in \cite{Hawking:1974sw}). If we accept this fact, the zeroth law of black hole mechanics tells us that black holes satisfy the zeroth law of thermodynamics: the temperature of a stationary black hole (i.e. in thermodynamic equilibrium) is constant. Furthermore, this picture is reinforced if we consider also the second law of thermodynamics. In General Relativity, the entropy of a black hole is defined on any cross-section $C$ (see appendix \ref{app:binsur} for the definition) of the event horizon as $S[C]=\frac{A[C]}{4}$, where $A[C]$ is the area of $C$, i.e. the area of the black hole at the ``time'' when its ``spatial volume'' is contained within $C$. The area theorem ensures that, under certain hypotheses, the area of two cross-sections at two different ``times'' is non-decreasing. This shows that black hole spacetimes satisfy the second law of thermodynamics: the entropy of a black hole is non-decreasing. Both this result and the area theorem are called the second law of black hole mechanics.

So far, we stated results valid in General Relativity (both in vacuum and in the presence of suitable matter fields, such as the Maxwell field), since this has been the most successful theory to describe gravitational phenomena as well as the first environment where such problems were initially investigated. However, General Relativity is only one of many possible theories of gravity. Any theory whose physical results do not change if we substitute the manifold, the metric and the matter fields with their image under an arbitrary diffeomorphism is a potential candidate for a sensible model of gravitational interactions. These theories have an arbitrary number of covariant derivatives of the metric and the matter fields in the Lagrangian. In addition to being of interest in their own right, these theories may be physically relevant when we investigate the physics of gravitational phenomena at high energies (close to the Planck scale). In fact, due to the undesired non-renormalisability of General Relativity, many attempts have been made at showing that General Relativity is an effective theory, i.e. a low-energy approximation, of a more complicated renormalisable diffeomorphism invariant theory (see, e.g., \cite{tHooft:1974toh}). In other words, General Relativity would be a low-energy approximation of another theory of gravity and corrections to General Relativity would be more and more important as the energy scale of the physical processes increases. In support of this hypothesis, it can be shown that, in string theory, the Einstein-Hilbert action is just the first term in an infinite series of gravitational corrections built from powers of the curvature tensor and its derivatives. Also, in \cite{Stelle:1976gc} it was shown that, if we add all possible quadratic curvature scalars to the Einstein-Hilbert Lagrangian, we obtain a renormalisable diffeomorphism invariant theory. It is thereby clear that being able to show that the three laws of black hole mechanics hold (under certain hypotheses) for black hole solutions predicted by a certain class of theories of gravity would be an important confirmation for the assumption that the complete theory of gravity is a diffeomorphism invariant generalisation of General Relativity. Moreover, we could also identify the most general theory in this hypothetical class as a good candidate for a complete theory of quantum gravity. As we will argue in the second part of this essay, it turns out that this is possible for the zeroth law and for the first law, but it is still unknown what class of theories of gravity satisfies also the second law.

In the second part \ref{sec:20}, we will discuss the proof of the first law for diffeomorphism invariant theories by Iyer and Wald, by closely following the original steps of \cite{Iyer:1994ys} that improve the result of \cite{Wald:1993nt}.
We will first need to introduce some useful results about the covariant phase space formulation~\cite{Lee:1990nz}, which will be done in section \ref{sec:2}. Then, in section \ref{sec:charges}, we will see how a conserved Noether charge can be associated with every diffeomorphism generated by a vector field, and how this definition reduces to familiar results in General Relativity. In section \ref{sec:hamenang}, we will define the Hamiltonian of such theories and, by analogy with what we did in section~\ref{sec:enchan}, we will use it to define the canonical energy and canonical angular momentum. In section \ref{sec:entfirst}, we define the entropy of a stationary black hole, explore some of its properties, and prove the first law of black hole mechanics for diffeomorphism invariant theories.

In the last part of the essay \ref{sec:second}, using the analogous discussion in \cite{Iyer:1994ys} as a starting point, we will spend a few words on attempts to find a definition of entropy of a dynamical, i.e. non-stationary black hole, that satisfies the second law of black hole mechanics. This is still an open problem which will need to be addressed in the future.

\section{The first law of black hole mechanics for Einstein-Maxwell theory}
\label{sec:first1}

In this first part of the essay, we wish to define the concept of energy, electric charge, and angular momentum for a solution of an asymptotically flat solution of the Einstein-Maxwell equations of motion, and review in detail the proof of the first law of black hole mechanics originally presented by Sudarski and Wald in \cite{Sudarsky:1992ty} and \cite{Wald:1993ki}. 

\textbf{Notation and conventions} Let us explain the notation and conventions used in the first part of this work. To make contact with the notation of \cite{Sudarsky:1992ty}, $\mu,\nu,\rho,\sigma,\dots$ will be abstract indices which denote abstract tensors on the spacetime manifold; $\alpha,\beta,\gamma,\delta,\dots$ will indicate tensor components in some coordinate basis;
$a,b,c,d,\dots$ will be abstract indices for tensors defined on a hypersurface; $i,j,k,l,\dots$ will indicate the components of tensors defined on a hypersurface in some coordinate basis.

We will consider only orientable and time-orientable spacetimes. We will also assume that the spacetimes are globally hyperbolic, i.e. they can be completely determined by initial data on a surface, called a Cauchy surface (see \cite{Wald:1984rg} for a rigorous definition). Let $M$ be the spacetime manifold and $g$ be the Lorentzian metric on $M$.
We will always assume that a suitable definition of asymptotic flatness can be given, so that there exists (at least one set of) \emph{asymptotically inertial coordinates}, i.e. a coordinate system $x^\alpha$ on $M$ such that
\begin{align}
g_{\alpha\beta}&=\eta_{\alpha\beta}+\mathcal{O}\biggl(\frac{1}{r}\biggr),\label{eq:asyfla1}\\
\partial_\gamma g_{\alpha\beta}&=\mathcal{O}\biggl(\frac{1}{r^{2}}\biggr),\label{eq:asyfla2}
\end{align}
in the limit $r\equiv\sqrt{x^i x^i}\to\infty$,
where $\eta_{\alpha\beta}=$diag$(-1,1,1,1)$. These conditions are imposed for physical reasons.  Here, we are restricting ourselves to 4-dimensional spacetimes, but the proof of the first law of black hole mechanics that we will present can be obviously extended to an $n$-dimensional spacetime once we generalised the notions of ``asymptotically flat spacetime'' and ``asymptotically flat end'' (defined in section \ref{sec:Hamfor}).

We will work with Einstein-Maxwell theory in 4-dimensions. This is the theory of a 4-dimensional spacetime $(M,g_{\mu\nu})$ in the presence of an electromagnetic field described by a 1-form $A_\mu$, called the Maxwell field. The equations of motion for Einstein-Maxwell theory can be derived from the action
\begin{equation}
\label{eq:SEM}
S_{EM}=\frac{1}{16\pi}\int_{M}{\epsilon(R-F_{\mu\nu}F^{\mu\nu})},
\end{equation}
where $\epsilon$ is the volume form of the spacetime $(M,g_{\mu\nu})$ (i.e., chosen an orientation for $M$, in any right-handed (RH) chart $x^\alpha=(x^0,x^1,\dots,x^{n-1})$ we have $\epsilon=\sqrt{-g}dx^0\wedge dx^1\wedge\dots\wedge dx^{n-1}$, where $g$ is the determinant of the matrix $g_{\alpha\beta}$ in coordinates $x^\alpha$),  $R$ is the Ricci scalar built from the Levi-Civita connection associated with $g_{\mu\nu}$ and $F_{\mu\nu}=(dA)_{\mu\nu}=2\nabla_{[\mu}A_{\nu]}$ is the field strength tensor of the electromagnetic field. The equations of motion are the Einsten equation
\begin{equation}
\label{eq:Eeq}
R_{\mu\nu}-\frac{1}{2}Rg_{\mu\nu}=2\biggl(F_\mu^{\;\;\rho}F_{\nu\rho}-\frac{1}{4}g_{\mu\nu} F^{\rho\sigma}F_{\rho\sigma}\biggr)
\end{equation}
and the Maxwell equations
\begin{equation}
\label{eq:Meq}
\nabla^\nu F_{\mu\nu}=0 \quad\quad\quad \nabla_{[\mu}F_{\nu\rho]}=0.
\end{equation}
This theory has an ``electromagnetic gauge freedom'', in the sense that we are free to replace $A_\mu$ with $A_\mu+(d\Lambda)_\mu$, where $\Lambda$ is a smooth function on $M$, without changing $F_{\mu\nu}$ and so the physical results of the theory. Furthermore, the theory has the usual ``gauge freedom'' of General Relativity, i.e. diffeomorphism invariance.

We define stationary spacetimes as those spacetimes with a Killing vector field, $k$, timelike at infinity. We also require that it is possible to find an electromagnetic gauge such that $k$ is also a symmetry of $A_\mu$, i.e. $\mathcal{L}_k A_\mu=0$. The spacetime is said to be static if $k$ is orthogonal to a family of hypersurfaces. We define a stationary and axisymmetric spacetime with stationary Killing field $k$ and axial Killing field $m$ in the usual way (see, e.g., \cite{Reall:2017bhnotes}), but we also require that there exists an electromagnetic gauge such that $\mathcal{L}_k A_\mu=\mathcal{L}_m A_\mu=0$.

\subsection{Hamiltonian formulation of Einstein-Maxwell theory}
\label{sec:Hamfor}

In this section, we will review the Hamiltonian formulation of Einstein-Maxwell theory. Many results are analogous to the Hamiltonian formulation of Einstein theory, which is presented, e.g., in \cite{Wald:1984rg} and \cite{Reall:2017bhnotes}, so they will not be reproduced here in detail and familiarity with the concepts of the initial value problem and the Hamiltonian formulation will be assumed. We will discuss more thoroughly the differences arising from the presence of the Maxwell field $A_\mu$ in the theory.

We assume that the spacetime is globally hyperbolic, hence it is always possible \cite{Wald:1984rg} to find a global time function $t$: a map $t\colon M\to \mathbb{R}$ such that \begin{enumerate}[(i)] \item $-(dt)^\mu$ (normal to surfaces of constant $t$) is timelike and future-directed, \item surfaces of constant $t$ are Cauchy surfaces, denoted by $\Sigma_t$. \end{enumerate} Since Cauchy surfaces are homeomorphic to each other \cite{Wald:1984rg}, then all the Cauchy surfaces have the same topological structure $\Sigma_t$. So $M$ can be foliated by Cauchy surfaces, parameterised by $t$, and the topology of $M$ is $\mathbb{R}\times \Sigma_t$.

Let $n^\mu$ be the past-directed unit normal vector field to the hypersurfaces $\Sigma_t$ (i.e. the one used in the divergence theorem). The spacetime metric, $g_{\mu\nu}$, induces a metric $h_{\mu\nu}$ on each $\Sigma_t$ by the formula
\begin{equation}
\label{eq:h}
h_{\mu\nu}=g_{\mu\nu}+n_\mu n_\nu.
\end{equation}
$h_{\mu\nu}$ is invariant under the projection $h^\mu_\nu$ onto $\Sigma_t$, therefore it can be identified with the tensor $h_{ab}$ defined on $\Sigma_t$, the Riemannian metric induced on $\Sigma_t$ by $g_{\mu\nu}$, under the natural isomorphism defined in \cite{Hawking:1973uf} and denoted by $\tilde{\theta}^\ast$ (where $\theta$ is the inclusion map that embeds $\Sigma_t$ into $M$). Let $t^\mu$ be the vector field on $M$ dual of $dt$, i.e. $t^\mu=\big(\frac{\partial}{\partial t}\big)^\mu$. This is called \emph{time evolution vector field} (despite its name, it is not necessarily timelike; this property strictly depends on the foliation chosen). We decompose $t^\mu$ into its parts normal and tangential to $\Sigma_t$ and this decomposition defines the \emph{lapse function}, $N$, and the \emph{shift vector}, $N^\mu$, of $t^\mu$ by\footnote{The lapse and shift function are defined as in \cite{Sudarsky:1992ty} and \cite{Wald:1993ki}; their definition is also the same as \cite{Wald:1984rg}, where the future-directed normal to $\Sigma_t$ is used.}
\begin{align}
N&= t^\mu n_\mu,\\
N^\mu&= h^\mu_\nu t^\nu.
\end{align}
$N^\mu$ is invariant under projection onto each $\Sigma_t$ so it can be identified with a vector defined on $\Sigma_t$, $N^a$, whose index is lowered with $h_{ab}$. More precisely, $N^a$ is the vector whose push-forward on $M$ is $N^\mu$. Using these definitions, we easily obtain $-Nn^\mu+N^\mu=t^\mu$, so
\begin{equation}
\label{eq:nvec}
n^\mu=-\frac{1}{N}(t^\mu-N^\mu),
\end{equation}
hence the inverse spacetime metric can be written as
\begin{equation}
\label{eq:ginv}
g^{\mu\nu}=h^{\mu\nu}-n^\mu n^\nu=h^{\mu\nu}-\frac{1}{N^2}(t^\mu-N^\mu)(t^\nu-N^\nu).
\end{equation}

We can define coordinates at each point $p\in M$ as follows. Consider the Cauchy surface at $t=0$, $\Sigma$, and define coordinates $\{x^i\}, i=1,\dots,n-1$ on this surface. Now consider the integral curve $\gamma(t)$ of $t^\mu$ through $p$. This curve intersects $\Sigma$ at the point $q$ with $t=0$. We define $\{x^i(p)\}$ as the value of the coordinates $x^i$ at $q\in\Sigma$. Finally, we use $t$ as the $x^0$ coordinate. In this chart, integral curves of $t^\mu$ have constant $x^i$. Let us consider the components of the above defined quantities in coordinates $x^\alpha=(t,x^1,\dots,x^{n-1})$. We have $t^\alpha=\big(\frac{\partial}{\partial t}\big)^\alpha=\delta^\alpha_0$ and $n_\alpha=\frac{(dt)_\alpha}{\sqrt{-g^{\beta\gamma}(dt)_\beta (dt)_\gamma}}=\frac{\delta^0_\alpha}{\sqrt{-g^{00}}}$ (note that $g^{00}<0$ because $dt$ is timelike by construction). We denote the components of the metric $h_{ab}$ on Cauchy surfaces of constant $t$ by $h_{ij}(t,x)$, the lapse function by $N(t,x)=t^\alpha n_\alpha=1/\sqrt{-g^{00}}$ and the components of the shift vector $N^a$ by $N^i(t,x)$. In these coordinates we recover the 3+1 form (also called \emph{ADM decomposition}) of the spacetime metric
\begin{equation}
\label{eq:met}
g=-(N^2-h_{ij}N^iN^j)dt^2+2h_{ij}N^jdtdx^i+h_{ij}dx^idx^j.
\end{equation}
From this form of the metric we see that the information contained in $g_{\mu\nu}$ is the same as the information contained in $(N,N^a,h_{ab})$.
The components of $n^\mu$ and $g^{\mu\nu}$ can be easily obtained from \eqref{eq:nvec} and \eqref{eq:ginv}. In particular, we have $g^{00}=-\frac{1}{N^2}$, and, using the formula $g^{00}=\vartriangle^{00}/g$ where $\vartriangle^{00}=\det h_{ij}$ is the cofactor of the element $g_{00}$ of the matrix $g_{\mu\nu}$, we find 
\begin{equation}
\label{eq:detg}
\sqrt{-g}=N\sqrt{h},
\end{equation}
 where $h=\det h_{ij}$.

We also define the \emph{extrinsic curvature} of each $\Sigma_t$ as $K_{\mu\nu}=-h_\mu^\rho h_\nu^\sigma \nabla_\rho n_\sigma=-\frac{1}{2}\mathcal{L}_n h_{\mu\nu}$\footnote{This definition of extrinsic curvature is the same as the one given in eq. (10.2.13) of \cite{Wald:1984rg}, where the future-directed unit normal is used.}. This can be identified with the tensor $K_{ab}$ on $\Sigma_t$ by the natural isomorphism $\tilde{\theta}^\ast$.

In \cite{Wald:1984rg} it is shown that, once we choose a ``time foliation'' (i.e. a global time function, and so a time evolution vector field), the ``Einstein-Hilbert'' part of the action \eqref{eq:SEM} can be written, dropping boundary terms, as
\begin{equation}
\label{eq:SE}
S_{EH}=\frac{1}{16\pi}\int_{M}{\epsilon R}=\frac{1}{16\pi}\int_{\mathbb{R}\times \Sigma_t}{\epsilon [^{(3)}R+K_{\mu\nu}K^{\mu\nu}-K^2]},
\end{equation}
where $^{(3)}R$ is the Ricci scalar of $\Sigma_t$ built with the Levi-Civita connection of $h_{\mu\nu}$, and
\begin{equation}
K_{\mu\nu}=\frac{1}{2N}[\dot{h}_{\mu\nu}-D_\mu N_\nu-D_\nu N_\mu],
\end{equation}
where $\dot{h}_{\mu\nu}=h_\mu^\rho h_\nu^\sigma \mathcal{L}_t h_{\rho\sigma}$ is the ``time derivative'' of $h_{\mu\nu}$, and $D_\mu$ is the covariant derivative on $\Sigma_t$ associated with the Levi-Civita connection of $h_{\mu\nu}$ (see \cite{Wald:1984rg} or \cite{Reall:2013grnotes} for the definition). Finally $K=h^{\mu\nu}K_{\mu\nu}$. Since in the RH coordinates $\{x^\alpha\}$ defined above $\epsilon=N\sqrt{h}dt\wedge dx^1\wedge\dots\wedge dx^{n-1}$, following the usual notation for integrals of differential forms (explained in \cite{Reall:2013grnotes}), we write \eqref{eq:SE} as
\begin{equation}
\label{eq:SE2}
S_{EH}=\int_{\mathbb{R}\times \Sigma_t}{dtd^3x \mathscr{L}_{EH}},
\end{equation}
where $\mathscr{L}_{EH}$ is a tensor density given by
\begin{equation}
\mathscr{L}_{EH}=\frac{1}{16\pi}\sqrt{h}N [^{(3)}R+K_{\mu\nu}K^{\mu\nu}-K^2].
\end{equation}

Let us now consider the ``Maxwell'' part of the action \eqref{eq:SEM}. Using the same notation, we have
\begin{equation}
\label{eq:SM}
S_{M}=-\frac{1}{16\pi}\int_{M}{dtd^3x\sqrt{h}N  F_{\mu\nu}F^{\mu\nu}}.
\end{equation}
Let us define 
\begin{align}
V&=n_\mu A^\mu,\\
^{(3)}A_\mu&=h_\mu^\nu A_\nu,
\end{align}
which are interpreted as, respectively, the electromagnetic scalar potential and the electromagnetic 3-vector potential\footnote{$V$ is defined as in \cite{Sudarsky:1992ty}, \cite{Wald:1993ki} and \cite{Wald:1984rg}.}
. $^{(3)}A_\mu$ can be identified with the tensor on $\Sigma_t$, $A_a$. We also define the electric and magnetic field measured by an observer with 4-velocity $-n_\mu$, respectively, by\footnote{The electric field is defined as in \cite{Sudarsky:1992ty} and \cite{Wald:1993ki}, except for a $\sqrt{h}$ factor; the definition of $E_\mu$ and $B_\mu$ is the one given in \cite{Wald:1984rg}, where the future-directed unit normal is used.}
\begin{align}
E_\mu&=-F_{\mu\nu}n^\nu,\\
B_\mu&=\frac{1}{2}\epsilon_{\mu\nu\rho\sigma}F^{\rho\sigma}n^\nu=(\star F)_{\mu\nu} n^\nu,
\end{align}
where $\star$ denotes the Hodge dual defined by the volume form of $(M,g_{\mu\nu})$, $\epsilon$. They can be identified with tensors on $\Sigma_t$, $E_a$ and $B_a$. Using the formula for the contraction of $\epsilon_{\mu\nu\rho\sigma}$ in terms of the Kronecker delta, it is easy to show that $B_\mu B^\mu=\frac{1}{2}F_{\mu\nu}F^{\mu\nu}+E_\mu E^\mu$, so
\begin{equation}
S_{M}=\int_{M}{dtd^3x \mathscr{L}_{M}},
\end{equation}
where
\begin{equation}
\mathscr{L}_{M}=\frac{1}{16\pi}2\sqrt{h}N(E_\mu E^\mu -B_\mu B^\mu).
\end{equation}
Moreover, we can write $E_\mu$ and $B_\mu$ in terms of quantities invariant under projection onto $\Sigma_t$ as follows.
Since $E_\mu$ is invariant under projection onto $\Sigma_t$,
\begin{equation}
\begin{split}
E_\mu&=h_\mu^\rho E_\rho=-h_\mu^\rho(\nabla_\rho A_\nu -\nabla_\nu A_\rho)n^\nu\\
&=-h_\mu^\rho(\nabla_\rho(A_\nu n^\nu)-A_\nu\nabla_\rho n^\nu -n^\nu\nabla_\nu A_\rho)=-D_\mu V+h_\mu^\rho\mathcal{L}_n A_\rho.
\end{split}
\end{equation}
Then, using $A_\mu=-n_\mu V+\,^{(3)}A_\mu$, we obtain
\begin{equation}
\begin{split}
h_\mu^\rho\mathcal{L}_n A_\rho&=h_\mu^\rho\mathcal{L}_n\; ^{(3)} A_\rho-h_\mu^\rho\mathcal{L}_n(n_\rho V)\\
&=h_\mu^\rho\mathcal{L}_n \;^{(3)} A_\rho-h_\mu^\rho V\mathcal{L}_n n_\rho=h_\mu^\rho\mathcal{L}_n\; ^{(3)} A_\rho-Vn^\rho\nabla_\rho n_\mu,
\end{split}
\end{equation}
where in the last equality we used the formula for the Lie derivative in terms of the covariant derivative, the fact that $n^\mu\nabla_\rho n_\mu=0$ (because $n^\mu n_\mu=-1$ everywhere on $M$) and the definition of $h_\mu^\rho=\delta_\mu^\rho+n_\mu n^\rho$. Now, using $n_\mu=-N\nabla_\mu t$ multiple times, we obtain
\begin{equation}
\begin{split}
n^\rho\nabla_\rho n_\mu&=-n^\rho \nabla_\rho(N\nabla_\mu t)=-n^\rho\nabla_\rho N \nabla_\mu t -N n^\rho \nabla_\mu\nabla_\rho t=\frac{1}{N}n_\mu n^\rho\nabla_\rho N+Nn^\rho\nabla_\mu\biggl(\frac{1}{N}n_\rho\biggr)\\
&=\frac{1}{N}(n_\mu n^\rho\nabla_\rho N+\nabla_\mu N)=\frac{1}{N}h_\mu^\rho\nabla_\rho N=\frac{1}{N}D_\mu N.
\end{split}
\end{equation}
Then, using \eqref{eq:nvec} and $^{(3)}A_\mu(t^\mu-N^\mu)=-N\;^{(3)}A_\mu n^\mu=0$, 
\begin{equation}
h_\mu^\rho\mathcal{L}_n \;^{(3)} A_\rho=h_\mu^\rho\frac{1}{N}(\mathcal{L}_N ^{(3)} A_\rho-\mathcal{L}_t ^{(3)} A_\rho).
\end{equation}
We define $\dot{^{(3)} A_\mu}\equiv h_\mu^\rho\mathcal{L}_t \,^{(3)} A_\rho$, while
\begin{equation}
\label{eq:LieN}
h_\mu^\rho\mathcal{L}_N \,^{(3)} A_\rho=N^v D_\nu \,^{(3)}A_\mu+\,^{(3)}A_\nu D_\mu N^\nu.
\end{equation}
Therefore,
\begin{equation}
\label{eq:EE}
E_\mu=-\frac{1}{N}[D_\mu(N V)+\dot{^{(3)} A_\mu}-N^v D_\nu ^{(3)}A_\mu-^{(3)}A_\nu D_\mu N^\nu].
\end{equation}
In this expression we can replace all the Greek abstract indices with Latin abstract indices, because every quantity is invariant under projection onto $\Sigma_t$.

Moving on to the magnetic field, we can write
\begin{equation}
\label{eq:Blevi}
B_\mu=\frac{1}{2} \;^{(3)}\epsilon_{\mu\rho\sigma} \,^{(3)}F^{\rho\sigma},
\end{equation}
where $^{(3)}\epsilon_{\mu\rho\sigma}=-n^\nu \epsilon_{\nu\mu\rho\sigma}$ is the volume form on $(\Sigma_t,h_{ab})$ induced by the volume form of $M$ (i.e. the volume form of $(\Sigma_t,h_{ab})$ in the orientation class used in Stokes' theorem; see \cite{Wald:1984rg} and \cite{Reall:2013grnotes}), and $^{(3)}F^{\rho\sigma}=h^\rho_\gamma h^\sigma_\delta F^{\gamma\delta}$, which is a tensor invariant under projection that can be identified with the tensor $F^{ab}$ on $\Sigma_t$.
Decomposing $A_\mu$ into its tangential and normal parts,
\begin{equation}
\begin{split}
^{(3)}F^{\rho\sigma}&=h^\rho_\gamma h^\sigma_\delta(\nabla^\gamma \,^{(3)}A^\delta-\nabla^\gamma(Vn^\delta)-\nabla^\delta \,^{(3)}A^\gamma+\nabla^\delta(Vn^\gamma))\\
&=D^\rho\,^{(3)}A^\sigma-D^\sigma\,^{(3)}A^\rho -K^{\rho\sigma}+K^{\sigma\rho}=D^\rho\,^{(3)}A^\sigma-D^\sigma\,^{(3)}A^\rho,
\end{split}
\end{equation}
where in the last equality we have used the fact that the extrinsic curvature is symmetric.
So,
\begin{equation}
B_\mu=\frac{1}{2}\,^{(3)}\epsilon_{\mu\nu\rho}( D^\nu\,^{(3)}A^\rho-D^\rho\,^{(3)}A^\nu).
\end{equation}
In this expression we can replace all the Greek abstract indices with Latin abstract indices.

In this way we have also expressed the Einstein-Maxwell action \eqref{eq:SEM} as a functional of $(N,N^\mu,h_{\mu\nu},^{(3)}A_\mu,V)$:
\begin{equation}
S_{EM}=\int_{M}{dtd^3x\mathscr{L}_{EM}},
\end{equation}
where
\begin{equation}
\mathscr{L}_{EM}=\mathscr{L}_{EH}+\mathscr{L}_{M}=\frac{1}{16\pi}\sqrt{h}N [^{(3)}R+K_{\mu\nu}K^{\mu\nu}-K^2 +2(E_\mu E^\mu -B_\mu B^\mu)].
\end{equation}
 However, we notice that there are no ``time derivatives'' of $N, N^\mu$ and $V$, so their conjugate momenta vanish. This suggests that we should view $N,N^\mu,V$ as non-dynamical variables that can be arbitrarily specified. They are unphysical degrees of freedom of the theory.

We now want to develop the Hamiltonian formalism. Hence, we need to define the conjugate momenta of the dynamical variables of the theory, $h_{\mu\nu}$ and $^{(3)}A_\mu$.
The momentum conjugate to $h_{\mu\nu}$ is defined by
\begin{equation}
\label{eq:pih}
\frac{1}{16\pi}\pi^{\mu\nu}\equiv\frac{\partial \mathscr{L}_{EM}}{\partial \dot{h_{\mu\nu}}}=\frac{1}{16\pi}\sqrt{h}[K^{\mu\nu}-K h^{\mu\nu}].
\end{equation}
The momentum conjugate to $^{(3)}A_\mu$, $\Pi^\mu$, is defined by\footnote{$\pi^{\mu\nu}$ is defined as in \cite{Sudarsky:1992ty} and \cite{Wald:1993ki}, while $\Pi^\mu$ is defined with a minus sign w.r.t. the definition used in \cite{Sudarsky:1992ty} and \cite{Wald:1993ki}; in \cite{Sudarsky:1992ty} the factors $1/16\pi$ are omitted.}
\begin{equation}
\label{eq:PiA}
\Pi^{\mu}\equiv \frac{\partial \mathscr{L}_{EM}}{\partial \dot{^{(3)}A_\mu}}=-\frac{4}{16\pi}\sqrt{h}E^\mu.
\end{equation}
These two momenta can be identified with tensor densities on $\Sigma_t$ and we can replace Greek abstract indices with Latin abstract indices in \eqref{eq:pih} and \eqref{eq:PiA}.

In the Hamiltonian formulation of Einstein-Maxwell theory, a point in the phase space is determined by the specification of initial data $(h_{ab},\pi^{ab},A_a,E^a)$ on an arbitrary initial Cauchy surface $\Sigma$ with $t=0$. Having identified tensors invariant under projection onto $\Sigma$ with tensors defined on $\Sigma$ (and so having replaced Greek abstract indices with Latin abstract indices), we define the Hamiltonian density on $\Sigma$ as
\begin{equation}
\label{eq:Hden}
\mathscr{H}_{EM}=\mathscr{H}_{EH}+\mathscr{H}_{M},
\end{equation}
where
\begin{equation}
\mathscr{H}_{EH}\equiv \pi^{ab\nu}\dot{h}_{ab}-\mathscr{L}_{EH}
\end{equation}
and
\begin{equation}
\mathscr{H}_{M}\equiv \Pi^a \dot{A_a}-\mathscr{L}_{M}.
\end{equation}
Then, we define the Hamiltonian on $\Sigma$ as
\begin{equation}
\label{eq:H}
H=\int_{\Sigma}{e \mathscr{H}_{EM}}=\int_{\Sigma}{d^3x \mathscr{H}_{EM}},
\end{equation}
where $e$ is the 3-form on $\Sigma$ defined in any RH chart $(x^1,\dots,x^{n-1})$ by $e=dx^1\wedge\dots\wedge dx^{n-1}$ (i.e., in such a chart, $^{(3)}\epsilon=\sqrt{h}e)$.

Integrating by parts and neglecting surface terms, we have
\begin{equation}
\int_{\Sigma}{d^3x\mathscr{H}_{EH}}=\int_{\Sigma}{d^3x\sqrt{h}(N\mathcal{H}_{EH}+N^a \mathcal{H}_{EHa})},
\end{equation}
where
\begin{align}
\mathcal{H}_{EH}&=\frac{1}{16\pi}\biggl[-\,^{(3)}R+\frac{1}{h}\biggl(\pi^{ab}\pi_{ab}-\frac{1}{2}\pi^2\biggr)\biggr],\\
\mathcal{H}_{EHa}&=-\frac{1}{8\pi}D_b(h^{-1/2}\pi^b_a).
\end{align}
where $\pi=h_{ab}\pi^{ab}$.

Let us consider the Maxwell part of the Hamiltonian density,
\begin{equation}
\mathscr{H}_{M}=\frac{1}{16\pi}\sqrt{h}[-4E^a\dot{A}_a-2N(E^aE_a-B_aB^a)].
\end{equation}
We have
\begin{equation}
B_aB^a=\frac{1}{4}\,^{(3)}\epsilon_{abc}\,^{(3)}\epsilon^{ade}F^{bc}F_{de}=\frac{1}{4}2\delta^d_{[b}\delta^e_{c]}F^{bc}F_{de}=\frac{1}{2}F^{bc}F_{bc}.
\end{equation}
Then, we write $-4E^a\dot{A}_a-2NE^aE_a=-4E^a(\dot{A}_a+NE_a)+2NE^aE_a$. From \eqref{eq:EE} we have $\dot{A}_a+NE_a=-D_a(NV)+N^bD_bA_a+A_bD_aN^b=-D_a(NV)+N^bF_{ba}+D_a(A_bN^b)$, so
\begin{equation}
\begin{split}
-4E^a\dot{A}_a-2NE^aE_a=& 4E^aD_a(NV)-4E^aN^bF_{ba}-4E^aD_a(A_bN^b)+2NE^aE_a\\
=&-4NVD_aE^a-4N^aF_{ab}E^b+4N^bA_bD_aE^a+2NE^aE_a\\
&+4D_a(E^aNV)-4D_a(E^aN^bA_b).
\end{split}
\end{equation}
Using the divergence theorem on the two terms with 4-divergence and dropping boundary terms, we have
\begin{equation}
\int_{\Sigma}{d^3x\mathscr{H}_{M}}=\int_{\Sigma}{d^3x\sqrt{h}[(-NV+N^aA_a)\mathcal{C}+N\mathcal{H}_M+N^a\mathcal{H}_{Ma}]},
\end{equation}
where we have defined
\begin{align}
\label{eq:CM}
\mathcal{C}\equiv&\frac{1}{4\pi}D_aE^a,\\
\mathcal{H}_M\equiv&\frac{1}{16\pi}(2E^aE_a+F_{ab}F^{ab}),\\
\mathcal{H}_{Ma}\equiv&-\frac{1}{4\pi}F_{ab}E^b.
\end{align}
Therefore, dropping boundary terms, the Hamiltonian can be written in the form
\begin{equation}
\label{eq:Hcon}
H=\int_{\Sigma_t}{d^3x\sqrt{h}[N\mathcal{C}_{(0)}+N^a\mathcal{C}_a+(-NV+N^aA_a)\mathcal{C}]},
\end{equation}
where
\begin{align}
\mathcal{C}_{(0)}\equiv&\mathcal{H}_{EH}+\mathcal{H}_{M}=\frac{1}{16\pi}\biggl[-\,^{(3)}R+\frac{1}{h}\biggl(\pi^{ab}\pi_{ab}-\frac{1}{2}\pi^2\biggr)+2E^aE_a+F_{ab}F^{ab}\biggl],\\
\mathcal{C}_a\equiv&\mathcal{H}_{EHa}+\mathcal{H}_{Ma}=-\frac{1}{8\pi}[D_b(h^{-1/2}\pi^b_a)+2F_{ab}E^b], \label{eq:momcons}
\end{align}
and $\mathcal{C}$ is given by \eqref{eq:CM}. We see that, if we regard $N$, $N^a$ and $V$ as dynamical variables, their Hamilton's equations are $\frac{\delta H}{\delta N}=0$, $\frac{\delta H}{\delta N^a}=0$ and $\frac{\delta H}{\delta V}=0$ (because the conjugate momenta are 0). These three equations give, respectively, $\mathcal{C}_{(0)}-V\mathcal{C}=0$, $	\mathcal{C}_a+A_a\mathcal{C}=0$ and $\mathcal{C}=0$. So we find that the Hamilton's equations for $N$, $N^a$ and $V$ are equivalent to $\mathcal{C}=0$, $\mathcal{C}_{(0)}=0$ and $\mathcal{C}_a=0$ on $\Sigma$. The last three equations are the constraints of Einstein-Maxwell theory. In particular, the first one is the local form of Gauss' law on $\Sigma$, the second one is the Hamiltonian constraint and the third one is the momentum constraint, i.e. eqs. (10.2.41) and (10.2.42) of \cite{Wald:1984rg} (where the future-directed unit normal is used) with $T_{\mu\nu}=\frac{1}{4\pi}(F_{\mu\rho}F_\nu^{\;\;\rho}-\frac{1}{4}F^{\rho\sigma}F_{\rho\sigma}g_{\mu\nu})$. Therefore, $N$, $N^a$ and $V$ can be regarded as Lagrange multipliers. Once again we have found that there are no dynamical equations for $N$, $N^a$ and $V$, so they can be specified arbitrarily on each slice. They are not physical degrees of the theory, in fact, if we are given a specification of $(h_{ab},\pi^{ab},A_a,E^a)$ on any slice, we automatically obtain $N,N^a,V$ on the same slice by solving the constraint equations. 

Having chosen $N,N^a,V$, to find the Hamilton's equations given by $H$ we compute the variation $\delta H$ on $\Sigma$ due to a variation of the initial data $(\delta h_{ab},\delta \pi^{ab},\delta A_a,\delta(\sqrt{h}E^a))$ of compact support on $\Sigma$, then we integrate by parts and drop all the terms arising from integrals over the boundary of $\Sigma$ (they vanish because the variation is of compact support). In this way we obtain
\begin{equation}
\label{eq:varHnoboun}
\delta H=\int_{\Sigma}{d^3x\{\mathcal{P}^{ab}\delta h_{ab}+\mathcal{Q}_{ab}\delta \pi^{ab}+\mathcal{R}^a\delta A_a+\mathcal{S}_a\delta(\sqrt{h}E^a)\}},
\end{equation}
where
\begin{align}
\mathcal{P}^{ab}\equiv&\frac{\delta H}{\delta h_{ab}}=\frac{1}{16\pi}[\sqrt{h}N a^{ab}+\sqrt{h}(h^{ab}D^cD_cN-D^aD^bN)-\mathcal{L}_N\pi^{ab}],\\
\mathcal{Q}_{ab}\equiv&\frac{\delta H}{\delta \pi^{ab}}=\frac{1}{16\pi}\biggl[\frac{N}{\sqrt{h}}(2\pi_{ab}-\pi h_{ab})+\mathcal{L}_Nh_{ab}\biggr],\\
\mathcal{R}^a\equiv&\frac{\delta H}{\delta A_a}=\frac{1}{4\pi}[\sqrt{h}D_b(NF^{ab})+\mathcal{L}_N(\sqrt{h}E^a)],\\
\mathcal{S}_a\equiv&\frac{\delta H}{\delta(\sqrt{h} E^a)}=\frac{1}{4\pi}\biggl[\frac{N}{\sqrt{h}}(\sqrt{h}E_a)+D_a(NV)-\mathcal{L}_NA_a\biggr], \label{eq:evA}
\end{align}
with
\begin{equation}
\begin{split}
a^{ab}=&\frac{2}{h}\biggl[(\sqrt{h}E^a)(\sqrt{h}E^b)-\frac{1}{2}h^{ab}(\sqrt{h}E_c)(\sqrt{h}E^c)\biggr]+2\biggl(F^{ac}F_c^{\;\;b}+\frac{1}{4}h^{ab}F^{cd}F_{cd}\biggr) \\
&+\biggl(R^{ab}-\frac{1}{2}h^{ab}R\biggr)+\frac{1}{h}\biggl[2\pi^a_c\pi^{bc}-\pi \pi^{ab}-\frac{1}{2}\biggl(\pi^{cd}\pi_{cd}-\frac{1}{2}\pi^2\biggr)\biggr],
\end{split}
\end{equation}
and \footnote{\label{longnote}Here, $\pi^{ab}$ and $\sqrt{h} E^a$ are tensor densities on $\Sigma_t$ of weight 1. 
The Lie derivative of a ${r}\choose{s}$ tensor density field $\mathcal{T}$ w.r.t. a vector field $X$ can be defined in terms of the pull-back in the exact same way as in the case of standard tensors. The expression in terms of the usual Levi-Civita covariant derivative of tensors is $(\mathcal{L}_X \mathcal{T})^{\mu_1\dots \mu_r}_{\;\;\;\;\;\;\;\;\;\;\nu_1\dots\nu_s}\equiv(\sqrt{h})^\omega X^\rho D_\rho\bigl(\frac{1}{(\sqrt{h})^\omega}\mathcal{T}^{\mu_1\dots \mu_r}_{\;\;\;\;\;\;\;\;\;\;\nu_1\dots\nu_s}\bigr) - (D_\rho X^\mu_1)\mathcal{T}^{\rho\mu_2\dots \mu_r}_{\;\;\;\;\;\;\;\;\;\;\nu_1\dots\nu_s}-\dots -(D_\rho X^\mu_r)\mathcal{T}^{\mu_1\dots \mu_{r-1}\rho}_{\;\;\;\;\;\;\;\;\;\;\nu_1\dots\nu_s}+(D_{\nu_1} X^\rho)\mathcal{T}^{\mu_1\dots \mu_r}_{\;\;\;\;\;\;\;\;\;\;\rho\nu_2\dots\nu_s}+\dots+(D_{\nu_s} X^\rho)\mathcal{T}^{\mu_1\dots \mu_r}_{\;\;\;\;\;\;\;\;\;\;\nu_1\dots\nu_{s-1}\rho}+\omega(D_\rho X^\rho)\mathcal{T}^{\mu_1\dots \mu_r}_{\;\;\;\;\;\;\;\;\;\;\nu_1\dots\nu_s}$.}
\begin{align}
\mathcal{L}_N h_{ab}=&D_a N_b+D_b N_a,\\
\mathcal{L}_N A_a=&N^bD_bA_a+A_b D_a N^b,
\end{align}
\begin{align}
\mathcal{L}_N \pi^{ab}=& \sqrt{h}N^cD_c\biggl(\frac{\pi^{ab}}{\sqrt{h}}\biggr)-2\pi^{c(a}D_cN^{b)}+\pi^{ab}D_cN^c,\\
\mathcal{L}_N (\sqrt{h}E^a)=&\sqrt{h}N^cD_c\biggl(\frac{\sqrt{h}E^a}{\sqrt{h}}\biggr)-(\sqrt{h}E^c)D_cN^a+(\sqrt{h}E^a)D_c N^c.
\end{align}
The Hamilton's equations are then\footnote{$\dot{\pi}^{ab}$ and $\dot{(\sqrt{h} E^a)}$ are the tensor densities on $\Sigma_t$ identified with $\dot{\pi}^{\mu\nu}\equiv h^\mu_\rho h^\nu_\sigma \mathcal{L}_t\pi^{\rho\sigma}$ and $\dot{(\sqrt{h} E^\mu)}\equiv h^\mu_\rho\mathcal{L}_t(\sqrt{h} E^\rho)$, respectively. See the footnote \ref{longnote} for the expression of Lie derivative of tensor densities in terms of the Levi-Civita connection.} 
\begin{equation}
\label{eq:Hameq}
\dot{h}_{ab}=16\pi\mathcal{Q}_{ab}, \quad  \dot{\pi}^{ab}=-16\pi\mathcal{P}_{ab}, \quad \dot{A}_a=-\frac{16\pi}{4}\mathcal{S}_a, \quad \dot{(\sqrt{h} E^a)}=\frac{16\pi}{4}\mathcal{R}^a,
\end{equation}
on each slice $\Sigma_t$\footnote{I wrote the numerical coefficients in this way in order to make contact with the results of \cite{Sudarsky:1992ty}.}.

For any choice of $N,N^a,V$ on each slice $\Sigma_t$, the constraint equations on the initial slice $\Sigma$ together with the Hamilton's equations for the dynamical variables on each slice $\Sigma_t$ are equivalent to the equations of motion \eqref{eq:Eeq},\eqref{eq:Meq} (see Chapter 10 and Appendix E of \cite{Wald:1984rg}, where the case of Einstein theory in vacuum is discussed). The generalisation of a famous theorem by Choquet-Bruhat and Geroch \cite{ChoquetBruhat:1969cb} ensures that, for any choice of $N,N^a,V$, and for initial data satisfying the three constraints on $\Sigma$, there exists a unique (up to diffeomorphisms) globally hyperbolic solution $(h_{ab}(t),\pi^{ab}(t),A_a(t),E^a(t))$ of the evolution equations \eqref{eq:Hameq} on $M=\mathbb{R}\times\Sigma$, with Cauchy surface $\Sigma$, such that the solution on $\Sigma$ is the initial data set $(h_{ab},\pi^{ab},A_a,E^a)$. This tells us that the solution of the initial value problem of the Einstein-Maxwell theory can predict the spacetime arising from a certain configuration of the dynamical variables at a certain ``time'' that satisfies the constraints.

\eqref{eq:Hcon} is said ``pure constraint'' form of the Hamiltonian. Since $t^\mu A_\mu=(-N n^\mu+N^\mu)(-V n_\mu+\,^{(3)}A_\mu)=-NV+N^\mu\,^{(3)}A_\mu$, we can also write
\begin{equation}
\label{eq:cons}
H=\int_{\Sigma}{d^3x\sqrt{h}(t^\mu C_\mu+t^\mu A_\mu \mathcal{C})},
\end{equation}
where we have defined $\mathcal{C}_\mu=n_\mu \mathcal{C}_{(0)}+\,^{(3)}\mathcal{C}_\mu$\footnote{Sudarski and Wald obtain the same expression for the Hamiltonian (see eq. (17) of \cite{Sudarsky:1992ty} and eq. (6) of \cite{Wald:1993ki}, in which the factor $\sqrt{h}$ has been absorbed in the definitions of $\mathcal{C}$, $\mathcal{C}_0$ and $\mathcal{C}_a$; in \cite{Sudarsky:1992ty} the factor $1/16\pi$ is omitted). However, we see from \eqref{eq:momcons} that the expression of $\mathcal{C}_a$ differs from the one given by eq. (14) of \cite{Sudarsky:1992ty} and eq. (3) of \cite{Wald:1993ki} for a relative minus sign. Therefore, our expression for $H$ is not consistent with the one given by Wald and Sudarski due to this discrepancy in the definition of $\mathcal{C}_a$. On the other hand, with our definition of $\mathcal{C}_a$, the equation $\mathcal{C}_a=0$ is consistent with the momentum constraint given by eq. (10.2.41) of \cite{Wald:1984rg} (where the future-directed unit normal is used), as it should be.} ($^{(3)}\mathcal{C}_\mu$ is the vector invariant under projection onto $\Sigma$ associated with $\mathcal{C}_a$ under the natural isomorphism $\tilde{\theta}^\ast$). Notice that $H$ vanishes for initial data satisfying the constraint equations on $\Sigma$. This implies that, if we consider an arbitrary perturbation $(\delta h_{ab},\delta \pi^{ab},\delta A_a,\delta(\sqrt{h}E^a))$ of the initial data on $\Sigma$ that satisfies the linearised constraints (i.e. $(h_{ab}+\delta h_{ab},\pi^{ab}+\delta \pi^{ab},A_a+\delta A_a,\sqrt{h}E^a+\delta(\sqrt{h}E^a)$ satisfies the constraints to linear order in the perturbations, which means that it is possible to predict the spacetime that solves the evolution equations with the perturbed initial data set), we have $H+\delta H=0$, and so $\delta H=0$. We will need this result later.

To obtain $H$ we have dropped boundary terms, but we are allowed to do so only when these are vanishing. When this is not the case, $H$ does not provide the correct Hamiltonian; there may be contributions from the boundary of $\Sigma$ (called surface terms) that are non-vanishing for our choice of $t^\mu$ (i.e. $N$ and $N^a$) and $V$. To obtain the Hamiltonian that takes into account these non-vanishing contributions from boundary terms, instead of keeping track of all the boundary terms arising throughout our derivation of $H$, it is easier to compute the variation of $H$ due to a variation of the initial data $(h_{ab},\pi^{ab},A_a,E^a)$ on $\Sigma$ satisfying some ``natural'' boundary conditions. Then we modify the definition of $H$ so that the new Hamiltonian $H'$ differs from $H$ only by surface terms, and so that the variation of the new Hamiltonian $\delta H'$ is given by the RHS of \eqref{eq:varHnoboun} (without other contributions from surface terms). This will ensure that the Hamilton's equations of $H$ and $H'$ are the same.

If $\Sigma$ is compact (i.e. $M=\mathbb{R}\times\Sigma$ is closed) there are no surface terms, so \eqref{eq:cons} gives the correct Hamiltonian and \eqref{eq:varHnoboun} gives the correct variation of $H$ for a completely arbitrary variations of the dynamical variables.

We now want to consider the case where the initial data set $(\Sigma, h_{ab}, \pi^{ab}, A_a, \sqrt{h}E^a)$ is an asymptotically flat end. In Einstein-Maxwell theory in 4 dimensions, an \emph{asymptotically flat end} is an initial data set such as (a) $\Sigma$ is diffeomorphic to $\mathbb{R}^3/B$, where $B$ is a closed ball centred on the origin of $\mathbb{R}^3$, (b) if we pull-back the coordinates on $\mathbb{R}^3/B$ to define coordinates $x^i$ on $\Sigma$ then
\begin{equation}
\begin{split}
h_{ij}&=\delta_{ij}+\mathcal{O}\biggl(\frac{1}{r}\biggr),\\
\pi^{ij}&=\mathcal{O}\biggl(\frac{1}{r^2}\biggr),\\
A_i&=\mathcal{O}\biggl(\frac{1}{r}\biggr),\\
(\sqrt{h}E^i)&=\mathcal{O}\biggl(\frac{1}{r^2}\biggr),
\end{split}
\end{equation}
as $r\equiv\sqrt{x^ix^i}\to\infty$, and (c) any $k$-th order derivative of the dynamical variables (w.r.t. the coordinates $x^i$) falls off with $k$ powers of $r$ faster than the corresponding dynamical variable (e.g. $h_{ij,k}=\mathcal{O}\bigl(\frac{1}{r^2}\bigr)$, $h_{ij,kl}=\mathcal{O}\bigl(\frac{1}{r^3}\bigr)$, etc.).
The charts $x^i$ that satisfy these properties are called \emph{asymptotically Cartesian coordinates}.
Instead of studying perturbations of compact support on $\Sigma$, it is natural to consider variations of the dynamical variables $(\delta h_{ab},\delta \pi^{ab},\delta A_a,\delta(\sqrt{h}E^a))$ such that $(\Sigma, h_{ab}+\delta h_{ab}, \pi^{ab}+\delta \pi^{ab}, A_a+\delta A_a, \sqrt{h}E^a+\delta(\sqrt{h}E^a))$ is still an asymptotically flat end. This means that, in asymptotically Cartesian coordinates $x^i$,
$
\delta h_{ij}=\mathcal{O}\bigl(\frac{1}{r}\bigr),\;
\delta \pi^{ij}=\mathcal{O}\bigl(\frac{1}{r^2}\bigr),\;
\delta A_i=\mathcal{O}\bigl(\frac{1}{r}\bigr),\;
\delta (\sqrt{h}E^i)=\mathcal{O}\bigl(\frac{1}{r^2}\bigr),
$
when $r\equiv\sqrt{x^ix^i}\to\infty$, and also that the variation of any $k$th-order derivative of the dynamical variables (w.r.t. the coordinates $x^i$) falls off with $k$ powers of $r$ faster than the corresponding dynamical variable (e.g. $\delta h_{ij,k}=\mathcal{O}\bigl(\frac{1}{r^2}\bigr)$, $\delta h_{ij,kl}=\mathcal{O}\bigl(\frac{1}{r^3}\bigr)$, etc.). Such a variation will be called an \emph{asymptotically flat variation}. When we compute the variation of $H$ induced by an asymptotically flat variation, extra terms will arise (w.r.t. \eqref{eq:varHnoboun}) due to contributions from the boundaries of $\Sigma$. These new terms arise when we integrate by parts the total divergence terms that appear when we try to obtain the form \eqref{eq:varHnoboun}. We also assume, for now, that no interior boundaries are present on $\Sigma$.

Before trying to obtain the correct Hamiltonian in this case, we also need to specify the time foliation, i.e. $t^\mu$, and $V$ on the manifold $M$. For now, we will keep $V$ completely arbitrary. Later, in the case of spacetimes with a stationary or axial Killing vector field $X^\mu$ we will require that a choice of $V$ (together with an electromagnetic gauge choice of $A_a$) can be made so that $X^\mu$ is also a symmetry of $A_\mu$, i.e. $\mathcal{L}_X A_\mu=0$. We denote the limit of $V$ as $r\equiv\sqrt{x^ix^i}\to\infty$ on $\Sigma$, where $x^i$ are asymptotically Cartesian coordinates defined in the definition of asymptotically flat end, by $\mathcal{V}$.
With regards to $t^\mu$, we will consider two cases: \begin{enumerate}[(i)] \item $t^\mu$ approaches a time translation, \item $t^\mu$ approaches a rotation.\end{enumerate}

Let us define more precisely what an asymptotic time translation and an asymptotic rotation are. In asymptotically flat spacetimes, there exist coordinates $x^\alpha=(x^0,x^i)$, called \emph{asymptotically inertial coordinates}, such that, in the limit $r\equiv\sqrt{x^i x^i}\to\infty$,
\begin{align}
g_{\alpha\beta}&=\eta_{\alpha\beta}+\mathcal{O}\biggl(\frac{1}{r}\biggr),\label{eq:met2}\\
\partial_\gamma g_{\alpha\beta}&=\mathcal{O}\biggl(\frac{1}{r^{2}}\biggr),
\end{align}
where $\eta_{\alpha\beta}=diag(-1,1,1,1)$.
An asymptotic time translation is a vector field that is given by $\bigl(\frac{\partial}{\partial x^0}\bigr)^\mu$ as $r\equiv\sqrt{x^ix^i}\to\infty$ where $(x^0,x^i)$ are asymptotically inertial coordinates, i.e. a vector field which, at infinity, becomes a timelike Killing vector field (of the metric at infinity, $\eta_{\mu\nu}$). If we choose $t^\mu=\bigl(\frac{\partial}{\partial t}\bigr)^\mu$ to be an asymptotic time translation, we are choosing $t$ so that there exist coordinates $x^i$ on $\Sigma$ such that, after being extended to $M$ along integral curves of $t^\mu$ as explained before, the coordinates $x^\alpha=(t,x^i)$ are asymptotically inertial coordinates. 
A possible choice for such $x^i$ on $\Sigma$ is the choice of asymptotically Cartesian coordinates. In these coordinates, the past-directed normal to $\Sigma_t$ at infinity is $n_\alpha=\delta^0_\alpha$, so $n^\alpha=\eta^{\alpha\beta}n_\beta=-\delta^\alpha_0$. Since $t^\alpha=\delta^0_\alpha$, we have $t^\mu \propto n^\mu$. Hence, $t^\mu$ is normal to each $\Sigma_t$ at infinity, and $N=1$ and $N^i=0$ at infinity. Now, comparing the components of the metric \eqref{eq:met2} at $r\to\infty$ with those given by \eqref{eq:met}, we see that choosing $t^\mu$ to be an asymptotic time translation is equivalent to choosing $N=1+\mathcal{O}\bigl(\frac{1}{r}\bigr)$ and $N^i=\mathcal{O}\bigl(\frac{1}{r}\bigr)$ as $r\to\infty$ in all asymptotically inertial coordinates $(t,x^i)$, and in particular if $x^i$ are asymptotically Cartesian coordinates.

An asymptotic rotation is a vector field that goes to one of the three vector fields $(\phi_l)^\mu=\epsilon_{ljk}x_j\bigl(\frac{\partial}{\partial x^k}\bigr)^\mu$, where $\epsilon_{ljk}$ is the usual Levi-Civita symbol (e.g., $(\phi_3)^\mu=x_1\bigl(\frac{\partial}{\partial x^2}\bigr)^\mu-x_2\bigl(\frac{\partial}{\partial x^1}\bigr)^\mu$), as $r\equiv\sqrt{x^ix^i}\to\infty$. We can define coordinates $(t,r,\theta,\phi)$, where $r,\theta,\phi$ are defined from $x^i$ by the usual relations between spherical and Cartesian coordinates. Then, $(\phi_1)^\mu=-\sin\phi\bigl(\frac{\partial}{\partial \theta}\bigr)^\mu-\cot\theta\cos\phi\bigl(\frac{\partial}{\partial \phi}\bigr)^\mu$, $(\phi_2)^\mu=\cos\phi\bigl(\frac{\partial}{\partial \theta}\bigr)^\mu-\cot\theta\sin\phi\bigl(\frac{\partial}{\partial \phi}\bigr)^\mu$ and $(\phi_3)^\mu=\bigl(\frac{\partial}{\partial \phi}\bigr)^\mu$. If we choose $t^\mu=\bigl(\frac{\partial}{\partial t}\bigr)^\mu$ to be an asymptotic rotation, i.e. $t^\mu=(\phi_l)^\mu$ at infinity, then there exists a coordinate $x^0$ and coordinates $x^i$ on $\Sigma$ such that, when we extend $x^i$ outside $\Sigma$ along integral curves of $t$, $(x^0,x^i)$ are asymptotically inertial coordinates, and $t^\mu=\epsilon_{ljk}x_j\bigl(\frac{\partial}{\partial x^k}\bigr)^\mu$ as $r\to\infty$. Again, a possible choice for such $x^i$ on $\Sigma$ is the choice of asymptotically Cartesian coordinates. We see that $t^\mu$ is tangent to each $\Sigma_t$ at infinity. So its projection along the normal to $\Sigma_t$ is $N=0$ at infinity and the projection onto $\Sigma_t$ is $N^\mu=(\phi_l)^\mu$ at infinity. Notice that, regarding $\phi_l$ as vectors defined on $\Sigma$, they are Killing vector fields of the metric of the asymptotically flat end $\Sigma$ at infinity, i.e. Killing vector fields of $\delta_{ab}$. Comparing the components of the metric \eqref{eq:met2} at $r\to\infty$ with those given by \eqref{eq:met}, we find that choosing $t^\mu$ to be an asymptotic time translation is equivalent to choosing $N=\mathcal{O}\bigl(\frac{1}{r}\bigr)$ and $N^i=(\phi_l)^i+\mathcal{O}\bigl(\frac{1}{r}\bigr)$ as $r\to\infty$ in all asymptotically inertial coordinates $(x^0,x^i)$, and in particular if $x^i$ are asymptotically Cartesian coordinates.

In the next section we will be also interested in a combination of the cases (i) and (ii), where $t^\mu=k+\Omega m$ is a linear combination of a time translation $k$ (which will be chosen as a stationary Killing field) and a rotation $m$ (which will be chosen as an axial Killing field) at infinity ($\Omega$ is a constant that will be interpreted as the angular velocity of the black hole). In this case, $N=1+\mathcal{O}\bigl(\frac{1}{r}\bigr)$ and $N^i=\Omega (\phi_l)^i+\mathcal{O}\bigl(\frac{1}{r}\bigr)$ as $r\to\infty$ in all asymptotically inertial coordinates $(x^0,x^i)$, and in particular if $x^i$ are asymptotically Cartesian coordinates. Therefore, for now we consider both cases at the same time.

We use asymptotically Cartesian coordinates on $\Sigma$ to evaluate the integrals over $\Sigma$. Consider the region of $\Sigma$ contained within a sphere of constant $r$, with boundary $S_r^2$, and let us denote the ``volume form'' on $S_r^2$ by $dS$. We will then take the limit $r\to\infty$. The surface terms arising from the variation of $H$ w.r.t. $h_{ab}$, after using the divergence theorem on total divergence terms, are
\begin{equation}
\label{eq:Sr21}
\begin{split}
\frac{1}{16\pi}\int_{S_r^2}dS \biggl\{ &-n^iND^j\delta h_{ij} +n^k N D_k(h^{ij}\delta h_{ij}) \\
&+n^i(D^j N)\delta h_{ij} + n^k\biggl[-(D_k N)h^{ij}\delta h_{ij} -2 N^i\frac{\pi^{jl}}{\sqrt{h}}h_{lk}\delta h_{ij}+h_{kl}N^l \frac{\pi^{ij}}{\sqrt{h}}\delta h_{ij}\biggr] \biggr\},
\end{split}
\end{equation}
where $n^i$ is the outward unit normal on $S_r^2$ in the tangent space of $\Sigma$. Since $dS=\mathcal{O}(r^2)$, when we take the limit $r\to\infty$ and we use the asymptotically flat boundary conditions on the initial data (which imply $\sqrt{h}=\mathcal{O}(1)$) and its variation, only the terms of the first line may give contribution for $N$ and $N^i$ satisfying one of the two asymptotic conditions written above. Since $h_{ij}\to \delta_{ij}$ when $r\to\infty$, then $D_i\to \partial_i$ and we do not need to distinguish between upstairs and downstairs indices. Therefore, the contribution to $\delta H$ can be written as
\begin{equation}
\label{eq:S1}
\begin{split}
\delta S_1&\equiv -\frac{1}{16\pi}\lim_{r\to\infty}\int_{S_r^2}{dS n_iN[\partial_j\delta h_{ij} -\partial_i\delta h_{jj}]}\\
&=-\frac{1}{16\pi}\delta \bigl[\lim_{r\to\infty} \int_{S_r^2}{dS n_i(N\partial_jh_{ij} -N\partial_i h_{jj})}\bigr].
\end{split}
\end{equation}

When we apply the same procedure to total divergence terms arising from the variation of $H$ w.r.t. $\pi^{ab}$, we obtain
\begin{equation}
\label{eq:Sr22}
-\frac{1}{16\pi}\int_{S_r^2}{dS\biggl[2n^i h_{jk}N^k h_{il} \frac{\delta \pi^{lj}}{\sqrt{h}}\biggr]}.
\end{equation}
This term may give contribution to $\delta H$, in the limit $r\to\infty$, for $N$ and $N^i$ satisfying one of the two asymptotic conditions written above. The contribution is
\begin{equation}
\label{eq:S2}
\begin{split}
\delta S_2&=-\frac{1}{16\pi}\lim_{r\to\infty}\int_{S_r^2}{dS n_i2 N_j \delta \pi_{ij}} \\
&=-\frac{1}{16\pi}\delta\bigl[\lim_{r\to\infty}\int_{S_r^2}{dS n_i 2N_j\pi_{ij}}\bigr].
\end{split}
\end{equation}

When we vary $H$ with respect to $A_a$, the surface terms on $S_r^2$ are
\begin{equation}
\label{eq:Sr23}
\frac{1}{16\pi}\int_{S_r^2}{dSn_i 4\biggl[-N((D^j A^i-D^i A^j)\delta A_j)-N^i\frac{\sqrt{h}E^j}{\sqrt{h}}\delta A_j+N^j\frac{\sqrt{h}E^i}{\sqrt{h}}\delta A_j\biggr]}.
\end{equation}
This term does not give contribution in the limit $r\to\infty$ for both asymptotic conditions on $N$ and $N^i$.

Finally, when we vary $H$ w.r.t. $\sqrt{h}E^a$, the surface terms on $S_r^2$ are
\begin{equation}
\label{eq:Sr24}
\frac{1}{16\pi}\int_{S_r^2}{dSn_i 4(-NV+N^jA_j)\frac{1}{\sqrt{h}}\delta(\sqrt{h}E^i)}.
\end{equation}
When we take the limit $r\to\infty$, only the first term may contribute for $N$ and $N^i$ satisfying one of the two asymptotic conditions written above\footnote{Notice that in \cite{Sudarsky:1992ty} and \cite{Wald:1993ki} Sudarski and Wald keep also the second term. This term changes the expression of the canonical angular momentum defined below. However, in other papers such as \cite{Brown:1990fk}, this term does not appear in the formula for the angular momentum, as it seems to be the case for our asymptotically flat boundary conditions.}.
The contribution can be written as
\begin{equation}
\label{eq:S3}
\begin{split}
\delta S_3&\equiv-\frac{1}{16\pi}\lim_{r\to\infty}\int_{S_r^2}{dSn_i 4N\mathcal{V}\delta(\sqrt{h}E_i)}\\
&=-\frac{1}{16\pi}\delta\bigl[\lim_{r\to\infty}\int_{S_r^2}{dSn_i 4N\mathcal{V}E_i}\bigr].
\end{split}
\end{equation}

Therefore, the correct Hamiltonian $H'$ for the two choices of time foliation and $V$ introduced above, whose variation (taking into account also surface terms) w.r.t. an arbitrary asymptotically flat variation of the initial data set is the RHS of \eqref{eq:varHnoboun}, is defined by\footnote{Comparing this expression with eq.(32) of \cite{Sudarsky:1992ty}, we see that in Sudarski's and Wald's expression there is one more term due to the fact that they kept the second term of \eqref{eq:Sr24}.}
\begin{equation}
\label{eq:Hcor}
\begin{split}
H'\equiv &H-S_1-S_2-S_3\\
=&H+\frac{1}{16\pi} \lim_{r\to\infty} \int_{S_r^2}{dS n_i\{N[\partial_jh_{ij} -\partial_i h_{jj}]+ 2N_j\pi_{ij}+ 4N\mathcal{V}E_i\}}.
\end{split}
\end{equation}
We see that the value of the true Hamiltonian for initial data satisfying the constraints (so $H=0$), i.e. the Hamiltonian on the initial slice of a solution of the Einstein-Maxwell evolution equations, has contribution only from terms on the boundary of $\Sigma$. In section \ref{sec:hamenang} we will see that this is true for every diffeomorphism invariant theory for which an Hamiltonian exists, i.e. those theories for which there exists a function on the phase space of the theory such that (i) its variation induced by an asymptotically flat variation of dynamical variables has no surface terms, and (ii) its variation induced by a variation of the dynamical variables of compact support leads to ``Hamilton's'' equations, which, together with the constraints of the theory, are equivalent to the equations of motion.

In the next section, we will see the importance of $H'$ to define the notions of physical quantities and we will prove a relation between the variation of these quantities when a certain type of initial data set is varied, namely the first law of black hole mechanics.

\subsection{Energy, charge, angular momentum, and entropy, and the first law of black hole mechanics}
\label{sec:enchan}

In this section, we will explain how notions of energy, charge and angular momentum arise from the correct Hamiltonian of Einstein-Maxwell theory and we will prove some results for the variations of these quantities w.r.t. a variation of the initial data in three different cases. The last result will prove the first law of black hole mechanics generalised to arbitrary asymptotically flat perturbations of a stationary black hole.

We define the \emph{canonical energy}, $\mathcal{E}$, as the value of the correct Hamiltonian \eqref{eq:Hcor} for the time foliation (i), where $t^\mu$ approaches a time translation, and for initial data satisfying the constraints (so $H=0$):
\begin{equation}
\mathcal{E}=m+\frac{1}{4\pi} \lim_{r\to\infty} \int_{S_r^2}{dS n_i \mathcal{V}E_i},
\end{equation}
where $m$ is the ADM mass, defined by
\begin{equation}
\label{eq:ADMmass}
m=\frac{1}{16\pi} \lim_{r\to\infty} \int_{S_r^2}{dS n_i(\partial_j(h_{ij}) -\partial_i h_{jj})}.
\end{equation}
We see that the canonical energy differs from the ADM mass by the term
\begin{equation}
\mathcal{E}_M=\frac{1}{4\pi} \lim_{r\to\infty} \int_{S_r^2}{dS n_i \mathcal{V}E_i}.
\end{equation}
This quantity vanishes in the case where there are no boundaries on $\Sigma$ aside from infinity. In fact, taking the limit $r\to\infty$ on $\Sigma$ of the evolution equation \eqref{eq:Hameq} for $A_a$, whose RHS is given by \eqref{eq:evA}, we have $0=-\partial_i\mathcal{V}$. So $\mathcal{V}$ is constant. Hence, $\mathcal{E}_M=\mathcal{V} Q$, where
\begin{equation}
Q=\frac{1}{4\pi} \lim_{r\to\infty} \int_{S_r^2}{dS n_i E_i}=\frac{1}{4\pi}\int_{\Sigma}{d^3 x\sqrt{h} D^i E_i}=0,
\end{equation}
where in the second equality we used the divergence theorem and in the third equality we used the constraint equation $\mathcal{C}=0$. However, in the case (considered below) where other boundaries are present on $\Sigma$, $Q$ does not vanish. In general, $Q$ and $\mathcal{V}$ have the interpretation of being the electric charge and potential at infinity, respectively.

We have, thereby, obtained that the correct Hamiltonian for an asymptotically flat surface $\Sigma$ with no other boundaries and for the time foliation (i) is $H'=H+m$. Let us now consider an asymptotically flat stationary spacetime, with stationary Killing field $k^\mu$, that is solution of the evolution equations with asymptotically flat initial data $(h_{ab},\pi^{ab},A_a,\sqrt{h}E^a)$ satisfying the constraints. Let us choose $t^\mu=k^\mu$. This is an asymptotic time translation by definition of stationary Killing field, so $N\to 1+\mathcal{O}\bigl(\frac{1}{r}\bigr)$ and $N^i\to\mathcal{O}(1/r)$ in asymptotically Cartesian coordinates $x^i$. Since $k$ generates isometries, we have $\dot{g}_{\mu\nu}=0$ everywhere. We assume that a gauge choice has been made so that the diffeomorphism generated by $t^\mu$ is a symmetry of $A_\mu$, i.e. $\mathcal{L}_t A_\mu=0$. Thus, $\dot{A}_\mu=0$ everywhere. Therefore, $\dot{h}_{ab}=0$, $\dot{\pi}^{ab}=0$, $\dot{A}_a=0$, $\dot{(\sqrt{h}E^a)}=0$. Hence, the variation of $H'$ w.r.t an arbitrary asymptotically flat perturbation of the initial data (which is given by the RHS of \eqref{eq:varHnoboun}) is 0. We also assume that the perturbation $(\delta h_{ab},\delta \pi^{ab},\delta A_a,\delta(\sqrt{h}E^a))$ of the initial data satisfies the linearised constraints, i.e. $(h_{ab}+\delta h_{ab},\pi^{ab}+\delta \pi^{ab},A_a+\delta A_a,\sqrt{h}E^a+\delta(\sqrt{h}E^a)$ satisfies the constraints to linear order in the variations, (so $Q+\delta Q=0$, which implies $\delta Q=0$). Then, $\delta H=0$ as we noted above. In this way, we proved the following theorem.
\begin{theorem}{}{1}
Let $(\Sigma,h_{ab},\pi^{ab},A_a,\sqrt{h}E^a)$ be an asymptotically flat end with initial data satisfying the constraints $\mathcal{C}=\mathcal{C}_{(0)}=\mathcal{C}_a=0$, and let $\Sigma$ have no interior boundary. Let us assume that the solution of the evolution equations $\eqref{eq:Hameq}$ with this initial data is a stationary, asymptotically flat spacetime $(M=\mathbb{R}\times\Sigma,h_{ab}(t),\pi^{ab}(t),A_a(t),\sqrt{h}E^a(t))$. Consider an arbitrary asymptotically flat perturbation $(\delta h_{ab},\delta \pi^{ab},\delta A_a,\delta(\sqrt{h}E^a))$ (whose evolution is not necessarily a stationary spacetime) of the initial data that satisfies the linearised constraints. Then, the corresponding variation of ADM mass is
\begin{equation}
\delta m=0.
\end{equation}
\end{theorem}
Recalling the fact that $(\Sigma,h_{ab},\pi^{ab},A_a,\sqrt{h}E^a)$ and $(\Sigma,h_{ab}+\delta h_{ab},\pi^{ab}+\delta \pi^{ab},A_a+\delta A_a,\sqrt{h}E^a+\delta(\sqrt{h}E^a)$ are suitable initial data for the Einstein-Maxwell evolution equations (because they satisfy the constraints), we see that this theorem states that any arbitrary perturbation to an asymptotically flat spacetime of an asymptotically flat stationary solution of the Einstein-Maxwell equations, with Cauchy surface that has no interior boundary, satisfies $\delta m=0$. In other words, every asymptotically flat stationary solution of the Einstein-Maxwell equations is an extremum of ADM mass $m$. We can ask ourselves if the converse of this theorem holds, namely, whether, for an initial data set on an asymptotically flat Cauchy surface $\Sigma$ with no interior boundary that extremises $m$ (i.e. any arbitrary asymptotically flat perturbation of the data set satisfies $\delta m=0$), the evolution of the initial data set is a stationary spacetime. In \cite{Sudarsky:1992ty}, Sudarski and Wald investigate this problem for the more general case of the Einstein-Yang-Mills case, and they motivate (without giving a complete proof) the conjecture that this result is valid.   

In a similar way, we define the \emph{canonical angular momentum}, $\mathcal{J}$, as (minus) the value of the correct Hamiltonian \eqref{eq:Hcor} in the case (ii), where $t^\mu$ approaches a rotation, and for initial data satisfying the constraints (so $H=0$), i.e.
\begin{equation}
\label{eq:J}
\mathcal{J}=-\frac{1}{16\pi} \lim_{r\to\infty} \int_{S_r^2}{dS n_i2\phi_j\pi_{ij}}.
\end{equation}
This is also called \emph{ADM angular momentum}\footnote{Once again, comparing \eqref{eq:J} with eq. (52) of \cite{Sudarsky:1992ty} or eq. (13) of \cite{Wald:1993ki}, we see that in Sudarski's and Wald's expressions for the canonical angular momentum there is one more term. This arises because they keep the second term of \eqref{eq:Sr24}.}. The fall-off conditions for asymptotically flat initial data do not ensure that the integral has a finite limit as $r\to\infty$. Thus, we have to require stronger fall-off conditions on the initial data in order that $\mathcal{J}$ be well defined. To find one possible set of such conditions, we convert the expression for $\mathcal{J}$ back to a volume integral on a total divergence term by using the divergence theorem (this is possible because $\Sigma$ has only the boundary at infinity). There is obviously more than one volume integral of a total divergence term such that it is equal (via the divergence theorem) to the the RHS of \eqref{eq:J}. One possibility, in terms of quantities that we have already defined, is
\begin{equation}
\mathcal{J}=-\frac{1}{16\pi}\int_{\Sigma}d^3 x\sqrt{h}D_i\biggl(2\phi_j\frac{\pi^{ij}}{\sqrt{h}}\biggr),
\end{equation}
where $\phi^a$ is any vector field on $\Sigma$  which asymptotically approaches a rotational Killing field of $\delta_{ij}$ in asymptotically Cartesian coordinates. Notice that the second term does not give contribution at infinity. Using the Leibniz rule and the constraint $\mathcal{C}_a=0$, we obtain
\begin{equation}
\label{eq:Jcond}
\mathcal{J}=-\frac{1}{16\pi}\int_{\Sigma}{d^3 x\pi^{ij}\mathcal{L}_\phi h_{ij}}.
\end{equation}
Since $d^3 x\sqrt{h}=\mathcal{O}(r^3)$ as $r\to\infty$, from this expression we see that if $\phi^a$ can be chosen so that $\mathcal{L}_\phi h_{ij}$ decays slightly faster than required by the asymptotically flat conditions, i.e.
\begin{equation}
\mathcal{L}_\phi h_{ij}=\mathcal{O}\biggl(\frac{1}{r^{1+\epsilon}}\biggr),
\end{equation}
then $\mathcal{J}$ is well-defined. Other possible sets of fall-off conditions can be imposed, but we will not discuss them here. We will merely assume that a set of conditions for the existence of $\mathcal{J}$ has been imposed in each context in which $\mathcal{J}$ is used.

We have, thereby, obtained that the correct Hamiltonian for an asymptotically flat surface $\Sigma$ with no other boundaries and for the time foliation (ii) is $H'=H-\mathcal{J}$. We can now prove a theorem analogous to the one above by employing a similar argument. Let us consider an asymptotically flat axisymmetric spacetime, with axial Killing field $m^\mu$ (which, by definition, approaches a rotation at infinity), that is solution of the evolution equations with asymptotically flat initial data $(h_{ab},\pi^{ab},A_a,\sqrt{h}E^a)$ satisfying the constraints. Let us choose $t^\mu=m^\mu$, so $\dot{g}_{\mu\nu}=0$. We assume that a gauge choice has been made so that the diffeomorphism generated by $t^\mu$ is a symmetry of $A_\mu$, i.e. $\mathcal{L}_t A_\mu=0$, so $\dot{A}_\mu=0$. Then, we have $\dot{h}_{ab}=0$, $\dot{\pi}^{ab}=0$, $\dot{A}_a=0$, $\dot{(\sqrt{h}E^a)}=0$. Therefore, the variation of $H'$ w.r.t an arbitrary asymptotically flat perturbation of the initial data (which is given by the RHS of \eqref{eq:varHnoboun}) is 0. We also assume that the perturbation $(\delta h_{ab},\delta \pi^{ab},\delta A_a,\delta(\sqrt{h}E^a))$ of the initial data satisfies the linearised constraints. Then, $\delta H=0$ as we noted above. In this way, we proved the following theorem.
\begin{theorem}
Let $(\Sigma,h_{ab},\pi^{ab},A_a,\sqrt{h}E^a)$ be an asymptotically flat end with initial data satisfying the constraints $\mathcal{C}=\mathcal{C}_{(0)}=\mathcal{C}_a=0$, and let $\Sigma$ have no interior boundary. Let us assume that the solution of the evolution equations $\eqref{eq:Hameq}$ with this initial data be an axisymmetric, asymptotically flat spacetime $(M=\mathbb{R}\times\Sigma,h_{ab}(t),\pi^{ab}(t),A_a(t),\sqrt{h}E^a(t))$. Consider an arbitrary asymptotically flat perturbation $(\delta h_{ab},\delta \pi^{ab},\delta A_a,\delta(\sqrt{h}E^a))$ (whose evolution is not necessarily an axisymmetric spacetime) of the initial data that satisfies the linearised constraints. Then, the corresponding variation of ADM mass is
\begin{equation}
\delta \mathcal{J}=0.
\end{equation}
\end{theorem}
In particular, this theorem implies that every asymptotically flat axisymmetric solution of the Einstein-Maxwell equations, with Cauchy surface that has no interior boundary, is an extremum of $\mathcal{J}$. Moreover, \eqref{eq:Jcond} shows that, if $\Sigma$ has only the boundary at infinity, $\mathcal{J}=0$ for every axisymmetric solution.

We now analyse the case where $(\Sigma,h_{ab},\pi^{ab},A_a,\sqrt{h}E^a)$ is an asymptotically flat end, but $\Sigma$ has also a smooth interior boundary $S$. Although a true Hamiltonian does not exist for this case, we can still obtain a generalisation of Theorem 1 above with a similar argument. The result is the first law of black hole mechanics. We define $H$ by \eqref{eq:Hcon}. We compute the variation $\delta H$ w.r.t. an asymptotically flat variation $(\delta h_{ab},\delta \pi^{ab},\delta A_a,\delta(\sqrt{h}E^a))$ of the initial data. We remove the covariant derivatives from the variations in order to obtain the RHS of \eqref{eq:varHnoboun}. We use the divergence theorem on the resulting total divergence terms. In this way, for each integral over the boundary of $\Sigma$ at infinity, we have an integral (with the same integrand, except for a minus sign) over the internal boundary $S$. With the same argument used above, $\mathcal{V}\equiv\lim_{r\to\infty}V=costant$ on $\Sigma$, where $r\equiv\sqrt{x^ix^i}$ and $x^i$ are asymptotically Cartesian coordinates on $\Sigma$. With regards to the time foliation, we are interested in the case where $t^\mu$ is a linear combination of a time translation $k$ and a rotation $m$ as explained above, i.e. the case in which in asymptotically Cartesian coordinates, $N\to 1$ and $N^i\to \Omega\phi^i$ as $r\to\infty$, where $\phi^i$ is a rotation Killing field of $\delta_{ij}$ and $\Omega$ is a constant. With these choices, the integrals over the boundary of $\Sigma$ at infinity that contribute to $\delta H$ are, once again, given by $\delta S_1$, $\delta S_2$, $\delta S_3$ written above, and so they give $-(\delta m+\mathcal{V}\delta Q-\Omega\delta\mathcal{J})$. On the other hand, all the integrals over $S$ contribute to $\delta H$, since no fall-off conditions on the initial data at $S$ have been imposed. It is precisely because of the presence of these interior boundary terms that it is not possible to find a true Hamiltonian $H'$ by requiring, as we did in the case where $\Sigma$ has no interior boundary, that its variation induced by an asymptotically flat variation of the initial data has no contribution from the boundaries of $\Sigma$. The contributions to $\delta H$ from integrals over $S$ can be read from \eqref{eq:Sr21},\eqref{eq:Sr22},\eqref{eq:Sr23},\eqref{eq:Sr24} simply by substituting $S_r^2$ with $S$, $n^i$ with the outward unit normal $r^i$ on $S$ in the tangent space of $\Sigma$. We also need to multiply by a minus sign due to the fact that both $r^i$ and $n^i$ point out of $\Sigma$. (More precisely, the minus sign can be explained as follows. By convention, we choose the orientation of $S$ and $S_r^2$ in the orientation class determined by Stokes' theorem regarding $S$ and $S_r^2$ as the boundaries of the regions containing the points internal to $S$ and $S_r^2$. As pointed out in \cite{Reall:2013grnotes}, since these regions have Riemannian metric, when we choose the direction of the unit normal to $S$ so that we get the correct sign in the divergence theorem applied to the region with points internal to $S$, we end up with a unit normal that points inside of $\Sigma$, i.e. $-r^i$. On the other hand, if we regarded $S$ as a boundary of $\Sigma$, we would obtain precisely the unit normal $r^i$. In both cases we would obtain the same unit normal $n^i$ on $S_r^2$. Hence, with our convention for the orientation on $S$ and $S_r^2$, the minus sign is necessary in the integrals over $S$ to apply correctly the divergence theorem to the region $\Sigma$ in the form $\int_\Sigma d^3x\sqrt{h}D_i X^i=\lim_{r\to\infty}\int_{S_r^2}dSn_i X^i -\int_{S}dSr_i X^i$.) Since we do not want to impose fall-off conditions for the various quantities at $S$ in some coordinate chart, we can also replace Latin coordinate indices with Latin abstract indices in these terms. If we now assume that the perturbation $(\delta h_{ab},\delta \pi^{ab},\delta A_a,\delta(\sqrt{h}E^a))$ satisfies the linearised constraints, then $\delta H=0$. Therefore, (also multiplying by $16\pi$) we have the result\footnote{Except for some different minus signs, we obtain eq. (57) of \cite{Sudarsky:1992ty}. However, in that expression, Sudarski and Wald omit the two terms of the last line of \eqref{eq:mis2}. These terms will not play any role in the following, but in principle they should be present in \eqref{eq:mis2}.}
\begin{equation}
\label{eq:mis2}
\begin{split}
&16\pi(\delta m+\mathcal{V}\delta Q-\Omega\delta\mathcal{J})=16\pi\int_{\Sigma}{d^3x[\mathcal{P}^{ab}\delta h_{ab}+\mathcal{Q}_{ab}\delta \pi^{ab}+\mathcal{R}^a\delta A_a+\mathcal{S}_a\delta(\sqrt{h}E^a)]}\\
&-\int_{S}{dS[r^a(D^b N)\delta h_{ab}-r^c(D_c N)h^{ab}\delta h_{ab}]}\\
&-\int_{S}dS \biggl\{-r^aND^b\delta h_{ab}+ r^c N D_c(h^{ab}\delta h_{ab}) -2r^c \biggl[N^a\frac{\pi^{bd}}{\sqrt{h}}h_{cd}\delta h_{ab}+h_{cd}N^d \frac{\pi^{ab}}{\sqrt{h}}\delta h_{ab}\biggl]\\
& -2r^a h_{bc}N^c h_{ad} \frac{\delta \pi^{bd}}{\sqrt{h}}\biggr\}-\int_{S}dSr_a 4\biggl[-N((D^b A^a-D^a A^b)\delta A_b) +(-NV+N^bA_b)\frac{1}{\sqrt{h}}\delta(\sqrt{h}E^a) \\
&-N^a\frac{\sqrt{h}E^b}{\sqrt{h}}\delta A_b+N^b\frac{\sqrt{h}E^a}{\sqrt{h}}\delta A_b\biggr].
\end{split}
\end{equation}

Suppose, now, that $(\Sigma,h_{ab},\pi^{ab},A_a,\sqrt{h}E^a)$ is an initial data set for a spacetime describing a stationary Einstein-Maxwell black hole whose event horizon is a bifurcate Killing horizon w.r.t. the Killing vector field $\xi^\mu$. Such black holes are believed to be the maximal analytic extensions of the only physically relevant black hole regions predicted by Einstein-Maxwell equations, as motivated at the beginning of section \ref{sec:appfirst}. Let $k^\mu$ be the stationary Killing vector field and let $B$, the interior boundary of $\Sigma$, be the bifurcation 2-surface. A theorem proved in \cite{Bardeen:1973gs} ensures that, if $\xi^\mu$ is not proportional to $k^\mu$, then the spacetime is axisymmetric w.r.t. to an axisymmetric Killing field that is a linear combination of $\xi^\mu$ and $k^\mu$. We can normalise $\xi^\mu$ so that $\xi^\mu=k^\mu+\Omega m^\mu$. This relation defines $\Omega$, which has the interpretation of being the angular velocity of the horizon (it is 0 if $\xi^\mu$ is proportional to $k^\mu$, i.e if the spacetime is merely stationary and not axisymmetric). In this situation, we can choose $t^\mu=\xi^\mu$. Then, since $\xi^\mu$ is a Killing vector field of the solution, $\dot{h}_{ab}=\dot{\pi}^{ab}=\dot{A}_s=\dot{(\sqrt{h}E)}^a=0$. Therefore, using the evolution equations, the integral over $\Sigma$ of expression \eqref{eq:mis2} vanishes. Moreover, since the Killing field $\xi^\mu$ vanishes on $B$ (see Chapter 9 of \cite{Reall:2017bhnotes} for an explanation), then $N=0$ and $N^a=0$ on $B$. So, the only non-vanishing contribution to the RHS of \eqref{eq:mis2} is given by the two terms of the second line. In appendix \ref{app:dec} it is proved that the energy-momentum tensor of the Maxwell field satisfies the dominant energy condition, hence the zeroth law of black hole mechanics (see, e.g., \cite{Wald:1984rg} and \cite{Reall:2017bhnotes}) states that the surface gravity $\kappa$ is constant on the Killing event horizon, and, in particular, on $B$. Using this result, in appendix \ref{app:sur} we prove that the two remaining terms give
\begin{equation}
\label{eq:la}
-\int_B dS[ r^a(D^bN)\delta h_{ab}-r^c (D_cN)h^{ab}\delta h_{ab}]=2\kappa\delta A[B],
\end{equation}
where $A[B]=\int_B dS=\int_B\sqrt{\sigma}dy^1 dy^2$ is the area of $B$ ($(y^1,y^2)$ is an arbitrary RH chart on $B$ w.r.t. the orientation on $B$ given by Stokes' theorem, and $\sigma$ is the determinant of the metric on $B$ induced by $h_{ab}$ in this chart). Now, for any black hole with bifurcate Killing event horizon we define the \emph{entropy} of a cross-section $C$ of the Killing horizon (see appendix \eqref{app:binsur}) by
\begin{equation}
\label{eq:entgr}
S[C]=\frac{A[C]}{4}.
\end{equation}
Since the Maxwell field satisfies the Null Energy Condition (because it satisfies the Dominant Energy Condition which implies the Null Energy Condition) and we assume that physical black hole spacetimes are strongly asymptotically predictable, the area theorem (see, e.g., \cite{Reall:2017bhnotes}) tells us that $S[C]$ satisfies the second law of black hole mechanics, i.e. $S[C']-S[C]\geq 0$ for any two cross-sections $C,C'$ with $C'$ in the causal future of $C$. (See sectiona \ref{sec:entfirst} and \ref{sec:second} for more details about this quantity in the more general case of diffeomorphism invariant theories). We have thereby proved the following theorem.

\begin{theorem}{\textbf{[First law of black hole mechanics]}}
Let $(\Sigma,h_{ab},\pi^{ab},A_a,\sqrt{h}E^a)$ be an asymptotically flat end with initial data satisfying the constraints $\mathcal{C}=\mathcal{C}_{(0)}=\mathcal{C}_a=0$, whose evolution is a stationary black hole spacetime with bifurcate Killing event horizon and the bifurcation surface $B$ is the interior boundary of $\Sigma$. Let $(\Sigma,h_{ab}+\delta h_{ab},\pi^{ab}+\delta \pi^{ab},A_a+\delta A_a,\sqrt{h}E^a+\delta(\sqrt{h}E^a))$ be arbitrary perturbed asymptotically flat initial data satisfying the constraints to linear order. Then,
\begin{equation}
\frac{\kappa}{2\pi}\delta S[B]=\delta m+\mathcal{V} \delta Q-\Omega\delta\mathcal{J}.
\end{equation}
\end{theorem}
As anticipated, this result generalises the first law of black hole mechanics proved by Bardeen, Carter and Hawking for perturbations to other asymptotically flat stationary black hole spacetimes \cite{Bardeen:1973gs}. We showed, in fact, that the first law holds also for perturbations to arbitrary asymptotically flat spacetimes.

In particular, this theorem implies that any asymptotically flat perturbation of an asymptotically flat stationary black hole solution of the Einstein-Maxwell equations with bifurcate Killing horizon and bifurcation surface $B$ such that $\delta Q=\delta\mathcal{J}=\delta S[B]=0$ satisfies $\delta m=0$. In other words, every asymptotically flat black hole solution of the Einstein-Maxwell equations with bifurcate Killing horizon and bifurcation surface $B$ is an extremum of $m$ at fixed $S[B], Q$, and $\mathcal{J}$.

As we did for Theorem 1, we can ask ourselves if the converse of this theorem holds, namely, whether, for an initial data set on an asymptotically flat Cauchy surface $\Sigma$ with interior boundary $B$ that extremises $m$ at fixed $S[B],Q$ and $\mathcal{J}$ (i.e. any arbitrary asymptotically flat perturbation of the data set with $\delta Q=\delta\mathcal{J}=\delta S[B]=0$ satisfies $\delta m=0$), the evolution of the initial data set is a stationary black hole spacetime with $B$ being the bifurcation surface of a bifurcate Killing horizon. Once again, Sudarski and Wald analyse the problem in \cite{Sudarsky:1992ty} in the more general context of the Einstein-Yang-Mills theory, and they motivate (without giving a complete proof) the conjecture that this result is valid.

\subsection{An application of the first law of black hole mechanics}
\label{sec:appfirst}

In this section, we will show how the first law of black hole mechanics, proved in the previous section, can be used to discover an important feature of physical stationary black holes, which enables to close a gap in one of the black hole uniqueness theorems (for a review of the black hole uniqueness theorems see \cite{Reall:2017bhnotes}). This discussion is presented in \cite{Wald:1993ki} and uses some results of \cite{Sudarsky:1992ty}. We emphasise that the following discussion relies on the Einstein-Maxwell equations, and so it cannot be extended to more general theories, such as those considered in the second part of the essay. (In fact, we do not know whether the black hole spacetimes predicted by those theories are physical or not, so extending the following discussion to those cases may not be interesting.)

To explain why this gap arises we need to motivate the claim that every physically relevant stationary Einstein-Maxwell black hole has an event horizon that is a bifurcate Killing horizon. More precisely, every physically relevant stationary Einstein-Maxwell black hole region has a maximal analytic extension that contains a bifurcate Killing horizon. We start from a result, proved by Hawking \cite{Hawking:1971vc} for the Einstein theory, whose extension to Einstein-Maxwell theory says that, for a stationary, analytic, asymptotically flat black hole solution of the Einstein-Maxwell equations, the event horizon is a Killing horizon. In the following we will accept the unphysical analyticity condition, i.e. we will extend the results proved with this assumption to physical spacetimes (which are not analytic). Hence, we can argue that every physical stationary Einstein-Maxwell black hole has a Killing horizon (the asymptotic flatness condition is implicitly required for any physical spacetime). We always assume that the surface gravity $\kappa$ on the Killing horizon of all the stationary black holes that we will consider is non-vanishing (the degenerate case $\kappa=0$ is uninteresting from a physical point of view). Finally, the fact that $\kappa$ is constant on the Killing horizon for the Einstein-Maxwell theory (we argued above that the hypotheses of the zeroth law are satisfied in this theory) implies that the Killing horizon of every stationary Einstein-Maxwell black hole can be extended (if necessary) to a bifurcate Killing horizon with bifurcation surface (see \cite{Racz:1992bp}). Therefore, we can argue that every physical Einstein-Maxwell stationary blak hole has a bifurcate Killing event horizon.

Let now $\xi$ be the Killing field that generates the Killing horizon with bifurcation surface $B$. As explained above, Hawking proves in \cite{Hawking:1973uf} that if $\xi$ is not the stationary Killing field $k$, then the black hole is stationary and axisymmetric with axial Killing field $m$ given by a linear combination of $\xi$ and $k$. Moreover, $\xi$ can be normalised so that $\xi=k+\Omega m$, and this relation defines the angular velocity of the black hole $\Omega$ (if $\xi=k$, in which case the black hole is said \emph{non-axisymmetric} or \emph{non-rotating}, this relation obviously still holds with $\Omega=0$). Another theorem proved by Hawking (\cite{Hawking:1971vc},\cite{Hawking:1973uf}), then, ensures that a stationary, asymptotically flat analytic solution of the Einstein-Maxwell equations, that is suitably regular on and outside an event horizon (we will not explain precisely the meaning of this regularity assumption), is also axisymmetric if it is not static. So $\xi$ does not coincide with $k$, i.e. $\Omega\neq 0$. Accepting the (unphysical) analyticity assumption, the literature usually refers to this result as ``stationary, but not static, implies axisymmetric for black holes''. Furthermore, under the same assumptions, if the stationary black hole is non-axisymmetric (i.e. $\xi=k$, i.e. $\Omega=0$) AND $k$ is globally timelike outside of the black hole (recall that the stationarity condition only requires that $k$ is timelike at infinity), then the spacetime is static. The literature usually refers to the combination of these two results as ``a stationary black hole is either static or axisymmetric''. However, an important case has not been considered: the stationary black hole may be non-axisymmetric but $k$ may be non-globally timelike outside the event horizon. There may exist stationary black hole solutions of the Einstein-Maxwell equations that are neither axisymmetric nor static. Since for such spacetimes $k$ is not globally timelike outside the event horizon, there must exist a non-trivial ergoregion (i.e. the region of spacetime where $k$ is spacelike). Furthermore, since $\xi=k$ is timelike at infinity and it is null on the event horizon (because the horizon is Killing, if we accept the analyticity the condition), then the ergoregion must be disjoint from the horizon (to be rigorous we should use the fact, proved in \cite{Kay:1988mu}, that $\xi$ is always timelike in a neighbourhood of the horizon outside the black hole). In this case Hawking's theorem does not apply, so we cannot argue that the stationary black hole is static. As we will explain, this is the gap closed by the first law of black hole mechanics proved above. In fact, it can be shown, using the first law, that stationary, non-axisymmetric Einstein-Maxwell black hole spacetimes whose ergoregion is disjoint from the horizon must be static (i.e. the ergoregion is trivial), even if we do not assume that the stationary Killing field $k$ is globally timelike. Hence, the statement ``a stationary black hole is either static or axisymmetric'' is correct in Einstein-Maxwell theory (with the assumptions made above). 

Before moving on to the proof of this result, we notice that a consequence of this is that black hole solutions of the Einstein-Maxwell theory with ergoregion disjoint from the horizon cannot exist. Let us prove this by contradiction. Let us assume that we have a black hole with a non-trivial ergoregion disjoint from the horizon. To define a non-trivial ergoregion the black hole must have a stationary Killing field $k$, so it is either static (so the ergoregion is trivial, which is a contradiction) or axisymmetric. However, by the results of Carter \cite{Carter:1973rla}, Robinson \cite{Robinson:1975bv}, Mazur \cite{Mazur:1982db} and Bunting (described and compared by Carter \cite{carter1985}), a stationary, axisymmetric, black hole solution of the Einstein-Maxwell equations is a member of the Kerr-Newman family, whose ergoregion is not disjoint from the event horizon. So we have a contradiction.

Let us now present the application of the first law that closes the gap explained above. Let us consider a stationary, non-axisymmetric (i.e. $\Omega=0$) black hole spacetime. Accepting the analyticity condition, the event horizon will be a Killing horizon w.r.t. the stationary killing field $\xi=k$. In this case, Theorem 3 is valid with $\Omega=0$, i.e.
\begin{equation}
\frac{\kappa}{2\pi}\delta S[B]=\delta m+\mathcal{V}\delta Q.
\end{equation}
From this equation we see that if we perturb an initial data set for a stationary, non-axisymmetric black hole solution of the Einstein-Maxwell equations with an arbitrary asymptotically flat perturbation satisfing the linearised constraints and $\delta Q=\delta S[B]=0$, we have $\delta m=0$. The proof that a stationary, non-axisymmetric Einstein-Maxwell black hole spacetime with ergoregion disjoint from the horizon must be static can be obtained by contradiction: we assume that it is not static, and we show that it is then possible to find an asymptotically flat perturbation of an initial data set for the stationary, non-axisymmetric black hole that satisfies the linearised constraints and $\delta Q=\delta S[B]=0$, but $\delta m\neq 0$.

In \cite{Chrusciel:1993cv} it is shown that we can always choose an asymptotically flat initial slice $\Sigma$ such that the trace of its extrinsic curvature, $K$, vanishes, intersects the bifurcation surface $B$, and is asymptotically orthogonal to $k$ at infinity. Let us choose such an initial slice and initial data $(h_{ab},\pi^{ab},A_a,\sqrt{h}E^a)$ on $\Sigma$ whose Einstein-Maxwell evolution is a stationary non-axisymmetric black hole with bifurcate Killing horizon and bifurcation surface $B$.
The desired perturbation has the form
\begin{align}
\delta h_{ab}&=4\phi h_{ab},\\
\delta \pi^{ab}&=-4\phi \pi^{ab}-\pi^{ab},\\
\delta A_a&=-A_a,\\
\delta E_a&=-6\phi E^a \qquad \text{or, equivalently} \delta(\sqrt{h}E^a)=2\phi\sqrt{h}E^a,
\end{align}
where $\phi$ is the solution to
\begin{equation}
D^a D_a\phi-\mu\phi=\rho
\end{equation}
on $\Sigma$ determined by the boundary conditions $\phi\to 0$ at infinity (i.e. in the limit $r=\sqrt{x^ix^i}\to\infty$, where $x^i$ are the asymptotically Cartesian coordinates on $\Sigma$) and $\phi=0$ on $B$, where
\begin{equation}
\begin{split}
\mu&=\frac{1}{h}\pi^{ab}\pi_{ab}+E^a E_a+\frac{1}{2}F^{ab}F_{ab},\\
\rho&=\frac{1}{4}\biggl(\frac{1}{h}\pi^{ab}\pi_{ab}+F^{ab}F_{ab}\biggr).
\end{split}
\end{equation}
It can be shown with a direct computation that this variation of the initial data satisfies the linearised constraints and also $\delta Q=\delta S[B]=0$. However, in \cite{Sudarsky:1992ty} it is shown that $\delta m<0$ unless $\rho=0$. Hence, we have a contradiction of the corollary of the first law above unless $\pi^{ab}=0$ and $F_{ab}=0$. We will now prove that this is not possible since it would imply that the spacetime is static, which is not possible by our hypothesis.

Since $K=0$ on $\Sigma$, $\pi^{ab}=0$ would imply that $K^{ab}=\frac{\pi^{ab}}{\sqrt{h}}=0$ on $\Sigma$. But since $k$ is a Killing field, the other slices $\Sigma_t$ obtained by considering the integral curves $\gamma(t)$ of $k$ through each point of $\Sigma$ at fixed $t$ would also have vanishing extrinsic curvature (in fact, since the time-evolution vector field $k$ is a Killing field, we have $\dot{\pi^{ab}}(t)=0$, which implies $\pi(t)\propto K(t)=0$ for any $t$, so $\dot{\pi^{ab}}(t)=\dot{\sqrt{h(t)}}K^{ab}(t)+\sqrt{h(t)}\dot{K^{ab}}(t)=(1/2)\sqrt{h(t)}h^{cd}(t)\dot{h}_{cd}(t) K^{ab}(t)+\sqrt{h(t)}\dot{K^{ab}}(t)$; hence, since $\dot{h}_{cd}(t)=0$, we have $\dot{K^{ab}}(t)=0$, i.e. the extrinsic curvature is the same on each slice $\Sigma_t$), so $\pi^{ab}(t)=0$ on each slice $\Sigma_t$. Furthermore, from $F_{ab}=0$ on $\Sigma$ we would have $E_a=0$ on $\Sigma$. Since $k$ is a Killing field, we have $\dot{E}^a(t)=0$, so $E^a=0$ on each slice $\Sigma_t$.

Let $k'^\mu=-(k^\nu n_\nu)n^\mu$ be the part of $k$ normal to each $\Sigma_t$ (here $n^\mu$ is the past-directed unit vector field normal to each $\Sigma_t$), and let us consider now the time evolution given by $k'^\mu$ (so, for example, $\dot{h}_{\mu\nu}$ means now $h_\mu^\rho h_\nu^\sigma\mathcal{L}_{k'}h_{\rho\sigma}$). Since $\pi^{ab}$ vanishes everywhere, we have $\dot{\pi}^{ab}=0$ everywhere. Then, from the evolution equation for $h_{ab}$ w.r.t. $k'$ and the fact that $\pi^{ab}=0$ and $N^a=0$ everywhere ($k'$ has no tangent component to $\Sigma_t$), we also have $\dot{h}_{ab}=0$ everywhere. Similarly, since $E^a$ vanishes everywhere, if we choose a gauge for which $V=0$ everywhere, from the evolution equations for $A_a$ w.r.t. $k'$ we obtain $\dot{A}_a=0$ everywhere. So, we proved that there exists a gauge such that $k'$ is a symmetry of $g_{\mu\nu}$ and $A_\mu$. Finally, since $k'$ is timelike at infinity and is orthogonal to the family of hypersurfaces $\Sigma_t$ by construction, the spacetime would also be static. Since we started assuming that the spacetime is not static, then $\rho$ cannot be 0 and we find the desired contradiction.

\section{The first law of black hole mechanics for diffeomorphism invariant theories}
\label{sec:20}

In the second part of the essay, we will consider diffeomorphism invariant theories of gravity, i.e. those theories whose only property is that their action is invariant under (orientation-preserving) diffeomorphisms acting on the dynamical fields, i.e. the metric and the matter fields. They admit an arbitrary number of derivatives of the fields. Einstein theory in vacuum and Einstein-Maxwell theory are two examples. Racz and Wald argue in \cite{Racz:1992bp} that the zeroth law of black hole mechanics (i.e. the fact that under certain hypotheses the surface gravity of a stationary black hole is constant over the event horizon) is equivalent to the statement that the Killing event horizon is of bifurcate type \cite{Racz:1992bp} (except for the uninteresting case $\kappa=0$; at the end of the second paragraph of \ref{sec:appfirst} we used precisely one implication of this equivalence). They prove this result without using a specific form of the field equations, but only geometrical properties. Therefore, the proof is valid for any diffeomorphism invariant theory, so we have that the zeroth law is valid for stationary black holes with bifurcate Killing horizon predicted by arbitrary theories of gravity.

We will show that, for this type of black hole spacetimes, it is also possible to generalise Theorem 3, i.e. the first law of black hole mechanics. This proof is presented by Iyer and Wald in \cite{Iyer:1994ys} and it involves the machinery of the symplectic formulation for general field theories with local gauge symmetries (in the case of our interest, the gauge symmetries are diffeomorphisms) explained in \cite{Lee:1990nz}, which generalises the symplectic formalism for classical mechanics summarised in appendix \ref{app:symmech}. We will not present here the complete symplectic formalism, but we will just introduce the quantities that are essential to prove the first law of black hole mechanics, and we will provide an intuitive explanation for their conceptual meaning by using analogies with the symplectic formalism for classical mechanics. Furthermore, before arriving at the actual proof of the first law of black hole mechanics, we will list (without proof) some important results whose (rather technical) proves are given in \cite{Iyer:1994ys}.
Finally, we will briefly discuss about the possibility of find a definition of the entropy for a dynamical black hole and a generalisation of the second law of black hole mechanics.

\textbf{Notation and conventions} In the second part of this work, in order to make contact with the notation of \cite{Iyer:1994ys}, $a,b,c,d,\dots$ will be abstract indices used to denote abstract tensors on the spacetime manifold, while $\mu,\nu,\rho,\alpha,\beta,\gamma,\dots$ will indicate tensor components in some coordinate basis. We will also use $i,j,k,l,\dots$ for the components of tensors defined on a hypersurface in some coordinate basis. We will always denote differential forms on the spacetime manifold by boldface letters, e.g. $\bm{L},\bm{Q}$, etc.

We will again consider orientable and time-orientable, globally hyperbolic spacetimes. We will always assume that a suitable definition of asymptotic flatness can be given, so that there exists (at least one set of) \emph{asymptotically inertial coordinates}, $x^\alpha$, defined by a straightforward generalisation of \eqref{eq:asyfla1} and \eqref{eq:asyfla2}. Similarly, we assume that a suitable generalisation of the definition of an asymptotically flat end can be given, so that the the metric approaches the Minkowski metric and the matter fields, together with the derivatives of the dynamical fields, decay fast enough at infinity to ensure that quantities of interest are well-defined, but not so fast that a sufficiently large class of solutions fails to exist. The precise fall-off conditions for asymptotic flatness strictly depend on the details of the theory, and, thus, must be examined case by case.

Stationary and axisymmetric spacetimes are then defined by a straightforward generalisation of the definitions given in section \ref{sec:first1}. In particular, an $n$-dimensional spacetime is said to be \emph{stationary and axisymmetric} if it is stationary with stationary Killing vector field $k$ and there exists a family of Killing vector fields $m^a_{(\alpha)}$, $\alpha=1,2,\dots,\left\lfloor\frac{n-1}{2}\right\rfloor$ (where $\lfloor a\rfloor$ denotes the floor value of $a$, e.g., $\lfloor 1.5\rfloor=1$ and $\lfloor 2\rfloor=2$), called \emph{axial Killing vector fields}, and their commutator with each other and with $k$ vanishes. In these definitions we also require that the Killing vector fields are symmetries of the matter fields.

\subsection{Some properties of diffeomorphism invariant field theories}
\label{sec:2}

Let us consider a field theory on an $n$-dimensional orientable and time-orientable spacetime manifold $M$ with Lorentz signature metric $g_{ab}$. We need to specify the field content of the theory. Let $g_{ab}$ and a collection of other ``matter'' tensor fields $\psi$ on $M$ be dynamical fields. Let $\mathring{\nabla}$ be an arbitrary connection on $M$. We will refer to all the dynamical fields $g_{ab}$ and $\psi$ as $\phi$. We denote the collection of background fields, such as the Riemann tensor $\mathring{R}^a_{\;\;bcd}$ of $\mathring{\nabla}$, by $\mathring{\gamma}$. These fields do not depend on the dynamical fields so they do not change under a variation of $(g_{ab},\psi)$. We define the Lagrangian of the theory as an n-form $\bm{L}$, i.e. we include in $\bm{L}$ the volume form $\bm{\epsilon}$ of $M$, i.e. $\bm{L}=L\bm{\epsilon}$, where $L$ is the scalar Lagrangian. We assume that $\bm{L}$ depends on the above defined quantities in the following sense
\begin{equation}
\label{eq:Lform}
\bm{L}=\bm{L}(g_{ab},\mathring{\nabla}_{a_1}g_{ab},\dots,\mathring{\nabla}_{(a_1\dots}\mathring{\nabla}_{a_k)}g_{ab},\psi,\mathring{\nabla}_{a_1}\psi,\dots,\mathring{\nabla}_{(a_1\dots}\mathring{\nabla}_{a_l)}\psi,\mathring{\gamma}).
\end{equation}
We also assume that $\bm{L}$ is local, i.e. all the fields and the covariant derivatives are function of the same point in $M$ (see \cite{doi:10.1063/1.528839} for a rigorous definition).

Let us consider the space $\mathscr{F}$ of all possible field configurations. Any field configuration in the spacetime represents a point in $\mathscr{F}$ denoted by $\phi$. The set of field configurations that satisfy the equations of motion of the theory (called \emph{on-shell} configurations) forms a subset of $\mathscr{F}$, denoted by $\mathscr{\bar{F}}$, which we assume to be a manifold. An infinitesimal field perturbation $\delta \phi$  over a field configuration $\phi$ corresponds to a vector tangent to a smooth one-parameter family of field configurations at $\phi\in\mathscr{F}$, which is denoted by $(\delta\phi)^A$. In general, we will use capital letters, e.g. A,B,C,etc., for tensors on $\mathscr{F}$.

We will consider only diffeomorphism invariant theories. To ensure the invariance of the action under (orientation-preserving) diffeomorphisms the Lagrangian must be diffeomorphism covariant under any (orientation-preserving) diffeomorphism $f\colon M\to M$, i.e.
\begin{equation}
\label{eq:covcon}
\bm{L}(f^\ast(\phi))=f^\ast(\bm{L}(\phi)),
\end{equation}
where $f^\ast$ denotes the pull-back under $f$. We emphasize that $f^\ast$ acts only on the dynamical fields of the theory.
An important result, proved in \cite{Iyer:1994ys}, is
\begin{lemma}
Let $\bm{L}$ be a Lagrangian with dependance on the fields as in \eqref{eq:Lform}. If $\bm{L}$ is diffeomorphism covariant, then $\bm{L}$ can be re-expressed in a manifestly covariant way, and with is no dependance on the background fields. More precisely,
\begin{equation}
\label{eq:Lform2}
\bm{L}=\bm{L}(g_{ab},\nabla_{a_1}R_{bcde},\dots,\nabla_{(a_1\dots}\nabla_{a_m)}R_{bcde},\psi,\nabla_{a_1}\psi,\dots,\nabla_{(a_1\dots}\nabla_{a_l)}\psi),
\end{equation}
where $\nabla$ is the Levi-Civita connection of $g_{ab}$, $m=max(k-2,l-2)$, and $R_{abcd}$ denotes the Riemann tensor of $\nabla$ (which is determined by $g_{ab}$.
\end{lemma}
For an $\bm{L}$ with this form, the condition \eqref{eq:covcon} is obviously satisfied simply by using the definition of pull-back.
Notice that antisymmetrised covariant derivatives can always be written in terms of lower order derivatives and the Riemann tensor by the Ricci identity (since the Levi-Civita connection is torsion-free), therefore we can include the dependence of $\bm{L}$ on a finite number of covariant derivatives just by writing the dependence on symmetrised covariant derivatives.

Let us consider the variation $\delta\bm{L}$ w.r.t. an arbitrary variation $\delta \phi=(\delta g_{ab},\delta \psi)$ of the dynamical fields on $M$ to linear order in $\delta\phi$. Using the Leibniz rule to remove the covariant derivatives from $\delta g_{ab}$ and $\delta \psi$ at the cost of obtaining total divergence terms, we can write $\delta\bm{L}$ in the form
\begin{equation}
\label{eq:varL}
\delta \bm{L}=\bm{E}\delta\phi+d\bm{\Theta},
\end{equation}
with
\begin{equation}
\label{eq:eqmot}
\bm{E}\delta\phi=(\bm{E}_g)^{ab}\delta g_{ab}+\bm{E}_\psi\delta\psi,
\end{equation}
where a sum over the matter fields $\psi$ is understood, and it is also understood that for each matter field $\bm{E}_\psi$ has tensor indices dual to $\psi$, and these indices are contracted with those of $\delta\psi$ at the RHS of \eqref{eq:eqmot}. For a variation $\delta\phi$ of compact support on $M$, the $d\bm{\theta}$ term does not give contribution to the action (by Stokes' theorem), so we can read the equations of motion of the theory:
\begin{equation}
(\bm{E}_g)^{ab}=0 \qquad \text{and} \qquad \bm{E}_\psi=0.
\end{equation}
By construction, $(\bm{E}_g)^{ab}$ and $\bm{E}_\psi$ depend locally on the dynamical fields and their derivatives but not on the variations. The $(n-1)$-form $\bm{\Theta}$ is called the \emph{symplectic potential form}. By construction, it depends locally on the dynamical fields $\phi$ and their derivatives, and it is linear in $\delta\phi$.
$\bm{\Theta}$ is determined by \eqref{eq:varL} up to the addition of a closed $(n-1)$-form (i.e. a $(n-1)$-form $\bm{X}$ such that $d\bm{X}=0$ everywhere on $M$). A theorem proved in \cite{doi:10.1063/1.528839} ensures that for any $p$-form $\bm{\alpha}$ locally dependent (in the sense explained above for the Lagrangian) on the dynamical fields $\phi$, the background fields $\psi$ and finitely many of their covariant derivatives (w.r.t. an arbitrary connection), that it is closed for every field configuration $\phi$ and it is 0 if all the dynamical fields are 0, then there exists a $(p-1)$-form $\bm{\beta}$ such that $\bm{\alpha}=d\bm{\beta}$ locally, and $\bm{\beta}$ depends locally on $\phi$, $\psi$ and finitely many of their derivatives. In other words, closedness and exactness is equivalent for differential forms locally dependent on the dynamical and background fields. This theorem holds for all the forms that we will consider in the following (in the cases of our interest there will be no dependence on the background fields and the connection will be the Levi-Civita connection of $g_{ab}$).

The arbitrariness in $\bm{\Theta}$ can be partially fixed, as explained by the following Lemma.
\begin{lemma}
Given a diffeomorphism covariant Lagrangian of the form \eqref{eq:Lform2}, the variation $\delta \bm{L}$ w.r.t. an arbitrary variation of the dynamical fields can be written in the form \eqref{eq:varL}, where $\bm{\Theta}$ can be chosen so that it is a diffeomorphism covariant $(n-1)$-form of the form
\begin{equation}
\label{eq:canfor}
\bm{\Theta}=2\bm{E}_R^{bcd}\nabla_d\delta g_{bc}+\bm{\Theta'},
\end{equation}
where
\begin{equation}
\label{eq:24}
\bm{\Theta'}=\bm{S}(\phi)\delta g_{ab}+\sum_{i=0}^{m-1}\bm{T}_i(\phi)^{abcda_1\dots a_i}\delta\nabla_{(a_1\dots}\nabla_{a_i)}R_{abcd}+\sum_{i=0}^{l-1}\bm{U}_i(\phi)^{a_1\dots a_i}\delta \nabla_{(a_1\dots }\nabla_{a_i)}\psi.
\end{equation}
In other words, in the expression for $\bm{\Theta}$, the $\delta$'s can be put to the left of derivatives of the dynamical fields everywhere except for the single term $\bm{E}_R^{bcd}\nabla_d\delta g_{bc}$. Finally, $\bm{E}_R^{bcd}$ is given by
\begin{equation}
(\bm{E}_R^{bcd})_{b_2\dots b_n}=E_R^{abcd}\bm{\epsilon}_{ab_2\dots b_n},
\end{equation}
where $E_R^{abcd}\bm{\epsilon}_{ab_2\dots b_n}$ is the equation of motion form that would be obtained for $R_{abcd}$ if it were viewed as an independent field in the Lagrangian \eqref{eq:Lform2} rather than a quantity determined by the metric, i.e.
\begin{equation}
\label{eq:ER}
E_R^{abcd}\equiv\frac{\partial L}{\partial R_{abcd}}-\nabla_{a_1}\frac{\partial L}{\partial \nabla_{a_1}R_{abcd}}+\dots+(-1)^m\nabla_{(a_1\dots }\nabla_{a_m)}\frac{\partial L}{\partial \nabla_{(a_1\dots }\nabla_{a_m)}R_{abcd}}=0.
\end{equation}
\end{lemma}

The proof of this result is given by Iyer and Wald in \cite{Iyer:1994ys}. This lemma shows that we can always require that $\bm{\Theta}$ is diffeomorphism covariant. We will always assume that such a choice for $\bm{\Theta}$ has been made in the following. Furthermore, the lemma also gives the so-called canonical form \eqref{eq:canfor} of $\bm{\Theta}$. However, this general form does not determine $\bm{\Theta}$ uniquely, in fact the canonical form is preserved by adding an exact $(n-1)$-form of the same form as the RHS of \eqref{eq:24}, i.e.
\begin{equation}
\bm{\Theta}\to\bm{\Theta}+d\bm{Y}(\phi,\delta\phi),
\end{equation}
where $\bm{Y}$ is a diffeomorphism covariant $(n-2)$-form linear in the varied fields $\delta\phi$. The choice of $d\bm{Y}$ does not still fix $\bm{\Theta}$ uniquely, in fact we can replace $\bm{L}$ by $\bm{L}+d\bm{\mu}$, since the equations of motion are unaffected\footnote{In fact, to determine the equations of motion we consider a variation of $\delta \phi$ of compact support on $M$. Adding the $d\bm{\mu}$ term, the variation of the Lagrangian becomes $\delta\bm{L}\to\delta\bm{L}+\delta d\bm{\mu}$. Now, in any coordinate chart $\delta d\bm{\mu}_{\alpha_1\dots \alpha_n}=n\delta \partial_{[\alpha_1}\bm{\mu}_{\alpha_2\dots\alpha_n]}=n \partial_{[\alpha_1}\delta\bm{\mu}_{\alpha_2\dots\alpha_n]}= d\delta\bm{\mu}_{\alpha_1\dots \alpha_n}$. Since this is a relation between tensors, it holds in any chart, so $\delta d\bm{\mu}=d\delta\bm{\mu}$. Using Stokes' theorem and the fact that $\delta\bm{\mu}$ is of compact support on $M$, we have that the term $\int_M{ d\delta\bm{\mu}}=\int_{\partial M}\delta\bm{\mu}=0$. So, adding $d\bm{\mu}$ does not change the action of the theory, and hence the equations of motion.}, so the physical content of the theory is unchanged. However, this substitution changes
\begin{equation}
\bm{\Theta}\to\bm{\Theta}+\delta\bm{\mu}.
\end{equation}
We will present below the consequences of the freedom
\begin{equation}
\label{eq:thetaamb}
\bm{\Theta}\to\bm{\Theta}+\delta\bm{\mu}+d\bm{Y}(\phi,\delta\phi)
\end{equation}
in the choice of $\bm{\Theta}$. 
(To be precise, the proof of the lemma provides implicitly an algorithm that determines $\bm{\Theta}$ uniquely. However, there is no reason to use that particular algorithm for $\bm{\Theta}$, thus we do not assume that the algorithm has been used to determine $\bm{\Theta}$, so the two sources of ambiguity explained above are present.)

For globally hyperbolic spacetimes, we now define the \emph{symplectic current} $(n-1)$-form by
\begin{equation}
\label{eq:omega}
\bm{\omega}(\phi,\delta_1\phi,\delta_2\phi)=\delta_2\bm{\Theta}(\phi,\delta_1\phi)-\delta_1\bm{\Theta}(\phi,\delta_2\phi),
\end{equation}
where $\delta_1\phi$ and $\delta_2\phi$ are two independent variations of the configuration $\phi$ of the dynamical fields.
$\bm{\omega}$ has the property that, if $\delta_1\phi, \delta_2\phi$ are solutions of the linearised equations of motion, (i.e. $\delta_1\bm{E}(\phi)=\delta_2\bm{E}(\phi)=0$ at linear order in $\delta_1\phi,\delta_2\phi$), then $d\bm{\omega}=0$. In fact, since $d\delta\bm{\Theta}=\delta d\bm{\Theta}$ and $\delta_1$ commute with $\delta_2$, we have
\begin{equation}
\begin{split}
d\bm{\omega}(\phi,\delta_1\phi,\delta_2\phi)&=\delta_2 d\bm{\Theta}(\phi,\delta_1\phi)-\delta_1 d\bm{\Theta}(\phi,\delta_2\phi)\\
&=\delta_2 (\delta_1\bm{L}-\bm{E}(\phi)\delta_1\phi)-\delta_1(\delta_2\bm{L}-\bm{E}(\phi)\delta_2\phi)=0.
\end{split}
\end{equation}
Let $\Sigma$ be a Cauchy surface with orientation induced by the orientation on $M$, as prescribed by Stokes' theorem considering $\Sigma$ as the boundary of the causal past of $\Sigma$. We define the \emph{pre-symplectic form} $\Omega$ on $\mathscr{F}$ by
\begin{equation}
\label{eq:Omega}
\Omega(\phi,\delta_1\phi,\delta_2\phi)=\int_\Sigma{\bm{\omega}(\phi,\delta_1\phi,\delta_2\phi)}.
\end{equation}
Since $\Omega$ is linear in $\delta_1\phi,\delta_2\phi$, this expression defines a 2-form $\Omega_{AB}$ at $\phi\in\mathscr{F}$ by $\Omega_{AB}(\delta_1\phi)^A(\delta_2\phi)^B=\Omega(\phi,\delta_1\phi,\delta_2\phi)$ for any $(\delta_1\phi)^A,(\delta_2\phi)^B$. $\phi$ is arbitrary so we can regard $\Omega_{AB}$ as a 2-form field on $\mathscr{F}$.

The importance of this quantity is explained in \cite{Lee:1990nz}. In particular, it allows to build the phase space of our theory starting from the field configuration space.

\noindent
\textbf{Aside} It is possible to explain intuitively such a construction as follows. $\Omega_{AB}$ is degenerate, i.e. there exist non-zero vector fields $\psi^A$ on $\mathscr{F}$ such that $\Omega_{AB}\psi^A=0$. This means that we are describing the system by using too many degrees of freedom, some of which are unphysical (which is the underlying reason for the presence of diffeomorphism gauge invariance). Our purpose is to build a space, namely the phase space, where $\Omega_{AB}$ is non-degenerate. Consider the degeneracy vectors of the 2-form $\Omega_{AB}$ at $\phi$ for any $\phi\in\mathscr{F}$. Each set of degeneracy vectors is a subspace of the tangent space at $\phi$. Consider the union $W$  of all these subspaces for each $\phi\in\mathscr{F}$. 
We can regard $W$ as the space of degeneracy vector fields $\psi^A$ for the 2-form field $\Omega_{AB}$ by simply defining a vector field $\psi^A$ as a map that associates the point $\phi$ with one degeneracy vector of the tangent space at $\phi$ in a smooth way. It turns out that the commutator of each pair of degeneracy vector fields is still a degeneracy vector field. Therefore, by the vector form of Frobenius' theorem (see \cite{Wald:1984rg}), we have that $W$ possesses integral submanifolds, i.e. there exist submanifolds of $\mathscr{F}$ whose union of tangent spaces at each point forms precisely $W$ (in other words, integral submanifolds are the higher dimensional analog of the integral curve of a vector field). Thus, we can define an equivalence relation on $\mathscr{F}$ by setting $\phi_1\simeq\phi_2$ if $\phi_1,\phi_2$ lie on the same submanifold. The phase space $\Gamma$ is defined as the set of equivalence classes of $\mathscr{F}$ and it is assumed to be a manifold. Let $\pi\colon\mathscr{F}\to\Gamma$ be the map that assigns each element of $\mathscr{F}$ to its equivalence class. We define the \emph{symplectic form} on $\Gamma$ as the 2-form whose pull-back on $\mathscr{F}$ w.r.t. $\pi$ is $\Omega_{AB}$. We still denote the symplectic form on $\Gamma$ by $\Omega_{AB}$, which is non-degenerate on $\Gamma$. $(\Gamma,\Omega_{AB})$ is a symplectic manifold ( defined in appendix \ref{app:symmech}). With physical terminology, it is the \emph{phase space} of the theory. \textbf{End of aside}

It is beyond the purposes of this essay to explain in detail the symplectic formalism, however $\Omega$ will be relevant in the following since it is used to define the Hamiltonian. We will list now some properties that we expect a nice pre-symplectic form (or just symplectic form, if regarded as a form on the phase space) to have.
When $\Sigma$ has a boundary at infinity, which happens, e.g., for asymptotically flat spacetimes, we assume that fall-off conditions on the fields $\phi$ have been imposed so that $g_{ab}$ approaches the Minkowski metric $\eta_{ab}$ and the matter fields decay fast enough for $\Omega$ to exist. We can investigate the dependence of $\Omega$ on the Cauchy surface $\Sigma$.
We also want $\Omega$ to be independent of the choice of Cauchy surface when the equations of motion are imposed on $\phi$ and the variations $\delta\phi$. Since $d\bm{\omega}=0$ when $\delta_1\phi,\delta_2\phi$ satisfy the linearised equations of motion, then if $\Sigma,\Sigma'$ are two Cauchy surfaces, $\Omega[\Sigma']-\Omega[\Sigma]$ is given by the flux $\bm{\omega}$ through a timelike surface $C$ at the boundary of $M$, $\int_C{\bm{\omega}}$. Therefore, we choose our fall-off conditions by requiring also that such a flux vanishes for every timelike surface $C$. Finally, we can wonder if the freedom in the choice of $\bm{\Theta}$ can affect $\Omega$. Adding the $\delta\bm{\mu}$ terms to $\bm{\Theta}$ does not change $\bm{\omega}$ (because $\delta_1\delta_2\bm{\mu}=\delta_2\delta_1\bm{\mu}$), and so also $\Omega$. Nevertheless, shifting $\bm{\omega}$ by $d\bm{Y}$ may change $\Omega$ by a quantity $\Delta \Omega$. However, the fall-off conditions that ensure the existence of $\Omega$ typically imply that $\Delta\Omega=0$. So $\Omega$ is typically independent of the choice of $\bm{\Theta}$ for suitable asymptotic conditions on the dynamical fields (which strictly depend on the specific theory under examination). 

\subsection{Noether charges}
\label{sec:charges}

We will now define Noether charges associated with diffeomorphisms. Let the vector field $\xi^a$ be the generator of a diffeomorphism and let $\phi$ be any field configuration (not necessarily satisfying the equation of motion). Let us consider the variation of $\phi$ given by $\mathcal{L}_\xi\phi$ (which can be regarded as a vector on $\mathscr{F}$). We define the \emph{Noether current} $(n-1)$-form associated with $\xi^a$ at $\phi$ (regarded as a point of $\mathscr{F}$) by
\begin{equation}
\label{eq:cu}
\bm{J}=\bm{\Theta} (\phi,\mathcal{L}_\xi\phi)-\xi\cdot\bm{L},
\end{equation}
where $(\xi\cdot\bm{L})_{a_1 \dots a_{n-1}}=\xi^b\bm{L}_{b a_1\dots a_{n-1}}$. 
Notice that the ambiguity \eqref{eq:thetaamb} in the choice of $\bm{\Theta}$, one source of which comes from the freedom in shifting $\bm{L}\to\bm{L}+d\bm{\mu}$, induces an ambiguity in $\bm{J}$:
\begin{equation}
\label{eq:Jamb}
\begin{split}
\bm{J}\to&\bm{\Theta}(\phi,\mathcal{L}_\xi\phi)+\mathcal{L}_\xi\bm{\mu}+d\bm{Y}(\phi,\mathcal{L}_\xi\phi)-\xi\cdot \bm{L}-\xi\cdot d\bm{\mu}\\
&=\bm{J}+d(\xi\cdot\bm{\mu})+d\bm{Y}(\phi,\mathcal{L}_\xi\phi),
\end{split}
\end{equation}
where in the equality we used the definition of $\bm{J}$, and Cartan's magic formula, $\mathcal{L}_\xi \bm{\mu}=\xi\cdot d\bm{\mu}+d(\xi\cdot\bm{\mu})$.
Let us now compute $d\bm{J}$. Since $\bm{L}$ is covariant, the variation of $\bm{L}$ induced by $\mathcal{L}_\xi \phi$ is given by $\mathcal{L}_\xi \bm{L}$. So we have
\begin{equation}
\begin{split}
d\bm{J}&=d\bm{\Theta}(\phi,\mathcal{L}_\xi\phi)-d(\xi\cdot \bm{L})\\
&=\mathcal{L}_\xi \bm{L}-\bm{E}\mathcal{L}_\xi \phi-\mathcal{L}_\xi \bm{L}+\xi \cdot d\bm{L}\\
&=\bm{E}\mathcal{L}_\xi \phi.
\end{split}
\end{equation}
where in the second equality we used \eqref{eq:varL} and Cartan's magic identity, and in the third equality we used $d\bm{L}=0$ because $d\bm{L}$ is a $(n+1)$-form on a $n$-dimensional manifold.
Thus, if $\phi$ satisfies the equations of motion, $\bm{J}$ is closed for all $\xi^a$. Therefore, by the theorem proved in \cite{doi:10.1063/1.528839}, whenever the equations of motion are imposed, there exists a $(n-2)$-form $\bm{Q}$ that depends locally (in the sense explained above for the Lagrangian) only on $\phi$ and $\xi^a$, and such that
\begin{equation}
\label{eq:Q}
\bm{J}=d\bm{Q},
\end{equation}
which shows that $\bm{Q}$ is conserved (when the equations of motion are imposed).
$\bm{Q}$ is called the \emph{Noether charge} associated with $\xi^a$ and $\phi$. Note that, even if we fix $\bm{J}$, \eqref{eq:Q} defines $\bm{Q}$ only up to the addition of a closed (and hence exact, because of the result proved in \cite{doi:10.1063/1.528839}) $(n-2)$-form, $d\bm{Z}(\phi,\xi)$, locally dependent on $\phi$ and $\xi$. Moreover, if we consider the ambiguity in $\bm{J}$ given by \eqref{eq:Jamb}, we have the ambiguity in $\bm{Q}$
\begin{equation}
\label{eq:Qamb}
\bm{Q}\to \bm{Q}+\xi\cdot\bm{\mu}+\bm{Y}(\phi,\mathcal{L}_\xi\phi)+d\bm{Z}.
\end{equation}
The following lemma (presented and proved in \cite{Iyer:1994ys}) states that $\bm{Q}$ can be always chosen so that it can be written in a manifestly covariant way. 
\begin{lemma}
The Noether charge $(n-2)$-form can always be expressed in the form
\begin{equation}
\label{eq:Qform}
\bm{Q}=\bm{W}_c(\phi)\xi^c+\bm{X}^{cd}(\phi)\nabla_{[c}\xi_{d]}+\bm{Y}(\phi,\mathcal{L}_\xi\phi)+d\bm{Z}(\phi,\xi),
\end{equation}
where $\bm{W}_c$, $\bm{X}^{ab}$, $\bm{Y}$, and $\bm{Z}$ are diffeomorphism covariant quantities which depend locally from the indicated fields and their derivatives (with $\bm{Y}$ linear in $\mathcal{L}_\xi\phi$ and $\bm{Z}$ linear in $\xi$). This decomposition of $\bm{Q}$ is not unique in the sense that there are many different ways  of writing $\bm{Q}$ in the form \eqref{eq:Qform}, i.e. $\bm{W}_c$, $\bm{X}^{ab}$, $\bm{Y}$, and $\bm{Z}$ are not uniquely determined by $\bm{Q}$ (see below). However, $\bm{X}^{ab}$ may be chosen to be
\begin{equation}
\label{eq:Xno}
(\bm{X}^{cd})_{c_3\dots c_n}=-E_R^{abcd}\bm{\epsilon}_{abc_3 \dots c_n},
\end{equation}
where $E_R^{abcd}$ is defined by \eqref{eq:ER}, and we may choose $\bm{Y}=\bm{Z}=0$.
 \end{lemma}
To convince ourselves that, once we have chosen $\bm{Q}$ in the form \eqref{eq:Qform}, $\bm{W}_c,\bm{X}^{cd},\bm{Y},\bm{Z}$ are not uniquely determined by the choice of $\bm{Q}$, we notice that the term $d(\bm{U}_c(\phi)\xi^c)$, where $\bm{U}_c(\phi)$ is a $(n-3)$-form, can be written as the sum of a term linear in $\xi$ (which comes out simply from the expression of the exterior derivative in terms of the Levi-Civita connection and the Leibniz rule), a term linear in $\nabla_{[c}\xi_{d]}$ and a term linear in $2\nabla_{(c}\xi_{d)}=\mathcal{L}_\xi g_{cd}$ (which come out from the decomposition into symmetric and antisymmetric part of the second term arising from the use of the Leibniz rule). Hence, we can add a term $\bm{U}_c(\phi)\xi^c$ to $\bm{Z}$ and modify $\bm{W},\bm{X},\bm{Y}$ so that $\bm{Q}$ is unchanged. In the following, we will not impose conditions on the expression of the quantities at the RHS of \eqref{eq:Qform}, even if, as shown by the lemma, this is possible. In particular, we will \emph{not} require $\bm{Y}=\bm{Z}=0$ and $\bm{X}$ to be given by \eqref{eq:Xno}. The reason behind this choice is that the change $\bm{L}\to\bm{L}+d\bm{\mu}$ in the Lagrangian may affect $\bm{X}$. However, as we will see later, $\bm{X}$ is directly involved in the definition of the black hole entropy, which is a physical quantity that should not be affected from the change in the Lagrangian. Therefore, in the following we will just assume that $\bm{Q}$ is chosen to have the form \eqref{eq:Qform}.

\subsubsection{Symplectic potential, Noether current and Noether charge for Einstein theory in 4 dimensions}
\label{sec:eins4}

We will now see in detail how the symplectic potential, the Noether current and the Noether charge can be calculated in Einstein theory (i.e. General Relativity) in 4 dimensions. Given the ambiguity in the definition of these quantities, we will have to make a choice (which will be clear from the explicit calculation) to write their explicit expressions.

The Lagrangian of Einstein theory on a 4-dimensional manifold is
\begin{equation}
\bm{L}_{abcd}=\frac{1}{16\pi}\bm{\epsilon}_{abcd} R.
\end{equation}
The Lagrangian is manifestly covariant. The only dynamical field is the metric $g_{ab}$. We want to compute the variation of $\bm{L}$ induced by a variation $\delta g_{ab}$. We use the formulae $\delta\bm{\epsilon}=\frac{1}{2}\bm{\epsilon}g^{ab}\delta g_{ab}$ and $\delta R=-R^{ab}\delta g_{ab}+\nabla_a X^a$, with $X^a=g^{bc}\delta \Gamma^a_{\;\;bc}-g^{ab}\Gamma^c_{\;\;bc}=g^{ad}g^{bc}(\nabla_c\delta g_{bd}-\nabla_d\delta g_{bc})$ where $\delta\Gamma^a_{\;\;bc}=\frac{1}{2}g^{ad}(\nabla_c\delta g_{db}+\nabla_b\delta g_{dc}-\nabla_d\delta g_{bc})$ is the difference between the Christoffel-symbols of the Levi-Civita connection for $g_{ab}$ and for $g_{ab}+\delta g_{ab}$. Hence, we have
\begin{equation}
\begin{split}
\delta \bm{L} &=\frac{1}{16\pi}(R\delta \bm{\epsilon}+\bm{\epsilon}\delta R)\\
&= \frac{1}{16\pi}\bm{\epsilon}\biggl(\frac{1}{2}Rg^{ab}\delta g_{ab}-R^{ab}\delta g_{ab}+\nabla_a X^a\biggr).
\end{split}
\end{equation}
Comparing this expression with \eqref{eq:varL}, we read the equations of motion
\begin{equation}
(\bm{E}_g^{ab})=\frac{1}{16\pi}\bm{\epsilon}\biggl(\frac{1}{2}Rg^{ab}-R^{ab}\biggr),
\end{equation}
which is equivalent to $R_{ab}=0$,
and
\begin{equation}
d\bm{\Theta}=\frac{1}{16\pi}(\nabla_a X^a) \bm{\epsilon}.
\end{equation}
Hence, as proved in Appendix B of \cite{Wald:1984rg}, we can choose
\begin{equation}
\label{eq:sympotGR}
\bm{\Theta}_{abc}=\frac{1}{16\pi} X^d\epsilon_{dabc}=\epsilon_{dabc}\frac{1}{16\pi}g^{de} g^{fh}(\nabla_f\delta g_{eh}-\nabla_e \delta g_{fh}).
\end{equation}
From this, we obtain $\bm{J}=\bm{\Theta}(g_{ab},\mathcal{L}_\xi g_{ab})-\xi\cdot L=\frac{1}{16\pi}\bm{\epsilon}_{dabc}[g^{de}g^{fh}(\nabla_f\nabla_e\xi_h-\nabla_e\nabla_f\xi_h)+g^{de}g^{fh}\nabla_f\nabla_h\xi_e -g^{de}g^{fh}\nabla_e\nabla_h\xi_f-R \xi^d]$. Using the Ricci identity, the symmetries of the Riemann tensor of the Levi-Civita connection, and the equations of motion $R_{ab}=0$ (in this order), we have $g^{de}g^{fh}(\nabla_f\nabla_e\xi_h-\nabla_e\nabla_f\xi_h)=-g^{de}g^{fh}R_{khfe}\xi^k=g^{de}R_{ek}\xi^k=0$. Then, using the same formulae in the same order, $g^{de}g^{fh}\nabla_e\nabla_h \xi_f=g^{de} g^{fh}\nabla_h\nabla_e\xi_f$. Thus, imposing the equations of motion for the Einstein theory, we have
\begin{equation}
\bm{J}_{abc}=\frac{1}{16\pi}\bm{\epsilon}_{dabc}g^{de}g^{hf}(\nabla_h \nabla_f\xi_e-\nabla_h\nabla_e\xi_f)=\frac{1}{8\pi}\bm{\epsilon}_{dabc}\nabla_e(\nabla^{[e}\xi^{d]}).
\end{equation}
Since $\bm{J}$ is linear in the second covariant derivatives of $\xi$, we can compute $\bm{Q}$ from $\bm{J}$ using the algorithm provided by lemma 1 of \cite{doi:10.1063/1.528839}.  We need to write $\bm{J}$ in the form of eq. (2) of \cite{doi:10.1063/1.528839}. To this end, we write $\bm{J}_{abc}=\frac{1}{16\pi}\bm{\epsilon}_{dabc}(g^{ef}\delta^d_h-g^{df}\delta^e_h)\nabla_e\nabla_f\xi^h$. Using the Ricci identity, we have $\nabla_e\nabla_f\xi^h=\nabla_{(e}\nabla_{f)}\xi^h+\nabla_{[e}\nabla_{f]}\xi^h=\nabla_{(e}\nabla_{f)}\xi^h+\frac{1}{2}R^h_{\;\;mef}\xi^{m}$. So, the contribution to $\bm{J}$ linear in $\xi$, using also the symmetries of the Riemann tensor and the equations of motion in this order, is ($\frac{1}{16\pi}$ times) $\bm{\epsilon}_{dabc}(g^{ef}\delta^d_h-g^{df}\delta^e_h)\frac{1}{2}R^h_{\;\;mef}\xi^{m}=-\frac{1}{2}\bm{\epsilon}_{dabc}g^{df}R_{mf}\xi^m=0$. In this way, we have obtained the desired expression $\bm{J}=A^{(2)\;\;\;\;\;\;ef}_{\;\;\;\;\;abc\;\;\;\;h}\nabla_{(e}\nabla_{f)}\xi^h$, where $A^{(2)\;\;\;\;\;\;ef}_{\;\;\;\;\;abc\;\;\;\;h}=\frac{1}{16\pi}\bm{\epsilon}_{dabc}(g^{ef}\delta^d_h-g^{d(f}\delta^{e)}_h)$. So, using the algorithm explained in \cite{doi:10.1063/1.528839}, we read
\begin{equation}
\label{eq:QGR}
\begin{split}
\bm{Q}_{bc}&=\frac{1}{32\pi}\bm{\epsilon}_{dabc}(g^{ef}\delta^d_h-g^{d(f}\delta^{e)}_h)\nabla_f\xi^h\\
&=\frac{1}{32\pi}\bm{\epsilon}_{dabc}(\nabla^a\xi^d-\nabla^d\xi^a)=-\frac{1}{16\pi}\bm{\epsilon}_{bcad}\nabla^a\xi^d,
\end{split}
\end{equation}
which has the form \eqref{eq:Qform} with $\bm{W}=\bm{Y}=\bm{Z}=0$ and $\bm{X}$ given by \eqref{eq:Xno}. From this expression we see that, for a stationary spacetime with stationary Killing vector $\xi^a=k^a$, the Noether charge $\bm{Q}(k)$ is one-half of the Komar mass.

\subsection{Hamiltonian, canonical energy and canonical angular momentum}
\label{sec:hamenang}

We will now present the definition of Hamiltonian, canonical energy and canonical angular momentum for any diffeomorphism invariant theory of gravity.

Let $\phi$ be any solution of the equations of motion, and let $\delta\phi$ be any variation of the dynamical fields (not necessarily satisfying the linearised equations of motion) about $\phi$. Let $\xi$ be an arbitrary, fixed (i.e. $\delta\xi=0$) vector field on $M$. We have
\begin{equation}
\begin{split}
\delta\bm{J}&=\delta\bm{\Theta}(\phi,\mathcal{L}_\xi\phi)-\xi\cdot \delta\bm{L}\\
&=\delta\bm{\Theta}(\phi,\mathcal{L}_\xi\phi)-\xi\cdot d\bm{\Theta}(\phi,\delta\phi)\\
&=\delta\bm{\Theta}(\phi,\mathcal{L}_\xi\phi)-\mathcal{L}_\xi\bm{\Theta}(\phi,\delta\phi)+d(\xi\cdot \bm{\Theta}(\phi,\delta\phi)),
\end{split}
\end{equation}
where in the second equality we used the equations of motion $\bm{E}=0$, and in the third equality we used Cartan's magic formula. Since we require $\bm{\Theta}$ to have the canonical form, which is manifestly covariant, then $\mathcal{L}_\xi\bm{\Theta}$ is the same as the variation induced by $\mathcal{L}_\xi\phi$. Hence, $\bm{\omega}(\phi,\delta\phi,\mathcal{L}_\xi\phi)=\delta\bm{\Theta}(\phi,\mathcal{L}_\xi\phi)-\mathcal{L}_\xi\bm{\Theta}(\phi,\delta\phi)$. Therefore,
\begin{equation}
\label{eq:75}
\bm{\omega}(\phi,\delta\phi,\mathcal{L}_\xi\phi)=\delta\bm{J}-d(\xi\cdot\bm{\Theta}).
\end{equation}
Let us now assume that $\xi$ is a symmetry of all the dynamical fields, i.e. $\mathcal{L}_\xi\phi=0$ ($\xi$ is, in particular, a Killing field of $g_{ab}$) but not necessarily $\mathcal{L}_\xi\delta\phi=0$, and that $\delta\phi$ satisfies the equations of motion to linear order. Since $\bm{\omega}$ is linear in the variation of the fields (by definition) and $\mathcal{L}_\xi\phi=0$, we have $\bm{\omega}(\phi,\delta\phi,\mathcal{L}_\xi\phi)=0$. Moreover, since $\bm{J}=d\bm{Q}$ (because $\phi$ satisfies the equations of motion) and $\delta\phi$ satisfies the linearised equations of motion, $\delta\bm{J}=\delta d\bm{Q}=d\delta\bm{Q}$ (to linear order in $\delta\phi$). Thus, we obtain
\begin{equation}
d\delta\bm{Q}-d(\xi\cdot\bm{\Theta})=0.
\end{equation}
Integrating this equation over a hypersurface $\Sigma$ with boundary $\partial\Sigma$ and using Stokes' theorem, we obtain
\begin{equation}
\label{eq:77}
\int_{\partial\Sigma}{(\delta\bm{Q}-\xi\cdot\bm{\Theta}(\phi,\delta\phi))}=0.
\end{equation}
In order to prove the first law of black hole mechanics, we will be interested in the case where $\Sigma$ is a Cauchy surface with one asymptotically flat end and an interior boundary given by the bifurcation Killing surface of a black hole. In this case, $\partial \Sigma$ will be the union of a $(n-2)$-sphere at infinity and the bifurcation Killing surface $B$.

By analogy with what we did in section \ref{sec:Hamfor} and \ref{sec:enchan}, we want to study first the contribution from the boundary of $\Sigma$ at infinity. Therefore, we assume for now that $(M,\phi)$ is a globally hyperbolic, asymptotically flat spacetime and $\Sigma$ (together with the pull-back of the fields on $\Sigma$) is an asymptotically flat (with one end) Cauchy surface with no interior boundary. We will need \eqref{eq:75}, which holds without requiring that $\xi$ is a symmetry of $\phi$ and $\delta\phi$ satisfies the linearised equations of motion. We consider the integral of \eqref{eq:75} over $\Sigma$. We choose $\xi$ as time-evolution vector field on $M$. By analogy with the results of symplectic mechanics (whose concepts useful for our purposes are summarised in \ref{app:symmech}), the Hamiltonian $H$ that governs the dynamics of observables generated by the time evolution vector field $\xi$ on $M$ is given (if it exists) by the generalisation of \eqref{eq:HH} to a field theory, i.e. (recalling that the time-evolution vector field on the phase space is $T^A=(\mathcal{L}_\xi\phi)^A$)
\begin{equation}
\label{eq:ara}
\partial_A H=\Omega_{AB}(\mathcal{L}_\xi\phi)^B.
\end{equation}
$H$ is the true Hamiltonian of the theory, denoted by $H'$ in the first part \ref{sec:first1}.
Contracting this with $(\delta\phi)^A$, we obtain
\begin{equation}
\delta H(\phi,\delta\phi)=\Omega(\phi,\delta\phi,\mathcal{L}_\xi\phi),
\end{equation}
where $\delta H(\phi,\delta\phi)=(\partial_A H)(\delta\phi)^A$ is linear in $\delta\phi$. So, using the definition of $\Omega$ and the integral over $\Sigma$ of \eqref{eq:75}, we obtain
\begin{equation}
\begin{split}
\delta H&=\delta\int_\Sigma {\bm{J}}-\int_\Sigma{d(\xi\cdot\bm{\Theta})}\\
&=\delta\int_\Sigma{\bm{J}}-\lim_{r\to\infty}\int_{S_r^2}{\xi\cdot\bm{\Theta}},
\end{split}
\end{equation}
where $r=\sqrt{x^ix^i}$ and $x^i$ are the asymptotically Cartesian coordinates involved in the definition of asymptotically flat end.
From this, we see that an Hamiltonian associated with the time-evolution vector field $\xi$ exists if and only if we can find a (not necessarily diffeomorphism covariant) $(n-1)$-form $\bm{B}$ such that
\begin{equation}
\label{eq:B}
\delta \lim_{r\to\infty}\int_{S_r^2}{\xi\cdot\bm{B}}=\lim_{r\to\infty}\int_{S_r^2}{\xi\cdot\bm{\Theta}}.
\end{equation}
In this case, $H$ is
\begin{equation}
H=\int_\Sigma{\bm{J}}-\lim_{r\to\infty}\int_{S_r^2}{\xi\cdot\bm{B}}
\end{equation}
When $\phi$ is a solution of the equations of motion, we have $\bm{J}=d\bm{Q}$, so
\begin{equation}
H=\lim_{r\to\infty}\int_{S_r^2}{(\bm{Q}-\xi\cdot \bm{B})}.
\end{equation}
We proved that, for every diffeomorphism invariant theory, the value of $H$, if it exists, on solutions of the equations of motion is a boundary term. If $\Sigma$ has no boundary, i.e. $\Sigma$ is compact, then $H=0$ on solutions of the equations of motion. This is exactly the results that we have obtained at the end of section \ref{sec:Hamfor} for Einstein-Maxwell theory.

In analogy with the discussion of section \ref{sec:enchan}, we now define the canonical energy of any asymptotically flat end of an asymptotically flat spacetime as the value of the Hamiltonian associated with a time-evolution vector field $\xi^a=k^a$, where $k^a$ is an asymptotic time translation. We will assume that suitable asymptotic conditions have been imposed on the dynamical fields so that $\bm{B}$ exists and the limit $r\to\infty$ of integrals over $S_r^2$ is finite. So
\begin{equation}
\mathcal{E}\equiv\lim_{r\to\infty}\int_{S_r^2}{(\bm{Q}(k)-k\cdot \bm{B})}.
\end{equation}

Let us convince ourselves that, in Einstein theory in 4-dimensions, this definition reduces to the usual definition of canonical energy for an asymptotically flat end, i.e. the ADM mass \eqref{eq:ADMmass}. Let $t$ be the parameter along integral curves of $k$. Let $(x^i)$ be asymptotically Cartesian coordinates on $\Sigma$. Let us extend them outside $\Sigma$ by keeping them constant along integral curves of $k$. As we argued in section \ref{sec:Hamfor}, $x^\alpha=(t,x^i)$ are asymptotically inertial coordinates, i.e. in these coordinates we have $g_{\alpha\beta}=\eta_{\alpha\beta}+\mathcal{O}(1/r)$ and $\partial_\gamma g_{\alpha\beta}=\mathcal{O}(1/r^2)$ as $r\equiv\sqrt{x^ix^i}\to\infty$. From this chart we define a new chart $x^\mu=(t,r,\theta,\phi)$ on $M$, where $\theta$ and $\phi$ are defined from $x^i$ by the usual relations between Cartesian and spherical coordinates. Assume also that the orientation on $M$ is chosen so that $x^\mu$ is right-handed (RH).
The 2-sphere at infinity is the limit as $r\to\infty$ of the 2-dimensional submanifolds with $r,t=constant$ and it is parameterised by $\theta,\phi$.
In this chart $g_{\mu\nu}=diag(-1,1,r^2,r^2\sin^2\theta)+ corrections$ as $r\to\infty$, where by ``corrections'' we mean higher order terms in $1/r$ . Let $dS=\sqrt{\sigma}d\theta\wedge d\phi$ be the volume form on the 2-sphere at infinity in the orientation class given by Stokes' theorem (where $\sigma=r^4\sin^2\theta$ is the determinant of the metric induced on the 2-sphere in coordinates $(\theta,\phi)$). We use these coordinates to compute the expression for the Noether charge in General Relativity \eqref{eq:QGR}. We have $\bm{\epsilon}_{\theta\phi\mu\nu}\nabla^\mu k^\nu=\bm{\epsilon}_{tr \theta\phi}(\eta^{tt}\nabla_t k^r-\eta^{rr}\nabla_r k^t)+ corrections$. Then, $\nabla_\rho k^\nu=\Gamma^{\nu}_{t\rho}=\frac{1}{2}\eta^{\nu\sigma}(g_{t\sigma,\rho}+g_{\rho\sigma,t}-g_{t\rho,\sigma})+ corrections$. So, (since $\bm{\epsilon}_{t r\theta\phi}=\sqrt{-\det g_{\mu\nu}}=r^2 \sin\theta+ corrections$) we have $\bm{\epsilon}_{\theta\phi\mu\nu}\nabla^\mu k^\nu=r^2\sin\theta(g_{tt,r}-g_{rt,t})+ corrections$. Hence, we have
\begin{equation}
\begin{split}
\lim_{r\to\infty}\int_{S_r^2}{\bm{Q}(k)}&=-\frac{1}{16\pi}\lim_{r\to\infty}\int_{S_r^2}{d\theta d\phi\bm{\epsilon}_{\theta\phi \mu\nu}\nabla^\mu k^\nu}\\
&=-\frac{1}{16\pi}\lim_{r\to\infty}\int_{S_r^2}{dS(\partial_r g_{tt}-\partial_t g_{rr})} + corrections.
\end{split}
\end{equation}
Since the leading order terms is $\mathcal{O}(1)$, then the corrections vanish in the limit $r\to\infty$.

Now we want to compute the second contribution to $\mathcal{E}$. Let $h_{pq}$ be the metric on $\Sigma$ in coordinates $x^p=(r,\theta,\phi)$.  
Using \eqref{eq:sympotGR} (in coordinates $x^\mu$) where $\delta g_{ab}$ is required to preserve the asymptotic flatness of the spacetime, we obtain, with a similar calculation,
\begin{equation}
\begin{split}
\lim_{r\to\infty}\int_{S_r^2}{k^a\bm{\Theta}_{abc}}&=-\frac{1}{16\pi}\lim_{r\to\infty}\int_{S_r^2}{d\theta d\phi\bm{\epsilon}_{tr\theta\phi}\eta^{rr}\eta^{\sigma\gamma}(\nabla_\sigma\delta g_{r\gamma}-\nabla_r\delta g_{\sigma\gamma})}+ corr.\\
&=-\frac{1}{16\pi}\lim_{r\to\infty}\int_{S_r^2}{dS[\partial_r\delta g_{tt}-\partial_t\delta g_{rt}+h^{pq}(\partial_p\delta h_{rq}-\partial_r \delta h_{pq})]}+ corr,
\end{split}
\end{equation}
where $h_{pq}$ are the spatial components of $g_{\mu\nu}$, i.e. the metric on $\Sigma$ in coordinates $(r,\theta,\phi)$. The first term can be immediately be written as
\begin{equation}
-\frac{1}{16\pi}\delta \biggl[\lim_{r\to\infty}\int_{S_r^2}{dS(\partial_r g_{tt}-\partial_t g_{rt})}\biggr].
\end{equation}
To obtain a similar expression for the second term (which we need to do in order to obtain $\bm{B}$), it is easier to use the asymptotically Cartesian coordinates $x^i$, for which the asymptotic conditions for $h_{ij}$ and its derivatives are the usual asymptotic conditions for an asymptotically flat end, and the asymptotic conditions for $\delta h_{ij}$ are those which ensure that $(\Sigma,h_{ij}+\delta h_{ij})$ is still asymptotically flat (these are the conditions given explicitly in section \ref{sec:Hamfor}). In these coordinates, we see that $h^{ij}\partial_i\delta h_{rj}=\delta^k_r \delta^{ij}\partial_i\delta h_{kj}= n^i \partial_j\delta h_{ij}$, where $n^a=\bigl(\frac{\partial}{\partial r}\bigr)^a$ ($n^a$ is the outward pointing unit normal to the 2-sphere at infinity in the tangent space of $\Sigma$), in the limit $r\to\infty$. Similarly, $h^{ij}\partial_r\delta h_{ij}= n^i \partial_i\delta h_{jj}$ as $r\to\infty$. So, we have
\begin{equation}
\lim_{r\to\infty}\int_{S_r^2}{k^a\bm{\Theta}_{abc}}=-\frac{1}{16\pi}\delta \lim_{r\to\infty}\int_{S_r^2}{dS[\partial_r g_{tt}-\partial_t\delta g_{rt}+n^i(\partial_j h_{ij}-\partial_i h_{jj})]}+ corr.
\end{equation}
Once again, since the leading order term is $\mathcal{O}(1)$, the corrections vanish in the limit $r\to\infty$.
So, any 3-form $\bm{B}$ such that, in the limit $r\to\infty$,
\begin{equation}
k^a \bm{B}_{abc}=-\frac{1}{16\pi}\tilde{\bm{\epsilon}}_{bc}[\partial_r g_{tt}-\partial_t\delta g_{rt}+n^i (\partial_j h_{ij}-\partial_i h_{jj})]
\end{equation}
satisfies \eqref{eq:B}. Here, $\tilde{\bm{\epsilon}}_{bc}$ is the volume form for the 2-sphere at infinity, i.e. $\tilde{\bm{\epsilon}}=r^2\sin\theta d\theta\wedge d\phi$.
Hence, we have
\begin{equation}
\begin{split}
\mathcal{E}&=\lim_{r\to\infty}\int_{S_r^2}{(\bm{Q}(k)-k\cdot B)}=\frac{1}{16\pi}\lim_{r\to\infty}\int_{S_r^2}{dSn^i(\partial_j h_{ij}-\partial_i h_{jj})}\\
&=m,
\end{split}
\end{equation}
where $m$ is the ADM mass defined by \eqref{eq:ADMmass}. This result shows that the canonical energy defined for a general diffemorphism invariant theory reduces to the ADM mass for Einstein theory.

We also define the \emph{canonical angular momentum} in analogy to what we did in section \ref{sec:enchan}. Let $\xi^a=m^a$ be an asymptotic rotation. The canonical angular momentum is defined as minus the value of the Hamiltonian associated with $m^a$ evaluated on the solution of the equations of motion. By definition of asymptotic rotation, in the coordinate chart $x^\mu=(t,r,\theta,\phi)$ defined above $m^a$ is one of the vector fields $(\phi_1)^\mu=-\sin\phi\bigl(\frac{\partial}{\partial \theta}\bigr)^\mu-\cot\theta\cos\phi\bigl(\frac{\partial}{\partial \phi}\bigr)^\mu$, $(\phi_2)^\mu=\cos\phi\bigl(\frac{\partial}{\partial \theta}\bigr)^\mu-\cot\theta\sin\phi\bigl(\frac{\partial}{\partial \phi}\bigr)^\mu$, $(\phi_3)^\mu=\bigl(\frac{\partial}{\partial \phi}\bigr)^\mu$ as $r\to\infty$. Thus, we immediately see that the pull-back (i.e. the action of the embedding of the 2-sphere at infinity into $M$) of $m^a\bm{\Theta}_{abc}$ to the 2-sphere at infinity vanishes because in coordinates $(\theta,\phi)$ two of the indices of $\bm{\Theta}$ are the same (so the result vanishes since $\bm{\Theta}$ is completely antisymmetric). So, $\bm{B}$ can be any 3-form that vanishes on the 2-sphere at infinity. Hence,
\begin{equation}
\mathcal{J}=-\lim_{r\to\infty}\int_{S_r^2}{\bm{Q}(m)}.
\end{equation}
For the Einstein theory, in an axisymmetric spacetime with axial Killing vector $m$, \eqref{eq:QGR} shows that $\mathcal{J}$ is exactly the Komar angular momentum. For a generic asymptotically flat spacetime solution of the (vacuum) Einstein equations, it can be shown that $\mathcal{J}$ is the ADM angular momentum defined in section \ref{sec:enchan}.

\subsection{Entropy of a stationary black hole and the first law of black hole mechanics}
\label{sec:entfirst}

Having defined these quantities, we are now almost ready to present the generalisation of the first law of black hole mechanics for diffeomorphism invariant theories. Let us consider the case of a solution $\phi$ of the equations of motion that describes a stationary black hole spacetime with stationary Killing vector field $k$ and bifurcate Killing event horizon with bifurcation $(n-2)$-surface $B$. Let $\xi$ be the vector field that defines the bifurcate Killing horizon. In particular, $\xi$ vanishes on $B$. Hawking's theorem \cite{Hawking:1973uf} states that, if $\xi$ is not proportional to $k$, the spacetime is also axisymmetric and $\xi$ can be written as a linear combination of $k$ and a family of axial Killing vector fields $m^a_{(\alpha)}$. Moreover, we can choose the normalisation of $\xi$ so that
\begin{equation}
\label{eq:xikm}
\xi^a=k^a+\Omega^{(\alpha)}m^a_{(\alpha)},
\end{equation}
where a sum over $\alpha=1,2,\dots,\left\lfloor\frac{n-1}{2}\right\rfloor$ is understood. Notice that $k$ is an asymptotic time translation and $m^a_{(\alpha)}$ are asymptotic rotation. This relation defines the constants $\Omega^{(\alpha)}$, which are called the angular velocities of the horizon. Consider an asymptotically flat hypersurface $\Sigma$ having $B$ as the only interior boundary (so $\partial\Sigma$ is the union of $B$ with a 2-sphere at infinity). Thus, $\delta\int_{B}{\bm{Q}(\xi)}=-\delta\int_{\partial\Sigma}{\bm{Q}(\xi)}+\delta\lim_{r\to\infty}\int_{S_r^2}{\bm{Q}(\xi)}$ (the sign is obtained by taking into account that the outward unit normals to $B$ and the 2-sphere at infinity in $\Sigma$ point in opposite directions) . Using \eqref{eq:77} to rewrite the integral over $\partial \Sigma$ and the fact that $\xi=0$ on $B$, we have
\begin{equation}
\label{eq:varQ}
\begin{split}
\delta\int_{S}{\bm{Q}(\xi)}&=-\int_{\partial\Sigma}{\xi\cdot\bm{\Theta}}+\delta\lim_{r\to\infty}\int_{S_r^2}{\bm{Q}(\xi)}\\
&=-\lim_{r\to\infty}\int_{S_r^2}{\xi\cdot\bm{\Theta}}+\delta\lim_{r\to\infty}\int_{S_r^2}{\bm{Q}(\xi)}\\
&=-\delta\lim_{r\to\infty}\int_{S_r^2}{\xi\cdot\bm{B}}+\delta\lim_{r\to\infty}\int_{S_r^2}{\bm{Q}(\xi)}\\
&=-\delta\lim_{r\to\infty}\int_{S_r^2}{k\cdot\bm{B}}+\delta\lim_{r\to\infty}\int_{S_r^2}{\bm{Q}(k)}+\Omega^{(\alpha)}\delta\lim_{r\to\infty}\int_{S_r^2}{\bm{Q}(m_{(\alpha)})}\\
&=\delta\mathcal{E}-\Omega^{(\alpha)}\delta\mathcal{J}_{(\alpha)},
\end{split}
\end{equation}
where in the second equality we used the fact that $\xi=0$ on $S$, in the third equality we used the definition of $\bm{B}$, in the fourth equality the expression \eqref{eq:xikm}, the fact that $m_{(\mu)}\cdot\bm{Q}$ vanishes on the $(n-2)$-sphere at infinity and the fact that $\bm{Q}(\xi)=\bm{Q}(k)+\Omega^{(\alpha)}\bm{Q}(m_{(\alpha)})$ (because $\bm{Q}$ has the form \eqref{eq:Qform}, so it is linear in $\xi$), and in the fifth equality the definitions of $\mathcal{E}$ and $\mathcal{J_{(\mu)}}$.

The above result looks similar to the desired form of the first law of black hole mechanics, but we still need to rewrite the LHS in terms of the entropy of the black hole.
We define the \emph{entropy} for an arbitrary cross-section $C$ of the Killing horizon (see appendix \ref{app:binsur}) of a stationary black hole as
\begin{equation}
\label{eq:S}
S[C]=2\pi\int_{C} {\bm{X}^{cd}\bm{\tilde{\epsilon}}_{cd}},
\end{equation}
where $\bm{\tilde{\epsilon}}_{cd}$ is the binormal to $C$ defined in appendix \ref{app:binsur}. First of all, we notice that in Einstein theory in 4 dimensions, for our choice of $\bm{Q}$ \eqref{eq:QGR}, this definition reduces to \eqref{eq:entgr} for a stationary black hole solution of the Einstein theory in 4 dimensions. In fact, from our choice of $\bm{Q}$ we have $(\bm{X}^{cd})_{ab}=-\frac{1}{16\pi}\bm{\epsilon}_{ab}^{\;\;\;\;cd}$. Then, in appendix \ref{app:binsur} it is proved that 
\begin{equation}
\label{eq:SGR}
\bm{\epsilon}_{ab}^{\;\;\;\;cd}\bm{\tilde{\epsilon}}_{cd}=-2\bm{\mathring{\epsilon}}_{ab},
\end{equation}
where $\bm{\mathring{\epsilon}}_{ab}$ is the volume form of $C$. So, we find $S[C]=A[C]/4$, as expected.

It is also important to notice that for a stationary black hole $S[C]$ is independent of the choice of cross-section $C$. To prove this we show that $S[C]$ is equal to the entropy of the bifurcation surface, $S[B]$. Since by definition any cross-section is intersected by the integral curves of $\xi$ exactly once, then the action of the diffeomorphism $\chi_t$ (for any $t$) of the 1-parameter group of diffeomorphisms generated by $\xi$ on all points of $C$ gives a new cross-section $\chi_t[C]$. Let us consider $S[\chi_t[C]]$. For a stationary black hole $\xi$ is a symmetry of the dynamical fields, i.e. $\mathcal{L}_\xi\phi=0$, so $\mathcal{L}_\xi\bm{X}^{cd}(\phi)=0$ because $\bm{X}$ is covariant. So $\bm{X}^{cd}(\phi)$ does not change under $\chi_t$. Thus, we have $S[\chi_t[C]]=S[C]$. This holds for any $t$. When $t\to-\infty$, $\chi_t[C]$ continuously approaches the bifurcation surface $B$, so (since $\bm{X}$ is smooth) we have $S[C]=\lim_{t\to-\infty}S[\chi_t[C]]=S[B]$, as we wanted to prove. 

However, we are interested in proving the first law of black hole mechanics including the possibility of perturbations $\delta\phi$ to a non-stationary black hole. The above argument does not work for non-stationary black holes, hence the variation $\delta S[C]$ induced by the variation $\delta\phi$ of a stationary black hole to a non-stationary black hole will depend on the choice of the cross-section of the stationary black hole on which $S[C]$ is computed. As a consequence, the proof of the first law of black hole mechanics that we present below will be valid only if we consider the variation of the entropy of the bifurcation surface, $\delta S[B]$ (but it will not be true in general if we substitute $B$ with an arbitrary cross-section $C$, unless we restrict ourselves to perturbations to another stationary black hole).

Furthermore, we argued above that requiring that $\bm{Q}(\xi)$ has the form \eqref{eq:Qform} does not determine $\bm{X}^{cd}$ uniquely. However, it turns out that for a stationary black hole, $S[C]$ is independent of the possible choices of $\bm{X}$. In fact, we have seen above that for a stationary black hole $S[C]=S[B]$, and we can show that the value of $\bm{X}$ on $B$ is uniquely determined by $\bm{Q}(\xi)$. In fact, since $\xi=0$ on $B$, given a $\bm{Q}(\xi)$ in the form \eqref{eq:Qform}, the contribution from $\bm{W}$, i.e. $\bm{W}_c\xi^c$, vanishes. Then, since $\bm{Z}$ is linear in $\xi=0$, $\bm{Z}=0$. Moreover, since $\bm{Y}$ is linear in $\mathcal{L}_\xi\phi=0$ (here we are using the fact that the solution is stationary), we also have $\bm{Y}=0$. So, $\bm{Q}(\xi)=\bm{X}^{cd} \nabla_{[c}\xi_{d]}$ on $B$, which shows that $\bm{X}^{cd}$ on $B$ is completely determined by $\bm{Q}(\xi)$. This result shows that $S[C]=S[B]$ is independent of the choice of $\bm{X}$ outside $B$. Moreover, lemma 3 tells us that $(\bm{X}^{cd})_{a_3\cdots a_n}=-E_R^{abcd}\bm{\epsilon}_{aba_3\cdots a_n}$ on $B$.

We may ask ourselves whether the variation $\delta S[C]$ of the entropy of a stationary black hole due to the variation $\delta\phi$ to a non-stationary spacetime is indepedent of the choice of $\bm{X}$. This is not true in general, but it is true if we compute the variation of $S$ on the bifurcation surface $B$. This result is an obvious consequence of the first law of black hole mechanics.

\begin{theorem}{\textbf{[First law of black hole mechanics]}}
Let $\phi$ be an asymptotically flat stationary black hole solution of the equations of motion with a bifurcate Killing horizon and bifurcate Killing surface $B$. Let $\delta\phi$ be a (not necessarily stationary) asymptotically flat solution of the linearised equations about $\phi$. Then, we have
\begin{equation}
\label{eq:firstlaw}
\frac{\kappa}{2\pi}\delta S[B]=\delta\mathcal{E}-\Omega^{(\alpha)}\delta\mathcal{J}_{(\alpha)},
\end{equation}
where $\kappa$ is the surface gravity of the Killing horizon.
\end{theorem}
\noindent
(This result shows that $\delta S[B]$ does not depend on the choice of $\bm{X}$ because the RHS of \eqref{eq:firstlaw} depends only on $\bm{Q}$.)
\emph{Proof} We will use the fact that $\kappa$ is constant on every bifurcate Killing horizon of a stationary black hole \cite{Racz:1992bp}. If we can prove that
\begin{equation}
\delta \int_B{\bm{Q}(\xi)}=\frac{\kappa}{2\pi}\delta S[B],
\end{equation}
then the result follows directly from \eqref{eq:varQ}. So we need to compute $\delta\bm{Q}$ on $B$. We are requiring $\bm{Q}$ to have the form \eqref{eq:Qform}, so $\delta (\bm{W}_c\xi^c)$ does not contribute to $\delta\bm{Q}$ (since $\xi=0$ on $B$ and $\delta\xi=0$ everywhere). Then, $\delta d\bm{Z}(\phi,\xi)=d\bm{Z}(\delta\phi,\xi)$ to linear order, which vanishes on $B$ because it is linear in $\xi$ and $\xi=0$ on $B$. Moreover, 
\begin{equation}
\delta\bm{Y}(\phi,\mathcal{L}_\xi\phi)=\mathcal{L}_\xi\bm{Y}(\phi,\delta\phi)=\xi\cdot d\bm{Y}+d(\xi\cdot\bm{Y}),
\end{equation}
where we used $\mathcal{L}_\xi\phi=0$ (and $\delta\xi=0$) in the first and second equality, and Cartan's magic formula in the third equality. Since $\xi=0$ on $B$, it follows (using Stokes' theorem on the second term) that $\int_B{\delta\bm{Y}(\phi,\mathcal{L}_\xi\phi)}$ does not contribute to $\delta \int_B{\bm{Q}(\xi)}$. Hence,
\begin{equation}
\label{eq:ette}
\delta \int_B{\bm{Q}(\xi)}=\delta \int_B{\bm{X}^{cd}(\phi)\nabla_{[c}\xi_{d]}}.
\end{equation}
As we proved in appendix \ref{app:binsur}, in the stationary background we have
\begin{equation}
\label{eq:app}
\nabla_c\xi_d=\kappa \bm{\tilde{\epsilon}}_{cd} \qquad \text{on $B$.}
\end{equation}
Furthermore, since $\xi=0$ on $B$ and $\delta\xi=0$ everywhere, we have
\begin{equation}
\label{eq:varxi}
\delta\nabla_c\xi^d=\delta\Gamma^d_{bc}\xi^b=0 \qquad \text{on $B$.}
\end{equation}
Now we consider the variation $\delta \bm{\tilde{\epsilon}}_c^{\;\;d}$ of $\bm{\tilde{\epsilon}}_c^{\;\;d}=g^{bd}\bm{\tilde{\epsilon}}_{cb}$, the projection onto the subspace of the tangent space at $B$ containing the vectors orthogonal to $B$. By the definition of binormal (see appendix \ref{app:binsur}), $V^c\bm{\tilde{\epsilon}}_c^{\;\;d}=0$ for all vectors $V^c$ tangent to $B$, so $V^c\delta\bm{\tilde{\epsilon}}_c^{\;\;d}=0$. This implies that, defining a basis $(u^a,l^a,v_1^a,\cdots,v_{n-2}^a)$ for the tangent space at an arbitrary point $p\in B$, where $u^a$ is the tangent to the affinely parameterised null geodesics generators of the Killing horizon, $l^a$ is the other null normal normalised so that $u\cdot l=-1$ (see appendix \ref{app:binsur}), and $v_1^a,\cdots,v_{n-2}^a$ are vectors tangent to $B$, and expanding $\delta\bm{\tilde{\epsilon}}_c^{\;\;d}$ in this basis, there are no terms of the type $(v_i)_c(v_j)^d$. We say that $\delta\bm{\tilde{\epsilon}}_c^{\;\;d}$ has no ``tangential-tangential'' part w.r.t. the background metric $g_{ab}$. Then, since $\bm{\tilde{\epsilon}}_c^{\;\;d}\bm{\tilde{\epsilon}}_d^{\;\;c}=2$, we have $\bm{\tilde{\epsilon}}_d^{\;\;c}\delta\bm{\tilde{\epsilon}}_c^{\;\;d}=0$, i.e. $\bm{\tilde{\epsilon}}^{cd}g_{a[c}\delta\bm{\tilde{\epsilon}}_{d]}^{\;\;a}=0$. This means that the expansion of $g_{a[c}\delta\bm{\tilde{\epsilon}}_{d]}^{\;\;a}$ at any point $p\in B$ in terms of the basis $(u^a,l^a,v_1^a,\cdots,v_{n-2}^a)$ has no terms of the type $u_cl_d$, i.e. $g_{a[c}\delta\bm{\tilde{\epsilon}}_{d]}^{\;\;a}$ has no ``normal-normal'' part w.r.t $g_{ab}$. We now define the quantity
\begin{equation}
\label{eq:diffbin}
\omega_{cd}=\nabla_{[c}\xi_{d]}-\kappa\bm{\tilde{\epsilon}}_{cd}.
\end{equation}
From Killing's equation we have $\nabla_{[c}\xi_{d]}=\nabla_c\xi_d$, so from \eqref{eq:app} we obtain that $\omega_{cd}=0$ on $B$. Furthermore, using the Leibniz rule for $\delta$, \eqref{eq:app} and \eqref{eq:varxi}, we have
\begin{equation}
\delta \omega_{cd}=\delta\bigl(g_{a[d}(\nabla_{c]}\xi^a-\kappa\bm{\tilde{\epsilon}}_{c]}^{\;\;a})\bigr)=-\kappa g_{a[d}\delta\bm{\tilde{\epsilon}}_{c]}^{\;\;a} \quad \text{on $B$.}
\end{equation}
So $\delta \omega_{cd}$ on $B$ has only a ``normal-tangential'' part (w.r.t. $g_{ab}$), i.e. its decomposition in terms of the basis $(u^a,l^a,v_1^a,\cdots,v_{n-2}^a)$ has only terms of the type $u_c (v_i)_d$ and $l_c (v_i)_d$. Now we can express \eqref{eq:ette} as
\begin{equation}
\label{eq:otto}
\delta \int_B{\bm{Q}(\xi)}=\delta \int_B{\bm{X}^{cd}(\phi)(\kappa\bm{\tilde{\epsilon}}_{cd}+\omega_{cd})}=\frac{\kappa}{2\pi}\delta S[B]+\int_B{\bm{X}^{cd}\delta\omega_{cd}}.
\end{equation}
As noted above, in the stationary background $\bm{X}^{cd}(\phi)$ is invariant under the 1-parameter group of diffeomorphisms generated by $\xi$. Hence, by Lemma 2.3 of \cite{Kay:1988mu}, proved in appendix \ref{app:binsur}, at each point $p\in B$ $\bm{X}^{cd}$ is also invariant under the map $i_\ast$ (the push-forward of the map $i$ defined precisely in appendix \ref{app:binsur}) on the tangent space at $p$ which reverses the normal directions to $B$ but keeps the tangential directions unchanged, i.e. a reflection about the tangent ``plane'' at $p$. On the other hand, since $\delta\omega_{cd}$ is purely ``normal-tangential'', it reverses the sign under the reflection $i_\ast$ at each $p\in B$ (because it changes the sign of $u^a$ and $l^a$ but leaves the other basis vectors unchanged, as explained in appendix \ref{app:binsur}). However, in \eqref{eq:otto} we are integrating the pull-back of $\bm{X}^{cd}\delta\omega_{cd}$ on $B$, which is purely tangential, and, hence, invariant under the reflection $i_\ast$. Thus, the pull-back of $\bm{X}^{cd}\delta\omega_{cd}$ on $B$ must vanish, so the second term of \eqref{eq:otto} does not contribute. $\square$

\section{Entropy of a dynamical black hole and the second law of black hole mechanics for diffeomorphism invariant theories}
\label{sec:second}

In conclusion, we discuss the conditions that must be imposed to find a sensible definition of the entropy $S_{dyn}$ for a dynamical, i.e. non-stationary, black hole in a general diffeomorphism invariant theory. We will also briefly present some difficulties in finding a prescription for the dynamical entropy satisfying a form of the second law of black hole mechanics, and some interesting results that have been obtained in this sense. 

Let $C$ be a cross-section of the event horizon of a dynamical black hole (notice that the event horizon is not a Killing horizon in general, because the spacetime is non-stationary). The definition of entropy $S_{dyn}[C]$ must be consistent with the one used in the first law of black hole mechanics. Therefore, we require $S_{dyn}[C]$ to satisfy the following properties: (i) if the black hole is stationary, the dynamical entropy should reduce to $S[C]$ defined above for stationary black holes, i.e.
\begin{equation}
S_{dyn}[C]=S[C]=2\pi\int_C{\bm{X}^{cd}\bm{\tilde{\epsilon}}_{cd}},
\end{equation}
and (ii) the variation of $S_{dyn}[B]$ on the bifurcation surface $B$ induced by an arbitrary asymptotically flat perturbation of the stationary black hole should give
\begin{equation}
\delta S_{dyn}[B]=\delta S[B]=2\pi\delta\int_B{\bm{X}^{cd}\bm{\epsilon}_{cd}}.
\end{equation}
We also want that (iii) changing the Lagrangian by addition of an exact $n$-form, $\bm{L}\to\bm{L}+d\bm{\mu}$ (which does not affect the equations of motion, and so the physical predictions of the theory) leaves $S_{dyn}$ unchanged.
These conditions suggest that we seek a formula of the type
\begin{equation}
S_{dyn}[C]=\int_C{\bm{X}^{cd}_{dyn}(\phi)\bm{\tilde{\epsilon}}_{cd}},
\end{equation}
where $\bm{X}^{cd}_{dyn}$ is a diffeomorphism covariant $(n-2)$-form that depends locally from the dynamical fields $\phi$ and their derivatives, and it can be obtained from the Lagrangian $\bm{L}$ by some algorithm that has to be found.
There is also one important further condition: (iv) $S_{dyn}[C]$ should satisfy a form of the second law of black hole mechanics, i.e. under certain conditions $S_{dyn}$ should be a non-decreasing quantity when evaluated on successively later cross-sections of the event horizon of a dynamical black hole. The arguments presented in section \ref{sec:intro}  explain the importance of such a condition in order to restrict the choice of suitable generalisations of General Relativity that may describe gravitational phenomena at high energies (the Planck scale).  

In \cite{Wald:1993nt} Wald proposes an algorithm for $\bm{X}_{dyn}^{cd}$ that gives simply
\begin{equation}
(\bm{X}^{cd}_{dyn})_{a_3\cdots a_n}=-E_R^{abcd}\bm{\epsilon}_{aba_3\cdots a_n}.
\end{equation}
Although this prescription satisfies the conditions (i) and (ii) (because, as we argued above, for a stationary black hole with bifurcate surface $B$,  $(\bm{X}^{cd})_{a_3\cdots a_n}=-E_R^{abcd}\bm{\epsilon}_{aba_3\cdots a_n}$ on $B$, and the entropy is independent of the cross-section), it fails to satisfy (iii), in fact the addition of $d\bm{\mu}$ to $\bm{L}$ with a suitable dependence on the Riemann tensor can change $E_R^{abcd}$ in such a way that the change in $S_{dyn}$ is nonvanishing for non-stationary black holes. However, in \cite{Iyer:1994ys} Wald and Iyer present a (more complicated) algorithm that manages to satisfy the conditions (i), (ii) and (iii). I will not explain their prescription here in detail. In brief, Iyer's and Wald's entropy is given by ($2\pi$ times) the integral over the cross-section $C$ of the boost invariant part of the Noether charge associated with the vector field, defined in a neighbourhood of $C$, that generates Lorentz boosts. We just mention that this definition reduces to the expression \eqref{eq:entgr} for the entropy of a non-stationary black hole for Einstein theory in 4 dimensions. 

It is important to mention that Iyer's and Wald's prescription is not the only one that satisfies (i),(ii) and (iii). In particular, ambiguities in the algorithm have been identified \cite{Jacobson:1993vj}, which, in essence, arise from the fact that there are ambiguities in the choice of the Noether charge $\bm{Q}$. Moreover, it is still unknown if condition (iv) (i.e. the second law of black hole mechanics) is satisfied for a sufficiently large class of diffeomorphism invariant theories. An attempt at addressing this issue faces some serious difficulties. First, in order to prove the second law it is not possible to make use of the field equations because these belong to a specific theory (while we would like to consider a class of theories). Second, it is believed that the second law should hold for theories of gravity satisfying certain physically reasonable conditions, e.g. the predictability of a solution on and outside a black hole (more precisely, a property analogous to the strongly asymptotic predictability) and the positivity of total energy (for example, in General Relativity we require the matter fields to satisfy the Null Energy Condition). To convince ourselves of the importance of the second property we notice that, if we write $S_{dyn}$ as the integral of a Noether charge of a vector field, the change in entropy between two cross-sections $C$ and $C'$ is expected to be related to the flux of the corresponding Noether current through the horizon between $C$ and $C'$. Hence, the second law is expected to be valid if we can impose some conditions to ensure that this flux is positive. Unfortunately, we do not know the most general conditions that have to be required. Furthermore, determining whether a complicated theory satisfies these conditions or not may be difficult.

Nevertheless, some examples of the validity of the second law have been found. For example, it has been recently proved \cite{Wall:2015raa} that, if we allow for some modifications of Iyer's and Wald's prescription (due to the ambiguities mentioned above), there exists an entropy of dynamical black holes that satisfies a physical form of the second law for every diffeomorphism invariant theory in which an energy-momentum tensor of the matter fields can be identified. This form of the second law is given by starting with a stationary black hole with bifurcate Killing horizon at some initial time and assuming that the matter fields satisfy the Null Energy Condition. Then, we suppose that the black hole undergoes a linear perturbation $\delta\phi$, originated by throwing a small amount of matter inside the black hole. So the black hole is non-stationary at intermediate times. Eventually, it settles down to a final stationary state. It is shown that the difference in the entropy of the black hole between any two instants of this process, i.e. computed on any two cross-sections of the event horizon, is positive. 
Many other examples of specific theories with dynamical black hole entropy satisfying the second law of black hole mechanics have been discovered, but the most general conditions under which a class of theories of gravity satisfies the second law are yet to be found.

\section*{Acknowledgements}

I thank Harvey Reall for the final corrections, and for explaining the guidelines on the structure of the essay during three group meetings with the other Part III students interested in this topic. I also thank Aron Kovacs for many useful discussions. The General Relativity and Black Holes lecture courses taught by Maciej Dunajski, Harvey Reall and Jorge Santos during Part III helped my understanding of the topics discussed in this essay.
During Part III of the Mathematical Tripos, I was funded by the Sheepshanks studentship in Astronomy offered by Trinity College, Cambridge, UK.

\appendix

\section{The energy-momentum tensor of the Maxwell field satisfies the Dominant Energy Condition (DEC)}
\label{app:dec}

In this appendix, following the notation of \ref{sec:first1}, we show that the energy-momentum tensor of the Maxwell field,
\begin{equation}
T_{\mu\nu}=\frac{1}{4\pi}\biggl(F_{\mu\rho}F_\nu^{\;\;\rho}-\frac{1}{4}F_{\rho\sigma}F^{\rho\sigma}g_{\mu\nu}\biggr),
\end{equation}
satisfies the Dominant Energy Condition (DEC), i.e. the requirement that $j^\mu=-T^\mu_{\;\;\nu}V^\nu$ is a future-directed causal vector (or zero) for all future-directed causal vectors $V^\mu$, i.e.
\begin{equation}
j^2\leq 0 \quad \text{and} \quad j	\cdot V \leq 0,
\end{equation}
for any future-directed vector that satisfies $V^2\leq 0$.

To do so, we first prove that DEC is equivalent to $T_{00}\geq \sqrt{T_{0i}T_{0i}}$ in every orthonormal basis $e_0,e_1,e_2,e_3$ at a point of $M$.

\noindent
\emph{Proof.} Let us consider an arbitrary point $p\in M$. 

\noindent
(a)Let us assume that DEC is satisfied. Pick an arbitrary orthonormal basis $e_{\alpha}$, $\alpha=0,1,2,3$ at $p$, i.e. a basis such that $g_{\alpha\beta}=\eta_{\alpha\beta}$ at $p$, where $\eta_{\alpha\beta}=diag(-1,1,1,1)$. DEC holds for any arbitrary future-directed vector satisfying $V^2=-(V^0)^2+V^iV^i\leq 0$. In particular, DEC holds if we choose $V^0=1, V^i=0$. From DEC, we have
\begin{equation}
0\geq-T_{\alpha\beta}V^\alpha V^\beta=-T_{00} \Longrightarrow T_{00}\geq0,
\end{equation}
and
\begin{equation}
0\geq T_{\alpha\beta}V^\alpha T^\alpha_{\;\;\gamma}V^\gamma=-(T_{00})^2+T_{0i}T_{0i} \Longrightarrow (T_{00})^2\geq T_{0i}T_{0i}.
\end{equation}
Since $T_{00}\geq 0$, this implies $T_{00}\geq \sqrt{T_{0i}T_{0i}}$ in the orthonormal basis $e_{\alpha}$. Since the basis is arbitrary, we proved the first implication.

\noindent
(b) Let us now assume that $T_{00}\geq \sqrt{T_{0i}T_{0i}}$ in every orthonormal basis $e_{\alpha}$, $\alpha=0,1,2,3$ at $p$. Let us consider an arbitrary future-directed causal vector $V^\mu$ and pick the orthonormal basis such that $e_0^\mu$ is parallel to $V^\mu$, i.e. $V^\alpha=(V^0,0,0,0)$. Then $j^\alpha=-T^\alpha_{\;\;\beta}V^\beta=(-T^0_{\;\;0}V^0,-T^i_{\;\;0}V^0)=(T_{00}V^0,-T_{0i}V^0)$. Thus,
\begin{equation}
j^2=[-(T_{00})^2+T_{0i}T_{0i}](V^0)^2\leq 0,
\end{equation}
and
\begin{equation}
j\cdot V=-T_{00}(V^0)^2\leq 0,
\end{equation}
because $T_{00}\geq \sqrt{T_{0i}T_{0i}}\geq 0$. So DEC is satisfied. $\square$

Now, we study the case of the Maxwell field. Let us consider an arbitrary point $p\in M$ and pick an orthonormal basis $e_\alpha$ at $p$. Let us choose $e_0$ to be future-directed. Since $e_0\cdot e_0=\eta_00=-1$, we can regard $e_0$ as the 4-velocity of an observer. Notice that this choice does not fix the direction of $e_0$, so, given an arbitrary future-directed causal vector $V^\mu$, we can still choose $e_0^\mu$ to be parallel to $V^\mu$ (so the argument of point (b) can be repeated). The components of $e_0^\mu$ are $e_0^\alpha=\delta^\alpha_0$. So, the components of the electric field measured by an observer with 4-velocity $e_0^\mu$ are $E_\alpha=F_{\alpha\beta}e_0^\alpha=-F_{0\alpha}$ at $p$. Since $F_{\alpha\beta}$ is antisymmetric, the only non-vanishing components of $E_\mu$ are
\begin{equation}
E_i=-F_{0i}.
\end{equation}
On the other hand, the non-vanishing components of the magnetic field $B_\mu=-(\star F)_{\mu\nu}e_0^\nu$ measured by an observer with 4-velocity $e_0^\mu$ in our orthonormal chart are
\begin{equation}
B_i=\frac{1}{2}\epsilon_{ijk} F_{jk} \Longrightarrow F_{ij}=\epsilon_{ijk}B_k,
\end{equation}
where $\epsilon$ is the usual Levi-Civita symbol (e.g. $\epsilon_{123}=1,\epsilon_{213}=-1$) and we do not need to distinguish between upstairs and downstairs indices $i,j,k,\dots$. Using these expressions we find
\begin{equation}
T_{00}=\frac{1}{8\pi}(E_iE_i+B_iB_i) \quad \text{and} \quad T_{0i}=-\frac{1}{4\pi}\epsilon_{ijk}E_j B_k.
\end{equation}
Notice that $T_{00}\geq 0$. In 3-vector notation $(T_{00})^2=\frac{1}{16\pi^2}\frac{1}{4}(\vec{E}^2+\vec{B}^2)^2$ and $T_{0i}T_{0i}=\frac{1}{16\pi^2}[(\vec{E}^2 \vec{B}^2-(\vec{E}\cdot\vec{B})^2]$. Thus, $(T_{00})^2-T_{0i}T_{0i}=\frac{1}{16\pi^2}\bigl[\frac{1}{4}(\vec{E}^2-\vec{B}^2)^2+(\vec{E}\cdot\vec{B})^2]$, which is obviously positive. Since $T_{00}$ is positive, we have found that the energy-momentum tensor of the electromagnetic field satisfies $T_{00}\geq \sqrt{T_{0i}T_{0i}}$ in our orthonormal basis. Using the proof of the point (b) above, we see that this implies that DEC is satisfied for our basis. Since DEC is a basis-independent statement, we have proved that the energy-momentum tensor of the Maxwell field satisfies DEC.

\section{Formula for the surface gravity in terms of the lapse function and proof of eq. \eqref{eq:la}}
\label{app:sur}

Following the notation of section \ref{sec:first1}, in this appendix we will show first that the derivative of the lapse function along the normal to the bifurcation surface $B$ tangent to the Cauchy surface $\Sigma$ is the surface gravity $\kappa$ \cite{Brown:1995su}. Then, we will prove eq. \eqref{eq:la}.

Our starting point will be the expression for the surface gravity $\kappa$ on the horizon given by eq.(12.5.18) of \cite{Wald:1984rg}, i.e.
\begin{equation}
\label{eq:kappaapp}
\kappa=\lim(|\xi||a|),
\end{equation}
where $|\xi|\equiv\sqrt{-\xi^\mu \xi_\mu}$, $a^\rho\equiv\frac{\xi^\mu\nabla_\mu\xi^\rho}{|\xi|^2}$ is the acceleration of the integral curves of $\xi$, $|a|\equiv \sqrt{a^\mu a_\mu}$, and $\lim$ stands for the limit as one approaches the horizon. The limit is independent of the path along which we approach a certain point of the horizon. Here, we are interested in the value of $\kappa$ on $B$, which can be written as \eqref{eq:kappaapp} with $\lim$ meaning the limit as one approaches $B$ from within $\Sigma$. This will be the meaning of $\lim$ in the following. Using Killing's equation, we find $\nabla_\rho|\xi|=-\frac{1}{|\xi|}\xi^\mu\nabla_\rho \xi_\mu=\frac{1}{|\xi|}\xi^\mu\nabla_\mu \xi_\rho=|\xi|a_\rho$. Then, the surface gravity becomes
\begin{equation}
\label{eq:kappa1}
\kappa=\lim\sqrt{(\nabla^\mu|\xi|)(\nabla_\mu|\xi|)}.
\end{equation}
Let $r^\mu$ be the outward pointing unit normal to each surface in $\Sigma$ with constant $N$, e.g. $B$ where $N=0$. The metric induced on these surfaces is $\sigma_{\mu\nu}=h_{\mu\nu}-r_\mu r_\nu$. $\sigma^\mu_\nu$ on each surface of constant $N$ is a projection operator acting on the tangent space of $\Sigma$ and giving vectors tangent to the surface.
Using $\xi$ as time-evolution vector field, we can define $N$ and $N^\mu$ on $\Sigma$ such that $\xi^\mu=-Nn^\mu+N^\mu$, where $n^\mu$ is the past-directed unit normal on $\Sigma$. Since $\xi=0$ on $B$, we have $\lim N=0$ and $\lim N^\mu=0$ (as noted in the main body). 
Since the surface gravity of a bifurcate Killing horizon is non-vanishing \cite{Wald:1995yp}, it follows from \eqref{eq:kappa1} that $\nabla^\mu |\xi|\nabla_\mu |\xi|\geq 0$ at $B$, i.e. $\nabla^\mu |\xi|$ is spacelike at $B$. In particular, this implies that $\lim\nabla^\mu |\xi|$ cannot be parallel to the everywhere timelike unit normal $n_\mu\propto \nabla_\mu t$ to $\Sigma$ ($t$ is the global time function). Hence, $\lim h^\nu_\mu\nabla^\nu |\xi| \neq 0$. We now observe that the limit of $|\xi|=\sqrt{-\xi^\mu\xi_\mu}=\sqrt{N^2-N^\mu N_\mu}$ is a constant (namely zero) as $B$ is approached from within $\Sigma$, so $\lim\sigma_\mu^\nu\nabla_\nu|\xi|=0$. Thus, using $\lim h^\nu_\mu\nabla^\nu |\xi| \neq 0$, we have $\lim r^\nu\nabla_\nu |\xi|\neq 0$.

Now consider the limit of $\frac{N^\mu}{|\xi|}$ as $B$ is approached from within $\Sigma$. This is an indeterminate form $\frac{0}{0}$. We can apply l'Hopital's rule and differentiate both numerator and denominator along the normal $r^\mu$ direction within $\Sigma$. The derivative of the numerator, $r^\nu\nabla_\nu |\xi|$, has nonzero limit by the argument above. The derivative of the numerator is 
\begin{equation}
\label{eq:B7}
\begin{split}
r^\nu\nabla_\nu N^\mu&=r^\nu \nabla_\nu(h^\mu_\rho \xi^\rho)\\
&=r^\nu\xi^\rho\nabla_\nu h^\mu_\rho+r^\nu h^\mu_\rho\nabla_\nu \xi^\rho.
\end{split}
\end{equation}
Using $h^\mu_\rho=\delta^\mu_\rho+n^\mu n_\rho$, we can write the first terms as
\begin{equation}
\label{eq:A}
\begin{split}
r^\nu t^\rho \nabla_\nu h^\mu_\rho&=r^\nu \xi^\rho \nabla _\nu (n^\mu n_\rho)=r_\nu \xi^\rho n_\rho \nabla^\nu n^\mu+r^\nu\xi^\rho n^\mu\nabla_\nu n_\rho=Nr_\nu \nabla^\nu n^\mu+r^\nu N^\rho n^\mu\nabla_\nu n_\rho\\
&=-N K^{\mu\nu}r_\nu-n^\mu r^\nu K_{\nu\rho}N^\rho,
\end{split}
\end{equation}
where in the last equality we used the invariance of $r^\mu$ under projection $h^\mu_\nu$ to insert the projection and obtain the extrinsic curvature of $\Sigma$, $K_{\mu\nu}=-h_\mu^\rho \nabla_\rho n_\nu$. The two terms in \eqref{eq:A} vanish in the limit because $N$ and $N^\mu$ vanish. Using the relation $h^{\mu\nu}=\sigma^{\mu\nu}+r^\mu r^\nu$ and the fact that $\nabla_\mu \xi_\nu$ is antisymmetric (due to Killing's equation), we can write the second term of \eqref{eq:B7} as
\begin{equation}
\label{eq:B9}
\begin{split}
r^\nu h^{\mu\rho}\nabla_\nu \xi_\rho&= r^\nu \sigma^{\mu\rho}\nabla_\nu \xi_\rho=-r^\nu \sigma^{\mu\rho}\nabla_\rho \xi_\nu=-r^\nu \sigma^{\mu\rho}\nabla_\rho(-Nn_\nu+N_\nu)\\
&=-N\sigma^{\mu\rho}K_{\rho\nu}r^\nu -r^\nu\sigma^{\mu\rho}\nabla_\rho N_\nu.
\end{split}
\end{equation}
The first term in \eqref{eq:B9} vanishes in the limit due to the factor of $N$, and the second term vanishes in the limit since the derivative is projected along the bifurcation surface where $N^\mu$ vanishes. The result is that $\lim r^\nu \nabla_\nu N^\mu=0$, so by l'Hopital's rule we have $\lim \frac{N^\mu}{|\xi|}=0$. Since $|\xi|=\sqrt{N^2-N^\mu N_\mu}$, we also find $\lim\frac{N^\mu}{N}=\lim\frac{N^\mu}{\sqrt{|\xi|^2+N^\mu N_\mu}}=\lim\frac{N^\mu}{|\xi|}\frac{1}{\sqrt{1+\frac{N^\mu N_\mu}{|\xi|^2}}}=0$. Furthermore, $\lim\frac{N}{|\xi|}=\lim\frac{1}{\sqrt{1-\frac{N^\mu N_\mu}{N^2}}}=1$.
Now, we can write
\begin{equation}
\label{eq:B10}
\nabla_\mu |\xi|=\frac{1}{2|\xi|}\nabla_\mu (N^2-N^\nu N_\nu)=\frac{N\nabla_\mu N}{|\xi|}-\frac{N^\nu \nabla_\mu N_\nu}{|\xi|}.
\end{equation}
From $\lim \frac{N^\mu}{|\xi|}=0$ we have that the second term vanishes in the limit as $B$ is approached from within $\Sigma$. From $\lim\frac{N}{|\xi|}=1$, we have that $\lim\nabla_\mu |\xi|=\lim \nabla_\mu N$. Thus, from \eqref{eq:kappa1} we have
\begin{equation}
\label{eq:uu}
\kappa=\lim\sqrt{\nabla_\mu N \nabla^\mu N}.
\end{equation}
Since $B$ is the surface of $\Sigma$ with $N=0$, the outward pointing unit normal $r^\mu$ to $B$ in the tangent space of $\Sigma$ is $r^\mu=\frac{\nabla^\mu N}{\sqrt{h^{\rho\sigma} \nabla_\rho N\nabla_\sigma N}}$. Therefore, $r^\mu \nabla_\mu N=\sqrt{\nabla_\mu N \nabla^\mu N}$. Hence, we can write the surface gravity on the bifurcation surface $B$ as $\kappa=r^\mu \nabla_\mu N$. Since $r^\mu$ is a vector tangent to $\Sigma$, we can insert the projection operator and obtain the desired expression
\begin{equation}
\label{eq:sss}
\kappa=r^\mu D_\mu N.
\end{equation}

We will now need \eqref{eq:sss} to prove eq. \eqref{eq:la}. We will regard all the tensors invariant under projection onto $\Sigma$ as tensors defined on $\Sigma$ (so we use Latin abstract indices $a,b,c,d$,etc.). Our purpose is to write the variation of the determinant of the metric $\sigma_{ab}$ on $B$ (which appears at the RHS of eq. \eqref{eq:la}) in terms of the variation of the pull-back to $B$ of the determinant of the metric $h_{ab}$ on $\Sigma$ (which appears at the LHS of eq. \eqref{eq:la}) in a suitable set of adapted coordinates. Since $B$ is the surface of $\Sigma$ determined by $N=0$, it is natural to choose $N$ as the first coordinate. The outward unit normal field to $B$, and to each surface of constant $N$ in $\Sigma$, is $r_a=\frac{(dN)_a}{\sqrt{h^{ab (dN)_a (dN)_b}}}=\frac{D_a N}{\sqrt{h^{ab}D_a N D_b N}}$, where here $d$ denotes the exterior derivative on $\Sigma$. Let us denote the dual of $(dN)_a$ by $s^a$ (i.e. $s^a$ is the vector field such that $s^a (dN)_a=1$, i.e. $s^a=\bigl(\frac{\partial}{\partial N}\bigr)^a$). We build coordinates on $\Sigma$ as follows. Let $y^p=(y^1,y^2)$ be a RH chart on $B$ w.r.t. the orientation defined by Stokes' theorem. For every point $p\in\Sigma$ we define coordinates $x^i=(N,y^1,y^2)$ where $(y^1,y^2)$ are the coordinates of the point $q\in B$ at which the integral curve of $s^a$ through $p$ intersects $B$. We can now define lapse and shift function for the manifold $\Sigma$ using $s^a$ as the time evolution vector field. The metric induced on $B$ is $\sigma_{ab}=h_{ab}-r_ar_b$. The lapse function is $\tilde{N}=r_a s^a$. The shift vector is $\tilde{N}^a=\sigma^a_b s^b=h^a_bs^b+r^ar_b s^b=\tilde{N}r^a+s^a$, from which we read $r^a=-\frac{1}{\tilde{N}}(s^a-\tilde{N}^a)$. We want to find the components of these tensors in coordinates $x^i$ in terms of $\sigma_{pq}$ (i.e. the $(pq)$ components of $\sigma_{ab}$), $\tilde{N}$ and $\tilde{N}^i$. In coordinates $x^i$, $r_i=\frac{\delta_i^1}{\sqrt{h^{11}}}$, $s^i=\delta^i_1$, so $\tilde{N}=\frac{1}{\sqrt{h^{11}}}$, from which we also find $r_a=\tilde{N}D_a N$. Furthermore,
\begin{equation}
\begin{split}
h_{11}&=h_{ij}s^i s^j=\sigma_{ij}^i s^j+r^i s_i r^j s_j=\tilde{N}^2+\tilde{N}^i\tilde{N}_i,\\
h_{1p}&=h_{ij}s^i\biggl(\frac{\partial }{\partial y^p}\biggr)^j=\sigma_{ij}s^i\biggl(\frac{\partial }{\partial y^p}\biggr)^j=\sigma_{pi}\tilde{N}^i,\\
h_{pq}&=h_{ij}\biggl(\frac{\partial }{\partial y^p}\biggr)^i\biggl(\frac{\partial }{\partial y^p}\biggr)^j=\sigma_{pq},\\
\end{split}
\end{equation}
where we also used that $r^a$ is orthogonal to vectors tangent to $B$, such as $\bigl(\frac{\partial }{\partial y^p}\bigr)$. Therefore, the 11-component of the cofactor matrix of $h_{ij}$ is $\triangle^{11}= \det\sigma_{pq}\equiv\sigma$. We also need to find $h^{11}$ on $B$ in these coordinates. The inverse metric on $\Sigma$ evaluated at $B$ (i.e. the pull-back of $h_{ab}$ on $B$ under the inclusion map) can be written as 
\begin{equation}
\label{eq:hsigma}
h^{ab}=\sigma^{ab}+r^ar^b=\sigma^{ab}+\frac{1}{\tilde{N}^2}(s^a-\tilde{N}^a)(s^b-\tilde{N}^b).
\end{equation}
Now consider $\sigma^1_i=h^1_i-r^1r_1=h^1_i-h^{1j}r_j r_i=h^1_i-h^{1j}\frac{\delta^1_j\delta^1_i}{h^{11}}=h^1_i-h^1_i=0$. So we also obtain $\sigma^{1i}=h^{ij}\sigma^1_j=0$, from which we have $\tilde{N}^1=\sigma^1_i s^i=0$. Using these results in \eqref{eq:hsigma}, we have the useful equation $h^{11}=\frac{1}{\tilde{N}^2}$. Using this result and $\triangle^{11}=\sigma$ in the formula for $h^{ij}$ in terms of the cofactor matrix of $h_{ij}$, $h^{ij}=\frac{1}{\det h_{ij}}\triangle^{ij}$, we find $\frac{1}{\tilde{N}^2}=\frac{\sigma}{h}$ (where we denoted $\det h_{ij}$ by $h$). Hence, $\sqrt{\sigma}\frac{1}{\tilde{N}}\sqrt{h}$. We use this results to compute the variation of $\sqrt{\sigma}$ induced by the variation $\delta h_{ij}$. This will show the validity of eq \eqref{eq:la}. For this calculation we also need to recall the formulae for the variation of the inverse of the metric and the determinant: $\delta h^{ij}=-h^{ik}h^{il}\delta h_{kl}$ and $\delta h=-h h^{ij}\delta h_{ij}$. Our result is
\begin{equation}
\begin{split}
\delta\sqrt{\sigma}&=-\frac{\sqrt{h}}{\tilde{N}^2}\delta\tilde{N}+\frac{1}{\tilde{N}}\frac{1}{2\sqrt{h}}\delta h=-\frac{\sqrt{h}}{\tilde{N}^2}\biggl(-\frac{1}{h^{11}}\frac{1}{2\sqrt{h^{11}}}\delta h^{11}\biggr)+\frac{1}{\tilde{N}}\frac{1}{2\sqrt{h}}h h^{ij}\delta h_{ij}\\
&=-\frac{\sqrt{h}}{\tilde{N}^2}\biggl(\frac{1}{h^{11}}\frac{1}{2\sqrt{h^{11}}}h^{1i}h^{1j}\delta h_{ij}\biggr)+\frac{1}{\tilde{N}}\frac{1}{2\sqrt{h}}h h^{ij}\delta h_{ij}\\
&=-\tilde{N}^2\sqrt{\sigma}\frac{1}{2}h^{1i}h^{1j}\delta h_{ij}+\frac{1}{2}\sqrt{\sigma}h^{ij}\delta h_{ij}\\
&=-\tilde{N}\sqrt{\sigma}(\tilde{N}\delta^1_k)\delta^1_l \frac{1}{2}h^{ik}h^{jl}\delta h_{ij}+\frac{1}{2}\sqrt{\sigma}h^{ij}\delta h_{ij}\\
&=-\tilde{N}\sqrt{\sigma} r_kD_lN\frac{1}{2}h^{ik}h^{jl}\delta h_{ij}+\frac{1}{2}\sqrt{\sigma}h^{ij}\delta h_{ij}\\
&=-\frac{1}{2}\tilde{N}\biggl(\sqrt{\sigma} r_kD_lNh^{ik}h^{jl}\delta h_{ij}-\frac{1}{\tilde{N}}\sqrt{\sigma}h^{ij}\delta h_{ij}\biggr).
\end{split}
\end{equation}
Hence,
\begin{equation}
\label{eq:LR}
\frac{2}{\tilde{N}}\delta\sqrt{\sigma}=-\biggl(\sqrt{\sigma} r_kD_lNh^{ik}h^{jl}\delta h_{ij}-\frac{1}{\tilde{N}}\sqrt{\sigma}h^{ij}\delta h_{ij}\biggr).
\end{equation}
Now we notice that $\frac{1}{\tilde{N}}=\sqrt{h^{11}}=\sqrt{h^{ij}D_iN D_jN}=\kappa$, where in the last equality we used \eqref{eq:sss}. We use this result at the LHS of \eqref{eq:LR}. Moreover, $\frac{1}{\tilde{N}}=\sqrt{h^{11}}=\frac{h^{11}}{\sqrt{h^{11}}}=\frac{D^k ND_k N}{\sqrt{D^l ND_lN}}=r^k D_k N$. We use this result at the second term of the RHS of \eqref{eq:LR}. So, we obtain
\begin{equation}
2\kappa\delta\sqrt{\sigma}=-(\sqrt{\sigma} r_kD_lNh^{ik}h^{jl}\delta h_{ij}-r^k D_k N\sqrt{\sigma}h^{ij}\delta h_{ij}).
\end{equation}
Finally, multiplying by the volume form on $B$ given by $dS=\sqrt{\sigma}dy^1\wedge dy^2$, integrating over $B$, and using the fact that $\kappa$ is constant on $B$ (from the zeroth law of black hole mechanics), we obtain eq. \eqref{eq:la}.

\section{Symplectic mechanics}
\label{app:symmech}

In this appendix, we will explain how the Hamilton equations are written in the symplectic formulation of classical mechanics. Hopefully, this will be useful for the reader to understand intuitively the role of the symplectic form on the phase space and expression \eqref{eq:ara}, without employing the machinery of the covariant phase space formulation of field theories presented in \cite{Lee:1990nz}.

A \emph{symplectic manifold} (or a \emph{phase space} in physical terminology) is a manifold $\Gamma$ equipped with a symplectic 2-form $\Omega_{AB}$ such that
\begin{itemize}
\item $\Omega_{AB}$ is closed, i.e. $d\Omega=0$, where $d$ is the exterior derivative on $\Gamma$.\\
\item $\Omega_{AB}$ is non-degenerate, i.e. $\Omega_{AB}\psi_1^A\psi_2^B=0$ for all vectors $\psi_2^A$ on $\Gamma$ if and only if $\psi_1^A=0$.
\end{itemize}
It turns out that symplectic manifolds are the natural framework to formulate Hamiltonian mechanics.
For a finite dimensional symplectic manifold $\Gamma$ there exists a chart $(q_i,p^i)$ on $\Gamma$ (called the Darboux chart) for which the symplectic form is $\Omega=dq_i\wedge dp^i$ (the sum over $i$ is intended). $q_i$ and $p^i$ play the role of the usual canonical variables of classical mechanics. If we use such a chart, the expressions that we will write down in this appendix can be immediately recognised as well-known expressions in the Hamiltonian formulation of classical mechanics. However, in the second part of the essay \ref{sec:20}, following the formalism developed in \cite{Lee:1990nz}, we considered equations that hold in arbitrary charts on the phase space of field theories (which explains why this formalism is also called \emph{covariant} phase space formulation of field theories). Therefore, to make the analogy with the notions introduced in the main body of the essay explicit, we will consider here a generic coordinate chart $x^A$ on $\Gamma$.

The non-degeneracy of the symplectic form implies that its inverse $\Omega^{AB}$, defined by $\Omega_{AB}\Omega^{AB}=\delta_A^B$, exists. The inverse can be used to define Poisson brackets over $\Gamma$ as follows. Given two functions $f,g$ on $\Gamma$, we define
\begin{equation}
\{f,g\}\equiv \Omega^{AB}\partial_A f\partial_B g.
\end{equation}
It can be checked that this definition satisfies the properties of the Poisson bracket. In particular, the closedness of $\Omega$ ensures the validity of the Jacobi identity. We can check that in the Daurboux chart, the Poisson bracket reduces to $\{f,g\}=\frac{\partial f}{\partial q_i}\frac{\partial g}{\partial p^i}-\frac{\partial g}{\partial q_i}\frac{\partial f}{\partial p^i}$.

Once we have chosen a vector field $T^A$ on $\Gamma$, called time-evolution vector field (not to be confused with the time evolution vector field on the spacetime manifold; we will see how the two are related), the dynamics of any observable is the rate of change of the observable along the integral curves $\gamma(t)$ of $T^A$. We define the Hamiltonian $H$ associated with $T^A$, if it exists, as the function that satisfies
\begin{equation}
\label{eq:HH}
\partial_A H=\Omega_{AB}T^B.
\end{equation}
The generalisation of equation \eqref{eq:HH} to the case of field theories is equation \eqref{eq:ara} used in section \ref{sec:hamenang}, where the time evolution vector field on $\Gamma$ is $T^A=(\mathcal{L}_\xi \phi)^A$ and $\xi^a$ is the time-evolution vector field on $M$. \eqref{eq:HH} is usually referred to as the Hamilton's equations in the symplectic formalism. To understand why, notice that this is equivalent (contracting with $\Omega^{AB}$) to $T^A=\Omega^{AB}\partial_B H$. So, the dynamics of an observable $f$ is
\begin{equation}
\label{eq:HH1}
\frac{df}{dt}=t^A\partial_A f=\Omega^{AB}\partial_A f\partial_B H=\{f,H\}.
\end{equation}
Using the Daurboux chart and choosing alternatively $f=q_i$ and $f=p^i$, from \eqref{eq:HH1} we obtain the well-known form of the Hamilton's equations
\begin{equation}
\dot{q_i}=\{q_i,H\}=\frac{\partial H}{\partial p^i} \qquad \dot{p^i}=\{p^i,H\}=-\frac{\partial H}{\partial q_i}.
\end{equation}

\section{Binormal to a cross-section of the Killing horizon, and reflection invariance}
\label{app:binsur}

In this appendix, following the notation of section [], we will give the definition of binormal to a cross-section of the Killing horizon $\mathcal{H}$, we will prove eq. \eqref{eq:ette}, eq. \eqref{eq:SGR} and Lemma 2.3 of \cite{Kay:1988mu}.

Let us consider a black hole whose event horizon is a bifurcate Killing horizon $\mathcal{H}$. Unlike in the main body of the essay, we will be concerned with all the four parts of the Killing horizon and we will need to distinguish between them. The bifurcate Killing horizon is composed of two null hypersurfaces: the future event horizon $\mathcal{H}^+$ and the past event horizon $\mathcal{H}^-$, which intersects each other in the bifurcation surface $B$. $\mathcal{H}^+$ is composed of two subsets: the causal future of $B$in $\mathcal{H}^+$, denoted by $\mathcal{H}^+_R$, and the causal past of $B$ in $\mathcal{H}^+$, denoted by $\mathcal{H}^+_L$. Similarly, $\mathcal{H}^-$ is composed of two subsets: the causal future of $B$ in $\mathcal{H}^-$, denoted by $\mathcal{H}^-_L$, and the causal past of $B$ in $\mathcal{H}^-$, denoted by $\mathcal{H}^-_R$.

Let $u^a$ be the tangent to the affinely parameterised generators of the $\mathcal{H}^+$ (recall that $u^a$ is normal to $\mathcal{H}^+$). A \emph{cross-section} $C$ of $\mathcal{H}^+$ is an hypersurface of $\mathcal{H}^+$ (i.e. a $(n-2)$-dimensional submanifold of the $n$-dimensional spacetime manifold $M$) such that (i) $u^a$ is nowhere tangent to $C$, (ii) each null generator of $\mathcal{H}^+$ intersects $C$ exactly once. A cross-section of $\mathcal{H}^+$ is a spacelike submanifold (i.e. every non-zero vector tangent to $C$ is spacelike). To prove this, let us consider a point $p\in C$ and an arbitrary non-zero vector $V$ tangent to $C$ at $p$. $V^a$ is also tangent to $\mathcal{H}^+$, so it is orthogonal to $u^a$. Since $u^a$ is null, $V^a$ can be spacelike or parallel to $u^a$ at $p$. The latter case must be discarded by definition of cross-section, so $V$ is spacelike. 

Let $C$ be a generic cross section of $\mathcal{H}^+$. Since $C$ is spacelike, for any $p\in C$ there are precisely 2 independent future-directed null vectors orthogonal to $C$ (up to the freedom to rescale them). Since we assume that $C$ is time-orientable, we can define these two vectors continuously over $C$, so they can be regarded as vector fields on $C$. We can rescale them so that one of them coincides with the tangent to the affinely parameterised generators of $\mathcal{H}^+$, $u^a$. We denote the other one by $l^a$ and we rescale it so that $u\cdot l=-1$. We define the \emph{binormal} to $C$ by
\begin{equation}
\bm{\tilde{\epsilon}}_{ab}=(u\wedge l)_{ab}=u_a l_b-l_a u_b.
\end{equation}
Notice that $\bm{\tilde{\epsilon}}_a^{\;\;b}$ projects any vector of the tangent space at $p\in C$ onto the subspace of vectors orthogonal to $C$. In particular, $\bm{\tilde{\epsilon}}_a^{\;\;b}u^a=u^b$ and $\bm{\tilde{\epsilon}}_a^{\;\;b}l^a=-l^b$. Notice also that $\bm{\tilde{\epsilon}}_{ab}$ satisfies $\bm{\tilde{\epsilon}}_{ab}\bm{\tilde{\epsilon}}^{ab}=-2$. 

Let $\xi$ be the Killing vector field that generates $\mathcal{H}^+$. Since $\xi$ is normal to $\mathcal{H}^+$ by definition, from Frobenius' theorem we have $\xi\wedge d\xi=0$ on $\mathcal{H}^+$. This is equivalent to $d\xi=\xi\wedge \omega$ where $\omega$ is any 1-form field on $\mathcal{H}^+$. Now, we evaluate this relation on an arbitrary cross-section $C$ of $\mathcal{H}^+$. On $C$ $\omega$ can be decomposed in its part tangent to $C$ and its parts parallel to $u$ and $l$, i.e. as $\omega^a=\alpha u^a+\beta l^a+\gamma V^a$ for some functions $\alpha,\beta,\gamma$ on $C$ and a vector field $V^a$ tangent to $C$. Hence, since $\xi=f u^a$ for some function $f$ on $\mathcal{H}^+$, we have $d\xi=\beta' u\wedge l+\gamma \xi\wedge V=\beta' \bm{\tilde{\epsilon}}+\gamma \xi\wedge V$ (we have also defined $\beta'=f \beta$). In index notation, $\nabla_{[a}\xi_{b]}=\beta'\bm{\tilde{\epsilon}}_{ab}+\gamma \xi_{[a}V_{b]}$. Using Killing's equation, the LHS is $\nabla_a\xi_b$. So, if we contract this equation with $\xi^a$, we have (using also $\xi^a\bm{\tilde{\epsilon}}_{ab}=\xi^a$, $\xi^2=0$, and $\xi\cdot V= fu\cdot V=0$ because $V$ is tangent to $C$ while $u$ is orthogonal to $C$), $\nabla_a\xi_b=\beta'\xi^a$. Comparing this equation with the well-known expression $\nabla_a\xi_b=\kappa\xi^a$ on $\mathcal{H}^+$ for the surface gravity $\kappa$, we have that $\beta'$ is the surface gravity on $C$. So, on an arbitrary cross-section $C$ of $\mathcal{H}^+$
\begin{equation}
\nabla_a\xi_b=\kappa\bm{\tilde{\epsilon}}_{ab}+\gamma \xi_{[a}V_{b]},
\end{equation}
where $V$ is an arbitrary vector tangent to $C$. In particular, if $C$ is the bifurcation surface $B$ where $\xi=0$, we have
\begin{equation}
\nabla_a\xi_b=\kappa\bm{\tilde{\epsilon}}_{ab} \quad \text{on $S$.}
\end{equation}
This proves eq.\eqref{eq:app}.

Let us consider now the case in which the black hole spacetime $M$ has 4 dimensions and let $\bm{\epsilon}_{abcd}$ be the volume form on $M$. For an arbitrary cross-section $C$ of the future event horizon $\mathcal{H}^+$ with binormal $\bm{\tilde{\epsilon}}_{cd}$, let us consider the quantity
\begin{equation}
\label{eq:eqsu}
\bm{\epsilon}_{ab}^{\;\;\;\;cd}\bm{\tilde{\epsilon}}_{cd}=2\bm{\epsilon}_{abcd}u^c l^d.
\end{equation} As suggested in appendix B of \cite{Wald:1984rg}, we define the volume form $\bm{\epsilon'}_{abc}$ on $\mathcal{H}^+$ by requiring that $\bm{\epsilon'}_{abc}$ is in the orientation class defined by Stokes' theorem and satisfies
\begin{equation}
\frac{1}{4}\bm{\epsilon}_{abcd}=u_{[a}\bm{\epsilon'}_{bcd]} \quad \text{on $\mathcal{H}^+$.}
\end{equation}
This gives (see again appendix B of \cite{Wald:1984rg})
\begin{equation}
\label{eq:eq}
\begin{split}
l^a\bm{\epsilon}_{abcd}&=(l^a u_a)\bm{\epsilon'}_{bcd}\\
&=-\bm{\epsilon'}_{bcd} \quad \text{on $\mathcal{H}^+$.}
\end{split}
\end{equation}
Let $v^a$ be the unit normal to $C$ in $\mathcal{H}^+$ ($v^a$ is timelike w.r.t. to the metric induced on $\mathcal{H}^+$). Since it is a vector orthogonal to $C$, it must be a combination of $u^a$ and $l^a$. Using the antisymmetry of $\bm{\epsilon}_{abcd}$, we have 
$l^av^b\bm{\epsilon}_{abcd}=l^a u^b\bm{\epsilon}_{abcd}$ on $C$. But, from \eqref{eq:eq}, we also have $l^av^b\bm{\epsilon}_{abcd}=-v^b\bm{\epsilon'}_{bcd}$ on $C$. Hence,
\begin{equation}
\begin{split}
l^a u^b\bm{\epsilon}_{abcd}&=-v^b\bm{\epsilon'}_{bcd}\\
&=\bm{\mathring{\epsilon}}_{cd},
\end{split}
\end{equation}
where $\bm{\mathring{\epsilon}}_{cd}$ is the volume form on $C$ induced by the volume form on $\mathcal{H}^+$, given by $v^b\bm{\epsilon'}_{bcd}=-\bm{\mathring{\epsilon}}_{cd}$ (using again the results of appendix B of \cite{Wald:1984rg}). Rearranging the indices, we have $\bm{\epsilon}_{cdab}u^a l^b=-\bm{\mathring{\epsilon}}_{cd}$. Hence, from \eqref{eq:eqsu}, we find
\begin{equation}
\bm{\epsilon}_{ab}^{\;\;\;\;cd}\bm{\tilde{\epsilon}}_{cd}=-2\bm{\mathring{\epsilon}}_{ab} \quad \text{on $C$.}
\end{equation}
This proves \eqref{eq:SGR}.

\subsection{Invariance under reflections about the bifurcation surface}
\label{sec:inv}

We now want to introduce the concept of invariance under reflections about the bifurcation surface and prove Lemma 2.3 of \cite{Kay:1988mu}. We will need some preliminary results. Let us consider the bifurcation surface $B$ of a bifurcate Killing horizon $\mathcal{H}$ w.r.t. the vector field $\xi^a$. We choose our convention so that the integral curves of $\xi$ are future-directed on $\mathcal{H}^+_R$ and $\mathcal{H}^-_R$, and past-directed on $\mathcal{H}^+_L$ and $\mathcal{H}^-_L$ (we are free to make this choice since, by the definition of Killing horizon, the direction of $\xi$ on the various regions of $\mathcal{H}$ does not affect the fact that $\xi$ is the Killing field generating the Killing horizon). This convention implies that $\xi^a= f u^a$, where $u^a$ is the tangent to the affinely parameterised null geodesic generators of $\mathcal{H}^+$, with $f$ positive on $\mathcal{H}^+_R$ and negative on $\mathcal{H}^+_L$.
Let $u$ be the affine parameter of the generators of $\mathcal{H}^+$ (so $u^a=\bigl(\frac{\partial}{\partial u}\bigr)^a$ on $\mathcal{H}^+$) chosen so that $u=0$ on $B$. 

Consider for now the portion $\mathcal{H}^+_R$.
Let then $s$ be the parameter of integral curves of $\xi$ on $\mathcal{H}^+_R$ (so $\xi^a=\bigl(\frac{\partial}{\partial s}\bigr)^a$ on $\mathcal{H}^+_R$) chosen so that $s=0$ on the $(n-2)$-surface of the Killing horizon given by $u=1$. We also have that $s\to-\infty$ when we approach $B$ because $\xi=0$ on $B$, i.e. $\partial_s g=0$ on $B$ for any smooth function $g$, which means that every smooth function as an horizontal asymptote at the value of $s$ corresponding to $B$. This is possible only if $s$ has an infinite value at $B$. Since $s$ increases when we move away from $B$ along integral curves of $\xi$, we must have $s\to-\infty$ on $B$. Now, we know that $\xi=f u^a$ on $\mathcal{H}^+$, and, as shown in \cite{Reall:2017bhnotes}, the surface gravity $\kappa$ on the horizon satisfies $\kappa=\xi^a\partial_a\log |f|=\frac{\partial \log |f|}{\partial s}$. Notice that $\kappa$ is always positive. On $\mathcal{H}^+_R$ we have $f>0$. Assuming that a form of the zeroth law is valid, i.e. $\kappa$ is constant on the horizon, we have $f=Ce^{\kappa s}$ on $\mathcal{H}^+_R$, where $C$ does not depend on $s$, i.e. it is constant along integral curves of $\xi$. 
Comparing $\xi^a=f u^a$, i.e. $\bigl(\frac{\partial}{\partial s}\bigr)^a=f \bigl(\frac{\partial}{\partial u}\bigr)^a$, with $\bigl(\frac{\partial}{\partial s}\bigr)^a= \frac{\partial u}{\partial s}\bigl(\frac{\partial}{\partial u}\bigr)^a$, we have
\begin{equation}
\frac{\partial u}{\partial s}=f=Ce^{\kappa s},
\end{equation}
so
\begin{equation}
u=\frac{C}{\kappa}e^{\kappa s}+C',
\end{equation}
where $C'$ does not depend on $s$. The condition $u=0$ on $B$ (together with $s\to-\infty$ at $B$) gives $C'=0$. The condition $s=0$ when $u=1$ gives $C=\kappa$. So
\begin{equation}
\label{eq:us}
u=e^{\kappa s}.
\end{equation}

$u^a$ is defined everywhere on $\mathcal{H}^+$, so we can repeat the same construction for $\mathcal{H}^+_L$. 
Let $s'$ be the parameter of integral curves of $\xi$ on $\mathcal{H}^+_L$(so $\xi^a=\bigl(\frac{\partial}{\partial s'}\bigr)^a$ on $\mathcal{H}^+_L$) chosen so that $s'=0$ on the $(n-2)$-surface of the Killing horizon given by $u=-1$.
We again have $s'\to-\infty$ as we approach $B$. On $\mathcal{H}^+_L$ we have $f=-Ce^{\kappa s'}$ and
\begin{equation}
\label{eq:us'}
u=-e^{\kappa s'}.
\end{equation}

Above we have defined $l^a$ on $\mathcal{H}^+$ as the other future-directed null normal on the cross-sections of $\mathcal{H}^+$ normalised so that $l\cdot u=-1$. In particular, $l^a$ is defined on $B$. We can now extend $l$ to $\mathcal{H}^-$ by taking it to be tangent to affinely parameterised null geodesics generators of $\mathcal{H}^-$, i.e. the solution of $l^b\nabla_b l^a$ that coincides with the $l^a$ defined above at $B$. (We could also extend $u^a$ to $\mathcal{H}^-$ by taking the other future-directed null normal normalised so that $l\cdot u=-1$.) Then, we define the surface gravity $\kappa$ on $\mathcal{H}^-$ by
\begin{equation}
\nabla_a(\xi^b\xi_b)=+2\kappa \xi_a.
\end{equation}
On $\mathcal{H}^-$ we have $\xi^a=g l^a$, where, for our conventions, $g$ is negative on $\mathcal{H}^-_L$ and positive on $\mathcal{H}^-_R$.
Let $l$ be the affine parameter of the generators of $\mathcal{H}^-$ (so $l^a=\bigl(\frac{\partial}{\partial l}\bigr)^a$ on $\mathcal{H}^-$) chosen so that $l=0$ on $B$.
The surface gravity can be written as
\begin{equation}
\kappa=-\xi^a\partial_a\log |g|.
\end{equation}
For our conventions $\kappa$ is still always positive on $\mathcal{H}^-$.

Now we can repeat the argument above. Consider the portion $\mathcal{H}^-_L$.
Let $r$ be the parameter of integral curves of $\xi$ on $\mathcal{H}^-_L$(so $\xi^a=\bigl(\frac{\partial}{\partial r}\bigr)^a$ on $\mathcal{H}^-_L$) chosen so that $r=0$ on the $(n-2)$-surface of the Killing horizon given by $l=1$. $r$ goes to $+\infty$ as we approach $B$. We have $\kappa=-\xi^a\partial_a\log |g|=-\frac{\partial \log |g|}{\partial r}$. On $\mathcal{H}^-_L$ $g<0$ so, assuming that a form of the zeroth law is valid, we have $g=-Ce^{-\kappa r}$ on $\mathcal{H}^-_L$, where $C$ does not depend on $r$, i.e. it is constant along integral curves of $\xi^a$. 
Comparing $\bigl(\frac{\partial}{\partial r}\bigr)^a=g \bigl(\frac{\partial}{\partial l}\bigr)^a$, with $\bigl(\frac{\partial}{\partial r}\bigr)^a= \frac{\partial l}{\partial r}\bigl(\frac{\partial}{\partial l}\bigr)^a$, we have
\begin{equation}
\frac{\partial l}{\partial r}=g=-Ce^{-\kappa r},
\end{equation}
so
\begin{equation}
l=\frac{C}{\kappa}e^{-\kappa r}+C',
\end{equation}
where $C'$ does not depend on $r$. The condition $l=0$ on $B$ (together with $r\to\infty$ at $B$) gives $C'=0$. The condition $r=0$ when $l=1$ gives $C=\kappa$. So
\begin{equation}
\label{eq:lr}
l=e^{-\kappa r}.
\end{equation}

Finally, we repeat the reasoning for $\mathcal{H}^-_R$.
Let $r'$ be the parameter of integral curves of $\xi$ on $\mathcal{H}^-_R$(so $\xi^a=\bigl(\frac{\partial}{\partial r'}\bigr)^a$ on $\mathcal{H}^-_L$) chosen so that $r'=0$ on the $(n-2)$-surface of the Killing horizon given by $l=-1$.
We also have $r'\to\infty$ as we approach $B$. Then, $\kappa=-\xi^a\partial_a\log |g|=-\frac{\partial \log |g|}{\partial r'}$. On $\mathcal{H}^-_R$ we have $g>0$. So, assuming that a form of the zeroth law is valid, we have $g=Ce^{-\kappa r'}$ on $\mathcal{H}^-_R$, where $C$ does not depend on $r'$, i.e. it is constant along integral curves of $\xi$. 
Hence, we find $\frac{\partial l}{\partial r'}=g=Ce^{-\kappa r'}$, so
\begin{equation}
l=-\frac{C}{\kappa}e^{-\kappa r'}+C',
\end{equation}
where $C'$ does not depend on $r'$. The condition $l=0$ on $B$ (together with $r'\to\infty$ at $B$) gives $C'=0$. The condition $r=0$ when $l=-1$ gives $C=\kappa$. So
\begin{equation}
\label{eq:lr'}
l=-e^{-\kappa r'}.
\end{equation}

Now we are ready to prove Lemma 2.3 of \cite{Kay:1988mu}. Let us define the reflection map $i$ about $B$ as the diffeomorphism that maps each point on a geodesic $\gamma$ orthogonal to $B$ to the ``reflected'' point on $\gamma$ about $B$, i.e., for each point affine parameter distance $u$ from $B$ on $\gamma$, $i(p)$ is the point affine parameter $-u$ from $B$ on $\gamma$. $i$ is certainly well-defined in a neighbourhood of $B$. Let us then consider the 1-parameter family of diffeomorphisms $\chi_t$ generated by $\xi^a$. Consider its action, for example, on points of $\mathcal{H}^+_R$. A point $p$ along an integral curve of $\xi$ with parameter $s$ is mapped to the point $\chi_t(p)$ with parameter $s+t$ along the same curve. Since points on $B$ have $s\to-\infty$, these diffeomorphisms leave the points on $B$ untouched. 

\noindent
\textbf{Lemma}[Lemma 2.3 of \cite{Kay:1988mu}] For any tensor at $B$ the invariance under $(\chi_t)_\ast$ is equivalent to invariance under $i_\ast$.

\noindent
\emph{Proof.} Let us choose a basis for the tangent space at any point $p\in B$ composed of the two null normals $u^a$, $l^a$ and $n-2$ vectors $v_1,\dots,v_{n-1}$ tangent to $B$. From the results obtained above, we have
\begin{equation}
(\chi_t)_\ast u^a=e^{\kappa t}u^a \qquad (\chi_t)_\ast l^a=e^{-\kappa t}l^a
\end{equation}
at any point $p\in B$.
We prove only the first relation. The second one can be shown in a similar way. Let us consider the case in which $p$ is an arbitrary point in the intersection of $B$ and $\mathcal{H}^+_R$. We build a Gaussian null coordinate chart $\psi$ on a neighbourhood of $\mathcal{H}^+$ as follows. Let $y^i=(y^1,\dots,y^{n-2})$ be coordinates on $B$ such that $y^i(p)=0$. Associate coordinates $(u,y^i)$ to the point affine parameter distance $u$ from $B$ along the generator of $\mathcal{H}^+$ which intersects $B$ at the point with coordinates $(y^i)$. Then, associate coordinates $(l,u,y^i)$ to the point affine parameter distance $u$ from $\mathcal{H}^+$ along the null affinely parameterised geodesics starting at the point on $\mathcal{H}^+$ with coordinates $(u,y^i)$ and tangent vector $l^a$ there. In this chart, $x^\mu(p)=(0,0,0,\cdots,0)$. Let us consider the vector $u^a$ at $p$. $p$ lies on a geodesic generator of $\mathcal{H}^+$, $\lambda(u)$, whose tangent at $p$ is $u^a$.
In Gaussian null coordinates, this curve is $x^\mu(\lambda(u))=(0,u,0,\cdots,0)$. For any function $f\colon M\to\mathbf{R}$, we have
\begin{equation}
u (f)=\left.\frac{d(f\circ\lambda(u))}{du}\right\vert_{u=0}=\left.\frac{\partial F(x)}{\partial x^\mu}\right\vert_{\psi(p)}\left.\frac{d x^\mu(\lambda(u))}{du}\right\vert_{u=0}=\left.\frac{\partial F(x)}{\partial u}\right\vert_{\psi(p)},
\end{equation}
where $F(x)=f\circ\psi^{-1}$.
Now we consider $(\chi_t)_\ast u^a$. Since $p$ is not changed by $\chi_t$, this is still a vector at $p$. By definition of push-forward, we have
\begin{equation}
(\chi_t)_\ast u (f)=\left.\frac{d(f\circ\chi_t\circ\lambda(u))}{du}\right\vert_{u=0}.
\end{equation}
$\chi_t$ has the effect of shifting the parameter $s$ along the integral curves of $\xi$ by $t$. From eq. \eqref{eq:us}, we see that $u(s)$ becomes then $u(s+t)=e^{\kappa t}u(s)$, so we have $\chi_t(\lambda(u))=\lambda(e^{\kappa t} u)$. In Gaussian null coordinates, $x^\mu(\lambda(e^{\kappa t}u))=e^{\kappa t}(0,u,0,\cdots,0)$. Therefore,
\begin{equation}
(\chi_t)_\ast u (f)=\left.\frac{d(f\circ\chi_t\circ\lambda(e^{\kappa t} u))}{du}\right\vert_{u=0}=\left.\frac{\partial F(x)}{\partial x^\mu}\right\vert_{\psi(p)}\left.\frac{d x^\mu(e^{\kappa t}\lambda(u))}{du}\right\vert_{u=0}=\left.\frac{\partial F(x)}{\partial u}\right\vert_{\psi(p)}e^{\kappa t}.
\end{equation}
$f$ is arbitrary and $(\chi_t)_\ast u^a$ and $u^a$ are vectors at the same point, so we have $(\chi_t)_\ast u^a=u^a$ on any point at the intersection of $B$ and $\mathcal{H}^+_R$. We can repeat this argument also for any point $p$ at the intersection of $B$ and $\mathcal{H}^+_L$, choosing this time Gaussian null coordinates such that $y^i$ vanish at $p$. Employing also \eqref{eq:us'}, we obtain the same result. This proves the first relation. The second relation is proved in an analogous way considering points at $B\cap\mathcal{H}^-_{L,R}$, defining suitable Gaussian null coordinates and using \eqref{eq:lr} and \eqref{eq:lr'}.

Moreover, we can similarly prove that
\begin{equation}
(\chi_t)_\ast v_1=v_1, \quad (\chi_t)_\ast v_2=v_2, \quad \cdots, \quad (\chi_t)_\ast v_{n-2}=v_{n-2},
\end{equation} 
by choosing a curve with tangent $v_i$ and only non-vanishing coordinate $y^i$ and using the fact that $\chi_t$ does not move the points of a curve in $B$.

The action of $i_\ast$ on the basis vectors is
\begin{align}
&i_\ast u^a=-u^a, \quad i_\ast l^a=-l^a\\
&i_\ast v_1^a=v_1^a, \quad i_\ast v_2^a=v_2^a, \quad \cdots \quad  i_\ast v_{n-2}^a=v_{n-2}^a.
\end{align}
The proof can be simply obtained by the definition of $i$, the choice of a helpful set of Gaussian null coordinates, and following the steps explained above.

Let us consider now an ${r}\choose{s}$ tensor $T^{a_1\cdots a_r}_{\;\;\;\;\;\;\;\;\;\;b_1\cdots b_s}$ at a point $p\in B$ and expand this in terms of the basis vectors. From our results for the action of $(\chi_t)_\ast$ on the basis vectors, it follows immediately that $(\chi_t)_\ast T=T$ if and only if the only non-vanishing coefficients in the expansion are those that multiply terms with the same number of $u^a$ and $l^a$. Moreover, from the results for the action of $i_\ast$ on the basis vectors, it follows that $i_\ast T=T$ if and only if the only non-vanishing coefficients in the expansion are those that multiply terms with the same number of $u^a$ and $l^a$, i.e. the same terms as above.
Hence, for tensors on $B$ the invariance under $(\chi_t)_\ast$ is equivalent to invariance under $i_\ast$. $\square$

\addtocontents{toc}{\protect\enlargethispage*{1.5\baselineskip}}

\label{cap:biblio}
\bibliographystyle{JHEP}
\bibliography{Rossi-First_law_BH_mechanics}

\end{document}